\documentclass[extra,onecolumn]{gji}

\usepackage{timet}
\usepackage[fleqn]{amsmath}
\usepackage{amssymb}
\usepackage{bm}
\usepackage{ulem}
\usepackage{graphicx}
\usepackage{subfig}
\usepackage{xcolor}
\usepackage{microtype}

\usepackage{general_sty}
\usepackage{mathematical_shorthands}
\usepackage{continuum_mechanics_definitions}
\usepackage{equilibrium_stress_sty}

\numberwithin{equation}{section}

\begin{document}

\title[On the stress dependence of the elastic tensor]{On the stress dependence of the elastic tensor}

\author[M. A. Maitra \& D. Al-Attar]
  {Matthew Maitra \& David Al-Attar \\
  Department of Earth Sciences, Bullard Laboratories, University of Cambridge, \\
  Madingley Road, Cambridge CB3 OEZ, United Kingdom. \\
  Email: mam221@cam.ac.uk}

\date{\today}

\maketitle

\begin{summary}	
	
The dependence of the elastic tensor on the equilibrium stress is investigated theoretically. Using ideas from finite elasticity, it is  first shown that both the equilibrium stress and elastic tensor are given uniquely in terms of the equilibrium deformation gradient relative to a fixed choice of reference body. Inversion of the relation between the deformation gradient and stress might, therefore, be expected to lead neatly to the desired expression for the elastic tensor. Unfortunately, the deformation gradient can only be recovered from the stress up to a choice of rotation matrix. Hence it is not possible in general to express the elastic tensor as a unique function of the equilibrium stress. By considering material symmetries, though, it is shown that the degree of non-uniqueness can sometimes be reduced, and in some cases even removed entirely. These results are illustrated through a range numerical calculations, and we also obtain linearised relations applicable to small perturbations in equilibrium stress. Finally, we make a comparison with previous studies before considering implications for geophysical forward- and inverse-modelling.
\end{summary}

\begin{keywords}
	Theoretical seismology; elasticity and anelasticity; seismic anisotropy.
\end{keywords}

%%%%%%%%%%%%%%%%%%%%%%%%%%%%%%%%%%%%%%%%%%%%%%%%%%%%%%%%
\section{Introduction}
\label{sec:intro}

Approaches to seismic wave propagation within a pre-stressed Earth have a long and complicated history; see \citet[Chapter 1]{dahlen1998theoretical}. Early work on theoretical seismology \citep[e.g.][]{Thomson_1863a,Lamb_1881} was built on the theory of classical linear elasticity. But classical linear elasticity is founded upon an assumption of small deformations away from a stress-free equilibrium. Its applicability to seismology is therefore unclear, given the presence of large equilibrium stresses within the Earth. In fact, it was not until the work of \citet{dahlen_1972a} that a correct treatment was given.

\citet{dahlen_1972a} derived the equations of motion relevant to global seismology by direct linearisation of the equations of finite elasticity. It is a result of this approach that the elastic tensor can be written as the sum of two pieces: one without explicit stress dependence, and a second piece that depends on stress linearly. There is no question that this decomposition is valid and that Dahlen's equations of motion are correct, but the decomposition is not unique in that the elastic tensor's stress dependence is left (partially) implicit. 

Later that year \citet{dahlen_1972b} used his earlier results to study plane-wave propagation in the presence of an arbitrary equilibrium pre-stress. To do so, he theorised that the elastic tensor's stress dependence should take a particular functional form. He assumed specifically that the only stress dependence was what his earlier formulae had made \textit{explicit}. That theory has two important implications. Firstly, the elastic wave speeds display no explicit dependence on equilibrium pressure. Secondly, deviatoric stresses induced within an isotropic medium have no first-order effect on P-wave speeds, whilst S-waves are split to the same accuracy. This is illustrated in the left panel of Fig.\ref{fig:dahlen_slowness}; as with all the figures in this paper, the values of the physical quantities have been chosen for the sake of illustration and are not necessarily geophysically realistic.
\iffair\begin{figure}
  \includegraphics[width=\textwidth]{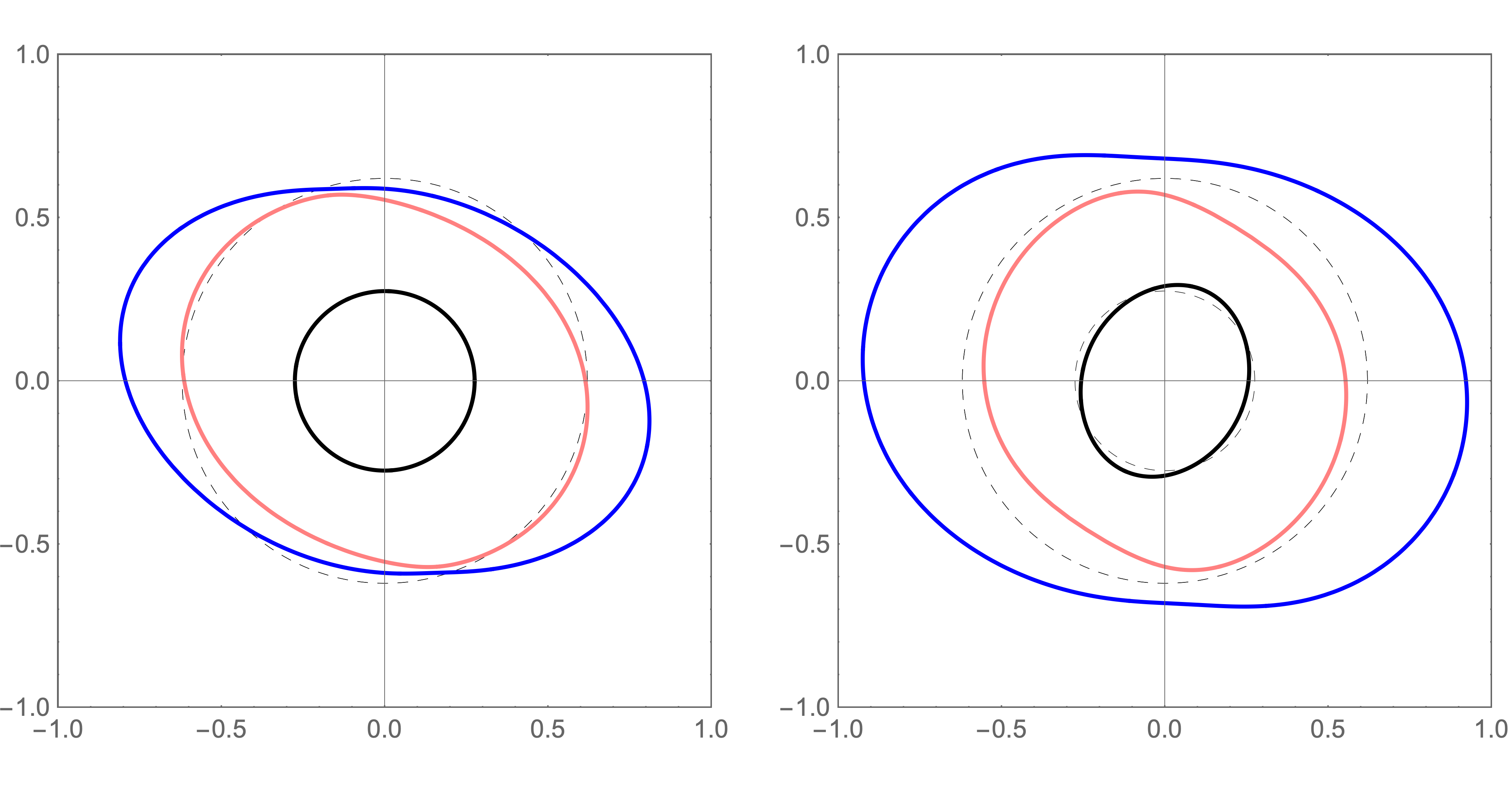}
  \caption{The effect of an induced deviatoric stress on plane-wave propagation within an initially isotropic body. Each panel shows a slice through the \( x \)--\( y \) plane of the slowness surface of an isotropic material in which a stress has been induced. The zero-stress result is shown dashed in the background for reference. The material on the left follows the theory of \citet{dahlen_1972b}. Whilst the S-waves are noticeably split the P-wave surface has not perceptibly  moved from its isotropic position, illustrating Dahlen's result that P-wave speeds are not changed to first-order by a small stress. By contrast, the material on the right obeys \citet{Tromp_2018}, and there we see that P-wave speeds are indeed perturbed to first-order in small stress. There is also a marked difference in the S-wave splitting pattern. In this figure we have taken \( \mu'=\kappa'=1 \), consistent with \citet{Stacey_1992}.}
  \label{fig:dahlen_slowness}
\end{figure}\fi

The problem of the elastic tensor's stress dependence has since been revisited by \citet{Tromp_2018} who were motivated by the possibility of using seismic data to image stresses within the Groningen gas field. Importantly, they arrived at a theory that predicts physical behaviour both quantitatively and qualitatively distinct from that derived by Dahlen. In direct contravention of Dahlen, \citeauthor{Tromp_2018} suggest that \textit{both} P- and S-wave speeds change to first-order if a small deviatoric stress is induced in an isotropic medium (Fig.\ref{fig:dahlen_slowness}, right panel).
\iffair\begin{figure}
  \includegraphics[width=\textwidth]{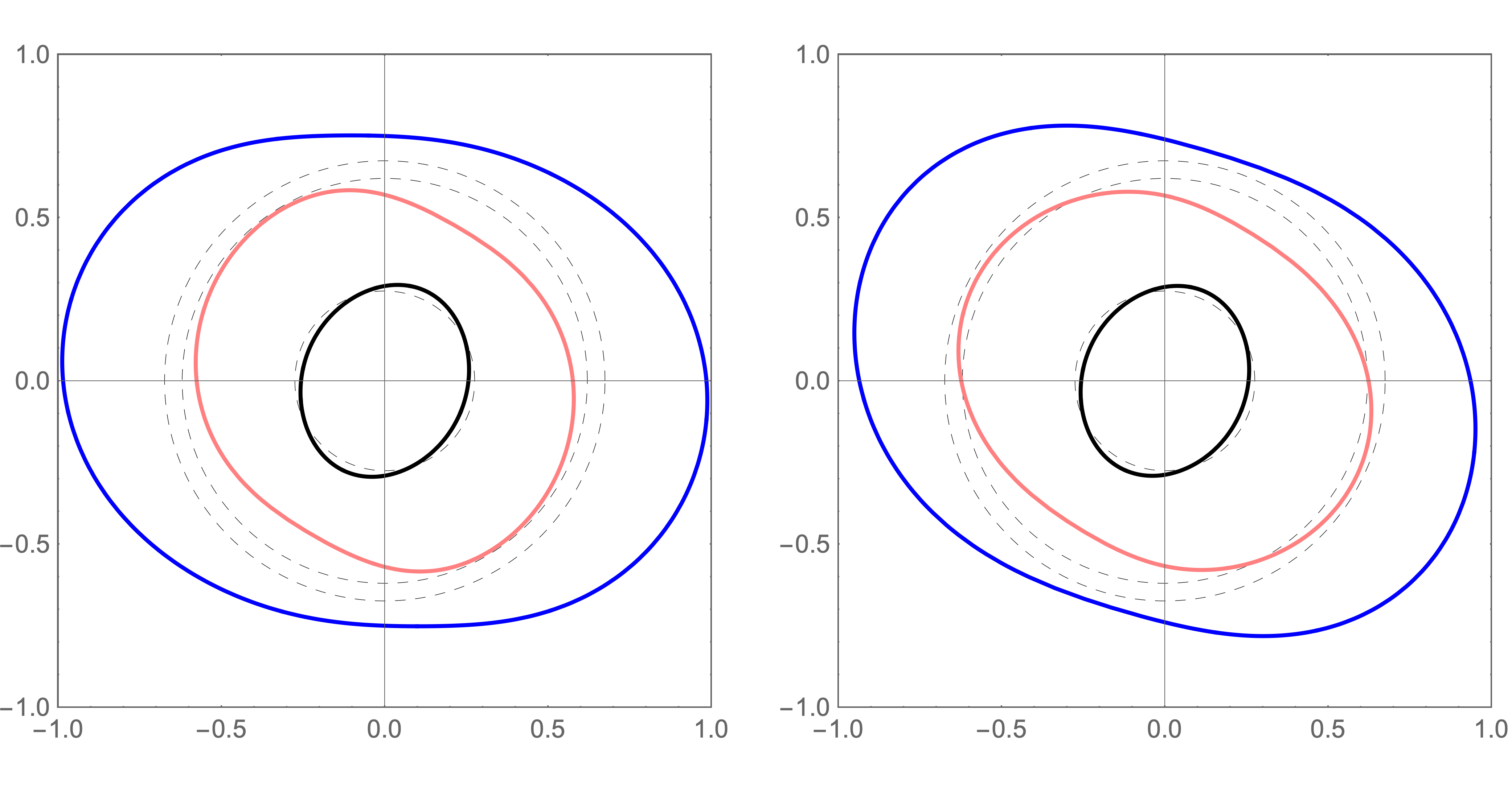}
  \caption{The effect of an induced deviatoric stress on plane-wave propagation within a body that is initially transversely-isotropic with its symmetry axis pointing out of the page. This figure is similar to Fig. \ref{fig:dahlen_slowness}, but now we are comparing the theories of \citet{Tromp_2018} (left) and \citet{Tromp_2019} (right). The theories give the same results for an isotropic material, so we have chosen a transversely-isotropic material in order to demonstrate that they predict distinct behaviour in general.}
  \label{fig:tromp_slowness}
\end{figure}\fi

The most recent work on the problem is that of \citet{Tromp_2019}. These authors not only generalised \citeauthor{Tromp_2018}'s results beyond the framework provided by \citet{dahlen1998theoretical} (see below) but also included comparisons between their new theoretical results and \textit{ab initio} calculations. Their results reduce to those of \citet{Tromp_2018} for an isotropic body, but make different predictions otherwise. Fig. \ref{fig:tromp_slowness} contrasts the two theories as applied to a transversely-isotropic body.

The seismological community is left with three distinct theories for the effect of equilibrium stress on seismic wave propagation. This has led us to undertake the present work, which revisits the elastic tensor's stress dependence and seeks to clarify it from the perspective of finite elasticity. It should be emphasised that the work of \citet{dahlen_1972a} -- which has underlain most of global seismology and related fields since its publication -- is fully correct and general. The problem that we wish to address concerns only the elastic tensor's dependence on equilibrium stress. To provide more specific context we will begin by presenting a heuristic approach to the problem that slightly generalises the previous \wording{discussions} and points towards their limitations.

\subsection{A heuristic linearised theory of the elastic tensor's stress dependence}
\label{intro:heuristic}

\subsubsection{Preliminaries}

In the notation of \citet{dahlen1998theoretical} the equations of motion governing global seismology are \citep[][p.84]{dahlen1998theoretical}
\begin{align}
\label{eq:momentum_equation_dt}
  \partial_{t}^{2}\sssss 
  -\frac{1}{\rho^{0}}\nabla\cdot\brak{\Lam :\! \nabla\sssss}  
  + 2\Omb\times\partial_{t}\sssss
  + \nabla\phi^{E1}
  +\sssss\cdot\nabla\nabla\brak{\phi^{0}+\psi} = 0 ,
\end{align}
for an Earth model initially at equilibrium with density \( \rho^{0} \) and gravitational potential \( \phi^{0} \), and which rotates at constant angular velocity \( \Omb \) giving rise to the centrifugal potential \( \psi \). The displacement from equilibrium is \( \sssss \), with \( \phi^{E1} \) the corresponding perturbation to the gravitational potential. Most importantly for our purposes, the Earth model is assumed to be pre-stressed, supporting a nonzero \textit{equilibrium Cauchy stress} \( \TT^{0} \), while \( \Lam \) is the \textit{elastic tensor}. (\wording{Henceforward} we will always refer to equilibrium Cauchy stress simply as \wording{``stress"} unless we state otherwise or the context offers no possibility of confusion; see Section \ref{constitutive_relations} for a discussion of different stress tensors.) As discussed at length in \citeauthor{dahlen1998theoretical}'s Section 3.6, \( \Lam \) is one of a number of elastic tensors that can be used depending on how the equations of motion are formulated. It is particularly relevant for us that there exists another elastic tensor \( \Xib \) related to \( \Lam \) by \citep[][eq.3.122]{dahlen1998theoretical}
\begin{align}
\label{eq:lambda_basic_decomposition}
  \Lambda_{ijkl} = \Xi_{ijkl} + T^{0}_{ik}\delta_{jl} .
\end{align}
\( \Xib \) is obtained as the second derivative of a strain-energy function. It therefore satisfies the \textit{hyperelastic symmetry}
\begin{align}
  \Xi_{ijkl} = \Xi_{klij}
\end{align}
and the \textit{minor symmetries}
\begin{align}
  \Xi_{ijkl} = \Xi_{jikl} = \Xi_{ijlk}
\end{align}
(collectively referred to as the \textit{classical elastic symmetries}) and thus possesses 21 independent components at most. We see that \( \Lam \) is decomposed into two parts, one with explicit, linear stress dependence, and another whose dependence on stress is \textit{a priori} unclear. The purpose of the present work is to establish how \( \Xib \) -- and hence \( \Lam \) -- depends on \( \TT^{0} \).

%\edit{Following sentence not technically true (amplitudes not done by Christoffel operator):} The propagation of small-amplitude seismic waves within the medium is described by the \textit{Christoffel operator} \( \mathbf{B} \), which is defined in terms of \( \rho \), \( \Lam \) and the propagation direction \( \hat{\mathbf{k}} \) as \citep[][eq.3.155]{dahlen1998theoretical}
%\begin{align}
%\label{eq:christoffel_intro}
%  \rho B_{jl} = \Lambda_{ijkl}\hat{k}_{i}\hat{k}_{k} .
%\end{align}
%Its eigenvalues (resp. eigenvectors) give the possible wave speeds (resp. polarisations) at each point.

For geophysical applications it seems reasonable to restrict attention to linearised stress dependence about a hydrostatic background; finding the linearised stress dependence of \( \Xib \) is necessary and sufficient to specify that of \( \Lam \). We regard \( \TT^{0} \) as being composed of a \textit{background hydrostatic stress} \( \TT^{B} \) described by a large pressure \( p^{0} \), with a small \textit{incremental stress} \( \Delta\TT^{0} \) superimposed. The total equilibrium stress is thus written as
\begin{align}
  \TT^{0} &= \TT^{B} + \Delta\TT^{0} \nonumber\\
  		  &= -p^{0}\id - \Delta p^{0}\id + \Delta\DD^{0} ,
\end{align}
where \( \Delta p^{0} \) and \( \Delta\DD^{0} \) are the hydrostatic and deviatoric components of \( \Delta\TT^{0} \). We will use this notation consistently for the rest of this section, which means that some of the expressions we quote will look slightly different from how they appear elsewhere. Now we ask what the most general linear stress dependence of \( \Xib \) could be. To that end we write \( \Xib \) as a Taylor series,
\begin{align}
\label{eq:pi_def}
  \Xi_{ijkl} 
  &= \Gamma_{ijkl} + \Pi_{ijklmn}\Delta T^{0}_{mn} + \mathcal{O}\brak{\left\|\Delta \TT^{0}\right\|^{2}} ,
\end{align}
and neglect all terms of quadratic and higher order in \( \left\|\Delta \TT^{0}\right\| \). \( \Gam \) is then the elastic tensor of the background hydrostatic state while \( \Pib \) represents the elastic tensor's stress-derivatives evaluated in that background state. To first-order accuracy a body's response to changes in incremental stress is thus determined entirely by \( \Pib \). Decomposing the stress into its hydrostatic and deviatoric parts,
\begin{align}
  \Delta T^{0}_{ij} = -\Delta p^{0}\delta_{ij}+\Delta\tau^{0}_{ij} ,
\end{align}
we can write
\begin{align}
  \Xi_{ijkl} = \Gamma_{ijkl} - \Pi_{ijklaa}\Delta p^{0} + \Pi_{ijklmn}\Delta\tau^{0}_{mn} ,
\end{align}
based on which we define the pressure-derivatives of the elastic tensor as
\begin{align}
\label{eq:xi_pressure_deriv}
  \frac{\partial\Xi_{ijkl}}{\partial p} \equiv \Xi'_{ijkl} = -\Pi_{ijklaa}
\end{align}
and its derivatives with respect to deviatoric stress as
\begin{align}
\label{eq:xi_deviatoric_stress_deriv}
    \frac{\partial\Xi_{ijkl}}{\partial \tau_{mn}} &= \Pi_{ijklmn}-\frac{1}{3}\Pi_{ijklaa}\delta_{mn} .
\end{align}
Both \( \Xib \) and \( \Gam \) must possess the full set of classical elastic symmetries, and the symmetry of the Cauchy stress means that \( \Pib \) may be taken as symmetric on its last two indices without loss of generality. Therefore \( \Pib \) is required at minimum to obey the relations
\begin{align}
\label{eq:pi_syms}
  \rlap{$\overbrace{\phantom{\Pi_{ijklnm} = \Pi_{ijklmn}}}^{\text{Symmetry of }\TT^{0}}$}\Pi_{ijklnm} = \underbrace{\Pi_{ijklmn} = \Pi_{jiklmn} = \Pi_{ijlkmn} = \Pi_{klijmn}}_{\text{Classical elastic symmetries}} ,
\end{align}
from which one can show that it possesses at most \( 21 \times 6 = 126 \) independent components.

\subsubsection{Isotropic materials}
\label{intro_isotropic}

In an isotropic, hydrostatically pre-stressed material neither \( \Gam \) nor \( \Pib \) can exhibit any preferred direction. This is true if \textit{and only if} the material is isotropic. No matter the value of the background hydrostatic pressure, \( \Gam \) therefore has just two independent components, the \textit{bulk modulus} \( \kappa \) and \textit{shear modulus} \( \mu \). According to convenience one can also use the relation
\begin{align}
  \kappa = \lambda+\frac{2}{3}\mu
\end{align}
to write \( \Gam \) in terms of the \textit{Lam\'{e} parameters} \( \lambda \) and \( \mu \). In terms of the elastic moduli \( \Gam \) takes the familiar form
\begin{align}
\label{eq:gamma_iso}
  \Gamma_{ijkl} = \brak{\kappa-\frac{2}{3}\mu}\delta_{ij}\delta_{kl}
  		+\mu\brak{\delta_{ik}\delta_{jl}
  		         +\delta_{il}\delta_{jk}} .
\end{align}
The tensor \( \Pib \) also has considerably fewer than 126 components. It too will only be composed of Kronecker deltas, and the most general such tensor that also satisfies the required symmetries (eq.\ref{eq:pi_syms}) has four independent components. Thus
\begin{align}
\label{eq:pi_iso_expression}
  \Pi_{ijklmn} &= 
   a \braksq{
   		 \delta_{ij}\delta_{k(m}\delta_{n)l}
   	    +\delta_{kl}\delta_{i(m}\delta_{n)j}
   	    -\frac{2}{3}\delta_{ij}\delta_{kl}\delta_{mn}}
  \nonumber\\
  &\qquad
  +b \braksq{
  		 \delta_{ik}\delta_{j(m}\delta_{n)l}
  		+\delta_{il}\delta_{j(m}\delta_{n)k}
  		+\delta_{jk}\delta_{i(m}\delta_{n)l}
  		+\delta_{jl}\delta_{i(m}\delta_{n)k}
  		-\frac{2}{3}\brak{\delta_{ik}\delta_{jl}
  		         +\delta_{il}\delta_{jk}}\delta_{mn}}
  \nonumber\\
  &\qquad
  -\braksq{
  		 \frac{c}{3}\,\delta_{ij}\delta_{kl}
  		+\frac{d}{3}\,\brak{\delta_{ik}\delta_{jl}
  		         +\delta_{il}\delta_{jk}}}\delta_{mn} ,
\end{align}
where we have explicitly symmetrised on \( m \) and \( n \) by writing parentheses around the indices, e.g.
\begin{align}
  \delta_{k(m}\delta_{n)l} = \frac{1}{2}\brak{\delta_{km}\delta_{nl}+\delta_{kn}\delta_{ml}} .
\end{align}
One can derive eq.\eqref{eq:pi_iso_expression} by splitting the incremental stress into its hydrostatic and deviatoric components, 
\begin{align}
  \Delta T^{0}_{ij} = -\Delta p^{0}\delta_{ij}+\Delta\tau^{0}_{ij} ,
\end{align}
and considering the pressure and deviatoric stress separately. By symmetry, the only possible effect of an induced incremental pressure is to alter the elastic moduli, which gives the terms in \( c \) and \( d \) above. Turning to the deviatoric stress, we must construct all possible tensors with the classical elastic symmetries that are permutations of \( \delta_{ij}\Delta\tau^{0}_{kl} \) \citep[see][p.79]{dahlen1998theoretical}. The first possibility is
\begin{align}
  \delta_{ij}\Delta\tau^{0}_{kl} + \delta_{kl}\Delta\tau^{0}_{ij} .
\end{align}
Each term possesses the minor symmetries trivially, but it is only when the two terms are added together that we gain the hyperelastic symmetry. The second such tensor is a little more complicated, taking the form
\begin{align}
  \delta_{ik}\Delta\tau^{0}_{jl} + \delta_{il}\Delta\tau^{0}_{jk} + \delta_{jk}\Delta\tau^{0}_{il} + \delta_{jl}\Delta\tau^{0}_{ik} .
\end{align}
These are the only two tensors that fulfil our requirements, and they lead respectively to the terms in \( a \) and \( b \) above. Note that we have defined \( a \), \( b \), \( c \) and \( d \) so that the terms in \( a \) and \( b \) only interact with deviatoric stress (they are traceless on \( (mn) \)), while those in \( c \) and \( d \) interact with pressure alone. The components of \( \Xib \) are finally
\begin{align}
\label{eq:xi_iso_expression_abcd}
  \Xi_{ijkl} &= \brak{\kappa-\frac{2}{3}\mu+\Delta p^{0}c}\delta_{ij}\delta_{kl}
  			  +\brak{\mu+\Delta p^{0}d}\brak{\delta_{ik}\delta_{jl}+\delta_{il}\delta_{jk}}
  			  \nonumber\\
  			  &\qquad
  			  +a\brak{\delta_{ij}\Delta\tau^{0}_{kl}+\delta_{kl}\Delta\tau^{0}_{ij}}
  			  +b\brak{\delta_{ik}\Delta\tau^{0}_{jl}+\delta_{il}\Delta\tau^{0}_{jk}
  			  		 +\delta_{jk}\Delta\tau^{0}_{il}+\delta_{jl}\Delta\tau^{0}_{ik}} .
\end{align}
%\edit{
%For later reference, the Christoffel operator's components are \edit{make sure Christoffel operator defined earlier if going to include this.}
%\begin{align}
%\label{eq:christoffel_abcd}
%  \rho B_{jl} &=
%  \braksq{\brak{\kappa+\frac{1}{3}\mu}-p^{0}\brak{c+d}}\hat{k}_{j}\hat{k}_{l}
%  + \braksq{\mu-p^{0}\brak{1+d}+(1+b)\hat{\mathbf{k}}\cdot\bm{\tau}^{0}\cdot\hat{\mathbf{k}}}\delta_{jl}
%  + b\tau^{0}_{jl} 
%  + (a+b)\braksq{\hat{k}_{j}\brak{\bm{\tau}^{0}\cdot\hat{\mathbf{k}}}_{l} + \hat{k}_{l}\brak{\bm{\tau}^{0}\cdot\hat{\mathbf{k}}}_{j}} .
%\end{align}
%}

We have arrived at the most general linearised theory consistent with an isotropic, hydrostatically stressed background. It should be noted that this heuristic theory is based on symmetry arguments alone and provides no further information about the four scalars it identifies. One could therefore just regard them as free parameters of the theory, to be fitted experimentally. However, it is \textit{not} clear that they actually represent four degrees of freedom. They could well be found to depend on some smaller (or larger) set of material-dependent parameters if constitutive behaviour were considered.

The theories of \citet{dahlen_1972b}, \citet{Tromp_2018} and indeed \citet[eq.3.135]{dahlen1998theoretical} can all be seen as special cases of eq.\eqref{eq:pi_iso_expression}. Each theory corresponds to choosing
\begin{align}
\label{eq:iso_c_d_choice}
	\begin{aligned}
		  c&=-2a \\
		  d&=-2b ,
	\end{aligned}
\end{align}
and they are then distinguished by their values of \( a \) and \( b \). In \citet[eq.3.135]{dahlen1998theoretical} \( a \) and \( b \) are regarded as free parameters, yielding the expression
\begin{align}
\label{eq:d_t_elastic_tensor}
  \Xi_{ijkl} = \brak{\kappa-\frac{2}{3}\mu}\delta_{ij}\delta_{kl}
  			  +\mu\brak{\delta_{ik}\delta_{jl}+\delta_{il}\delta_{jk}}
  			  +a\brak{\delta_{ij}\Delta T^{0}_{kl}+\delta_{kl}\Delta T^{0}_{ij}}
  			  +b\brak{\delta_{ik}\Delta T^{0}_{jl}+\delta_{il}\Delta T^{0}_{jk}
  			  		 +\delta_{jk}\Delta T^{0}_{il}+\delta_{jl}\Delta T^{0}_{ik}} .
\end{align}
The authors subsequently arrive at Dahlen's original expression \citep[e.g.][eq.3.139]{dahlen1998theoretical}
\begin{align}
\label{eq:dahlen_elastic_tensor}
  \Xi_{ijkl} = \brak{\kappa-\frac{2}{3}\mu}\delta_{ij}\delta_{kl}
  			  +\mu\brak{\delta_{ik}\delta_{jl}+\delta_{il}\delta_{jk}}
  			  +\frac{1}{2}\brak{\delta_{ij}\Delta T^{0}_{kl}+\delta_{kl}\Delta T^{0}_{ij}}
  			  -\frac{1}{2}\brak{\delta_{ik}\Delta T^{0}_{jl}+\delta_{il}\Delta T^{0}_{jk}
  			  		 +\delta_{jk}\Delta T^{0}_{il}+\delta_{jl}\Delta T^{0}_{ik}}
\end{align}
by making the ``convenient" choice
\begin{align}
\label{eq:dahlen_abcd_conditions}
  a&=-b=\frac{1}{2} .
\end{align} 
Given eq.\eqref{eq:iso_c_d_choice} this is the unique choice of \( a \) and \( b \) that ensures the aforementioned characteristic features of Dahlen's theory: that seismic wave speeds have no explicit pressure-dependence, and that P-wave speeds are unaffected by deviatoric stress to first order in perturbation theory. \citeauthor{Tromp_2018}, on the other hand, set
\begin{align}
\label{eq:t_t_ab_choice}
  \begin{aligned}
    a &= \frac{1}{2}+\frac{1}{3}\mu'-\frac{1}{2}\kappa' \\
    b &= -\frac{1}{2}-\frac{1}{2}\mu' ,
  \end{aligned}
\end{align}
with \( \kappa' \) and \( \mu' \) denoting the pressure-derivatives of the elastic moduli. Their elastic tensor is thus \citep[eq.44]{Tromp_2018} 
\begin{align}
\label{eq:t_t_elastic_tensor_full}
    \Xi_{ijkl} &= 
    \brak{\kappa-\frac{2}{3}\mu}\delta_{ij}\delta_{kl}
	+\mu\brak{\delta_{ik}\delta_{jl}+\delta_{il}\delta_{jk}} 
	\nonumber\\
	&\qquad
	+\frac{1}{2}\brak{1-\kappa'+\frac{2}{3}\mu'}\brak{\delta_{ij}\Delta T^{0}_{kl}+\delta_{kl}\Delta T^{0}_{ij}}
    -\frac{1}{2}\brak{1+\mu'}\brak{\delta_{ik}\Delta T^{0}_{jl}+\delta_{il}\Delta T^{0}_{jk} +\delta_{jk}\Delta T^{0}_{il}+\delta_{jl}\Delta T^{0}_{ik}} .
\end{align}
The motivation behind this particular combination of \( a \) and \( b \) is that it leads to the wave speeds
\begin{align}
\label{eq:tromp_trampert_isotropic_wave_speeds}
  \begin{aligned}
  	\rho c_{p}^{2} &= \brak{\kappa+\kappa'\Delta p^{0}} + \frac{4}{3}\brak{\mu+\mu'\Delta p^{0}} \\ 
    \rho c_{s}^{2} &= \brak{\mu+\mu'\Delta p^{0}}
  \end{aligned}
\end{align}
when an incremental hydrostatic stress \( \Delta p^{0} \) is induced in an isotropic medium of density \( \rho \). The authors argue that this is desirable on the grounds that these are the classical expressions for isotropic wave speeds, but with elastic constants that are \textit{explicitly} corrected for an incremental pressure.

\subsubsection{Anisotropic materials}
\label{intro_anisotropic}

\wording{The theory just discussed depended on the assumption that the body was isotropic.} It would be rendered inconsistent if \( \Gam \) were taken to be anything other than an isotropic tensor, as we did throughout Section \ref{intro_isotropic}. This is a consequence of the fact that the theory is based on eq.(\ref{eq:pi_iso_expression}); a more general form of \( \Pib \) must be used if materials of lower symmetry are to be considered.

The work of \citet{Tromp_2019}, mentioned earlier, has partially resolved this issue. They generalise the results of \citet{Tromp_2018} to obtain an elastic tensor that depends on \wording{the stress} linearly, but does \textit{not} implicitly assume that the material under study is isotropic. They take the tensor \( \Gam \) to have components \citep[][eq.8]{Tromp_2019}
\begin{align}
\label{eq:Gamma_aniso}
  \Gamma_{ijkl} = \brak{\kappa-\frac{2}{3}\mu}\delta_{ij}\delta_{kl}
  		+\mu\brak{\delta_{ik}\delta_{jl}
  		         +\delta_{il}\delta_{jk}}
  		+ \gamma_{ijkl} ,
\end{align}
where \( \bm{\gamma} \) satisfies the classical elastic symmetries but is \textit{not} an isotropic tensor. Their elastic tensor \( \Xib \) can be derived from our eq.(\ref{eq:pi_def}) by demanding that \( \Pib \) possess the symmetries
\begin{align}
\label{eq:pi_extra_symmetries}
  \Pi_{ijklmn} = \Pi_{ijmnkl} = \Pi_{klmnij} = \Pi_{mnijkl} = \Pi_{mnklij}
\end{align}
\textit{in addition} to those stated in eq.(\ref{eq:pi_syms}). By doing this, eq.(\ref{eq:xi_pressure_deriv}) can be solved uniquely to give \textit{all} of \( \Pib \)'s components in terms of those of \( \Xib' \):
\begin{align}
\label{eq:jeroen_pi_solution}
  \Pi_{ijklmn} = -\frac{1}{8}\brak{
  					 \delta_{in}\Xi'_{mjkl}
  					+\delta_{im}\Xi'_{njkl}
  					+\delta_{jn}\Xi'_{imkl}
  					+\delta_{jm}\Xi'_{inkl}
  					+\delta_{kn}\Xi'_{ijml}
  					+\delta_{km}\Xi'_{ijnl}
  					+\delta_{ln}\Xi'_{ijkm}
  					+\delta_{lm}\Xi'_{ijkn}
  				}.
\end{align}
%This expression may be combined with \citeauthor{Tromp_2019}'s eq.(26),
%\begin{align}
%  \Xi'_{ijkl} = \Gamma'_{ijkl} - \brak{\delta_{ij}\delta_{kl}-\delta_{ik}\delta_{jl}-\delta_{il}\delta_{jk}},
%\end{align}
%to yield their elastic tensor \citep[][eq.25]{Tromp_2019}
%\begin{align}
%\label{eq:elastic_tensor_tromp_et_al}
%  \Xi_{ijkl} &= \Gamma_{ijkl} 
%  			  + \Gamma'_{ijkl} \Delta p^{0}
%  			  - \Delta p^{0}\brak{\delta_{ij}\delta_{kl}-\delta_{ik}\delta_{jl}-\delta_{il}\delta_{jk}}
%  			  \nonumber\\
%  			  &\qquad
%  			  +\frac{1}{2}\brak{\Delta \tau^{0}_{ij}\delta_{kl}+\Delta \tau^{0}_{kl}\delta_{ij}}
%  			  -\frac{1}{2}\brak{\Delta \tau^{0}_{ik}\delta_{jl}+\Delta \tau^{0}_{jk}\delta_{il}+\Delta \tau^{0}_{il}\delta_{jk}+\Delta \tau^{0}_{jl}\delta_{ik}}
%  			  \nonumber\\
%  			  &\qquad
%  			  -\frac{1}{4}\brak{\Gamma'_{imkl}\Delta \tau^{0}_{mj}
%  			                   +\Gamma'_{jmkl}\Delta \tau^{0}_{mi}
%  			                   +\Gamma'_{kmij}\Delta \tau^{0}_{ml}
%  			                   +\Gamma'_{lmkl}\Delta \tau^{0}_{mk}} .
%\end{align}
The elastic tensor's dependence on incremental stress is thus parametrised by pressure-derivatives alone, analogously to the isotropic case (cf. eq.\ref{eq:iso_c_d_choice}). Under this theory \( \Pib \) possesses the symmetries that would be expected of the third strain-derivative of a strain-energy function, and is described by 21 independent components at most.

\subsection{Aims of this paper}

This completes our initial survey of the elastic tensor's linearised stress dependence. Working on the basis of symmetry arguments alone we have \wording{established} that the previous approaches to the problem \wording{might} not be sufficiently general. In particular, equations \eqref{eq:iso_c_d_choice} and \eqref{eq:jeroen_pi_solution} seem to imply that the stress dependence of the elastic tensor should be parametrisable solely in terms of the pressure dependence of the elastic moduli, a physical result that is not obvious to the present authors. Nevertheless, recall that \citet{Tromp_2019} tested the validity of their theory by carrying out \textit{ab initio} calculations. They found good, but not perfect, agreement between theory and experiment. Given the nature of such calculations it is difficult to make precise statements about the significance of the discrepancies, but the general agreement does provide clear support for their parametrisation. It is also worth noting that the approach of Section \ref{intro:heuristic} brings one to a usable theory rather quickly. The issue, however, is that it gives no clear sense of how \( \Pib \) is determined from the underlying constitutive relation. As the theory stands there appears to be no way to obtain \( \Pib \)'s components besides by performing experiments. Such experiments are already challenging for the cubic materials considered by \citet{Tromp_2019}, wherein three parameters were to be found, and they might become very difficult for more anisotropic materials. One might wonder if \( \Pib \) emerges from a more \wording{``fundamental"} source than its definition in eq.\eqref{eq:pi_def}. 

The present work thus has two main aims. The first is to better understand the previous theoretical work on the elastic tensor's linearised stress dependence. Having established the general characteristics of that theory in Section \ref{intro:heuristic} we now wish to ask whether the components of \( \Pib \) can be obtained more readily \wording{in some other way}. A secondary aim is to construct a nonlinear theory of the elastic tensor's stress dependence. This is not purely academic, despite the fact that deviatoric stresses within the present-day Earth are presumably rather small. Methodologically speaking, we feel that it is clearer to derive as much as possible without approximating any physical behaviour because it provides a firm foundation for subsequent, physically-motivated linearisation. In formulating a nonlinear theory we hope to gain greater insight into the linear theory.

In order to make progress we take a new approach rooted firmly in the theories of finite elasticity and constitutive behaviour. We present that argument in Section \ref{sec:lambda_of_sigma}, having first reviewed the necessary ideas in Section \ref{theory}. After deriving our main result we give some examples, both numerical and analytical, in Section \ref{sec:calculations}, and discuss the implications of our theory in Section \ref{sec:discussion}. In this introduction we have used as far as possible the notation of \citet{dahlen1998theoretical} in order to make close contact with previous work. Our theoretical developments owe a lot to \cite{marsden1994mathematical}, so from Section \ref{theory} onwards we switch to (approximately) their notation, which will allow the interested reader of Sections \ref{theory} and \ref{sec:lambda_of_sigma} to ``cross-reference" easily. The content of Sections \ref{theory}--\ref{sec:calculations} is necessarily quite technical, so the reader who is primarily interested in our results might wish to proceed straight to Section \ref{sec:discussion}. We have therefore restated \wording{some of} the paper's important results therein using the notation of \citeauthor{dahlen1998theoretical}, as well as including a complete ``translation table" between the two notation systems in Appendix \ref{notation_conventions}.

%%%%%%%%%%%%%%%%%%%%%%%%%%%%%%%%
\section{A review of elasticity}
\label{theory}

In this section we summarise the aspects of elasticity  pertinent to this paper. For more details see, for example, \cite{marsden1994mathematical}, \cite{dahlen1998theoretical} or \cite{Truesdell2004}. In order to make reference to formulae within the solid mechanics literature we will now follow the notation of \citeauthor{marsden1994mathematical} (henceforward MH) fairly closely. A modest innovation on our part is to use sans-serif bold font for fourth-rank tensors so as to contrast them with second-rank tensors. This allows us to distinguish between the second elastic tensor \( \CCC \) and the right Cauchy-Green deformation tensor \( \CC \) without having to resort to index notation. We also refer the reader to Appendix \ref{app:notations}: Section \ref{app_sub:groups} lists some standard results from group theory that we will refer to; \ref{app_sub:linear_ops} defines some non-standard notations and operators that we have found helpful; and \ref{app:sub:W_V_differentiation} finally illustrates the usage of these operators while presenting a calculation that is salient to the present work.

\subsection{Basic definitions and results}
\label{contmech}

\subsubsection{Equations of motion}
\label{eom}

The deformation of an elastic body is  described relative to a fixed \textit{reference configuration}, with each particle labelled by its position within the associated \textit{reference body} \( \MM \subseteq \RThree\), which is assumed to be connected, bounded, and have an open interior. At a time \( t \), the position in physical space of the particle at \( \mathbf{x} \in \MM \) is written \( \phib(\mathbf{x},t) \). In this manner, we define a mapping
\begin{align}
  \phib : \MM\times\reals \rightarrow \RThree, 
\end{align}
which is called the \textit{motion} of the body relative to the reference configuration. For a fixed time, \( t \), the image of the  mapping \( \phib\brak{\cdot,t} \) is written \( \MM_{t} \) and represents the region of physical space the body instantaneously occupies. It will be assumed that for each fixed time the mapping \( \phib\brak{\cdot,t}:\MM \rightarrow \MM_{t} \) is smooth with a smooth inverse. A fundamental object derived from the motion is its \textit{deformation gradient},
\begin{align}
  \FF = D_{\mathbf{x}}\phib ,
\end{align}
where \( D_{\mathbf{x}} \) denotes partial differentiation with respect to position as defined through
\begin{align}
  \phib \brak{\mathbf{x} + \delta \mathbf{x},t
  } =   \phib \brak{\mathbf{x},t
  } +  \brak{D_{\mathbf{x}}\phib}\brak{\mathbf{x},t} \cdot \delta \mathbf{x}
  + \mathcal{O}\brak{\|\delta \mathbf{x}\|^{2}} .
\end{align}
(We will generally neglect the subscript on \( D \) where it is unambiguous which variable we are differentiating with respect to.) Equivalently, the  Cartesian components of the deformation gradient are
\begin{align}
  F_{ij} = \frac{\partial \varphi_{i}}{\partial x_{j}}  .
\end{align}
The \textit{Jacobian} is then defined as
\begin{align}
  J = \det\FF  .
\end{align}
Due to our assumption that \( \phib\brak{\cdot,t}:\MM \rightarrow \MM_{t} \) has a smooth inverse, it follows from the inverse function theorem (MH, p.31) that the deformation gradient takes values in the general linear group \(\mathbf{GL}(3)\) (Appendix \ref{app_sub:groups}). We assume without loss of generality that \( J \) is everywhere positive, meaning that the motion is orientation preserving.

The \textit{density} at time \( t \) at the point \( \yy\in\MM_{t} \) in physical space is written \( \varrho\brak{\yy,t} \). From conservation of mass, we are led to define the \textit{referential density},
\begin{align}
	\rho(\xx) = J(\xx,t)\varrho\braksq{\phib(\xx,t),t} ,
\end{align}
a time-\textit{independent} function within the reference body (MH, p.87, Theorem 5.7). Cauchy's theorem implies that the traction \( \TT \) acting on a surface-element within \( \MM_{t} \) is related linearly to the  unit-normal \( \hat{\NN} \) of the corresponding reference surface element within \( \MM \) (MH, p.127). We can therefore define the \textit{first Piola-Kirchhoff  stress tensor} \( \PP \) through (MH, p.7)
\begin{align}
\label{eq:traction_FPK_definition}
  \TT = \PP\cdot\hat{\NN}  .
\end{align}
This expression is equivalent to eq.(2.41) of \cite{dahlen1998theoretical} but, as discussed in Appendix \ref{notation_conventions}, we place our indices according to a different convention. From Newton's second law we obtain the momentum equation
\begin{align}
\label{eq:phi_nonlinear}
	\rho\ddot{\phib} - \mathrm{Div}\PP = \mathbf{f} ,
\end{align}
where dots are used to  represent time differentiation, the divergence of a tensor field is given by
\begin{align}
  \brak{\Div\PP}_{i} = \partial_{j}P_{ij} , 
\end{align}
and \( \mathbf{f} \) denotes the body forces acting on \( \MM \).

\subsubsection{Constitutive relations}
\label{constitutive_relations}

To complete the equations of motion we need to relate \( \PP \) and \( \phib \) through a suitable constitutive relation. We follow \citet{dahlen_1972a} by restricting attention to \textit{hyperelastic materials}, in which case the first Piola-Kirchhoff stress depends on the motion through the expression
\begin{align}
  \PP(\xx,t) = \brak{D_{\FF}W}\braksq{\xx,\FF(\xx,t)},
\end{align}
where \( W:\MM \times \mathbf{GL}(3)\rightarrow \mathbb{R} \) is the \textit{strain-energy function} and \( D_{\mathbf{F}}W \) denotes its partial derivative with respect to \( \mathbf{F} \) (MH, p.190, Theorem 2.4).

The form of the strain-energy function is constrained by the \textit{principle of material-frame indifference}. Discussed at length by \citet{marsden1994mathematical} and \citet{Truesdell2004}, it requires that 
\begin{align}
	W(\xx,\RR\FF) &= W(\xx,\FF) 
\end{align}
for all rotation matrices \( \RR \in \mathbf{SO}(3) \) and \( \FF \in \mathbf{GL}(3) \) (see Appendix \ref{app_sub:groups}). It can be shown using the polar decomposition theorem (MH, p.8) that this condition holds if and only if for some auxiliary strain-energy function \( V \) we can write
\begin{align}
\label{eq:W_V_relationship}
  W(\xx,\FF) = V\brak{\xx,\CC} ,
\end{align} 
with the symmetric \textit{right Cauchy--Green deformation tensor} defined to be
\begin{align}
  \label{eq:cdef}
  \CC = \FF^{T}\FF  .
\end{align}
Applying the chain rule to differentiate eq.(\ref{eq:W_V_relationship}), we arrive at an alternative expression for the first Piola-Kirchhoff stress in terms of \( V\). It is readily established (see Appendix \ref{app:sub:W_V_differentiation}) that
\begin{align}
  D_{\FF}W\brak{\xx,\FF} =  2 \FF D_{\CC}V\brak{\xx,\CC}  .
\end{align}
From this expression we are led to define the symmetric \textit{second Piola-Kirchhoff stress tensor} (MH, p.196, Theorem 2.11)
\begin{align}
  \label{eq:spkdef}
  \SSSSS\brak{\xx,t} = 2 D_{\CC}V\braksq{\xx,\CC\brak{\xx,t}}  .
\end{align}
Hence we obtain the identity
\begin{align}
  \label{eq:fspid}
  \PP = \FF \SSSSS  .
\end{align}

A third useful stress tensor is the \textit{Cauchy stress}, \( \sig \). It relates the traction on a surface element at \( \yy\in\MM_{t} \) to the surface's instantaneous unit-normal \( \nhat \). From this definition it can be shown (MH, p.135) that 
\begin{align}
  \PP = J \brak{\sig\circ\phib} \FF^{-T}  .
\end{align}
Although it is perhaps not obvious from this expression, the Cauchy stress is symmetric. Using eq.(\ref{eq:fspid}) we obtain (MH, p.136, Definition 2.8)
\begin{align}
  \label{eq:sigid}
  \sig\circ\phib = \frac{1}{J}\FF\SSSSS\FF^{T} ,
\end{align}
and the symmetry of \( \sig \) follows from that of \( \SSSSS \).

\subsubsection{Linearisation of the equations of motion}
\label{linearisation}

For seismological purposes it is generally sufficient to study linearised elastodynamics. We define an \textit{equilibrium configuration} \( \phib_{e}:\MM \rightarrow \RThree \) to be a time-independent solution of the equations of motion subject to a time-independent body force \( \mathbf{f}_{e} \) and surface traction \( \TT_{e} \). The resulting equilibrium first Piola-Kirchhoff stress is given by
\begin{align}
  \PP_{e}(\xx) = D_{\FF}W\braksq{\xx,\FF_{e}(\xx)} ,
\end{align}
where \( \FF_{e} = D_{\mathbf{x}}\phib_{e} \). Using eq.(\ref{eq:fspid}) and (\ref{eq:sigid}), the three equilibrium  stress tensors are then related by
\begin{align}
\label{eq:various_eqm_stresses_general}
  \PP_{e} = \FF_{e}\SSSSS_{e} = J_{e}\brak{\sig_{e}\circ\phib_{e}}\FF_{e}^{-T}  .
\end{align}

If such a body is subject to a small disturbance from equilibrium, we can look for solutions of the form
\begin{align}
\label{eq:phi_pert}
	\phib(\xx,t) = \phib_{e}(\xx) + \epsilon\, \uu(\xx,t) + \mathcal{O}\brak{\epsilon^{2}} , 
\end{align}
where \( \mathbf{u} \) is the displacement vector and \(\epsilon \) a
dimensionless perturbation parameter. Under this ansatz the deformation gradient becomes
\begin{align}
  \FF(\xx,t) = \FF_{e}(\xx) + \epsilon\, D_{\mathbf{x}}\uu(\xx,t) + \mathcal{O}\brak{\epsilon^{2}} ,
\end{align}
while the first Piola-Kirchhoff stress expands to
\begin{align}
\label{eq:T_pert_expression}
  \PP(\xx,t) &= \PP_{e}(\xx) + \epsilon\,\AAA\brak{\xx}\cdot D_{\mathbf{x}}\uu(\xx,t) + \mathcal{O}(\epsilon^{2}) ,
\end{align}
where we have defined the \textit{first elastic tensor} (MH, p.209, Proposition 4.4b) 
\begin{align}
\label{eq:Lam_defn}
  \AAA(\xx) = D^{2}_{\FF}W\braksq{\xx,\FF_{e}(\xx)}  .
\end{align}
Note that the first elastic tensor possesses the so-called hyperelastic symmetry,
\begin{equation}
  \label{eq:hyp}
  \AAA^{T} = \AAA ,
\end{equation}
due to the equality of mixed partial derivatives. In index notation this relationship takes the familiar form \( \mathsf{A}_{ijkl} = \mathsf{A}_{klij} \). We will henceforward follow standard seismological terminology and refer to \( \AAA \) simply as ``the elastic tensor" unless that is likely to cause confusion.

As shown by eq.(\ref{eq:T_pert_expression}), the elastic tensor completely describes the linearised constitutive behaviour of the body.  In particular, at a point \( \mathbf{x} \in \MM \) there will be three possible elastic wave speeds in the direction of the unit vector \( \hat{\mathbf{p}} \). These wave speeds, \( c \),  are determined through the eigenvalue problem \citep[e.g.][Section 3.6.3]{dahlen1998theoretical}
\begin{align}
  \label{eq:chris}
  \mathbf{B}(\mathbf{x},\hat{\mathbf{p}}) \cdot \mathbf{a} = c^{2} \mathbf{a}, 
\end{align}
where \( \mathbf{a} \) is the corresponding polarisation vector, and
the Christoffel matrix has components
\begin{align}
\label{eq:christoffel_defn}
  B_{ik}(\mathbf{x},\hat{\mathbf{p}}) = \frac{1}{\rho(\mathbf{x})}\mathsf{A}_{ijkl}(\mathbf{x})
  \hat{p}_{j}\hat{p}_{l}.
\end{align}
This matrix is symmetric due to eq.(\ref{eq:hyp}), hence the \( c^{2} \) are real. Within an elastic solid it is conventionally assumed that these squared wave speeds are  positive, a necessary condition for well-posedness of the linearised equations of motion \citep[e.g.][]{marsden1994mathematical}. As the propagation direction \( \hat{\mathbf{p}} \) varies over the unit two-sphere, the three  positive wave-speeds define the so-called \textit{slowness surface} at \( \xx \). In general, this surface will be comprised of three distinct sheets, though they can sometimes touch due to  degenerate eigenvalues within eq.(\ref{eq:chris}).

Finally, it is useful to express the elastic tensor in terms of the auxiliary strain-energy function \( V \). We define the equilibrium second Piola-Kirchhoff stress as
\begin{align}
  \label{eq:spkdef}
  \SSSSS_{e}\brak{\xx} = 2 D_{\CC}V\braksq{\xx,\CC_{e}\brak{\xx}} ,
\end{align}
and introduce the \textit{second elastic tensor} (MH, p.209, Proposition 4.4a)
\begin{align}
  \CCC\brak{\xx} = 4 D^{2}_{\CC}V\braksq{\xx,\CC_{e}\brak{\xx}}.
\end{align}
Suppressing all spatial arguments to avoid clutter, it then follows from the results of Appendix \ref{app:sub:W_V_differentiation} (see also MH, p.209, Proposition 4.5) that
\begin{align}
  \label{eq:lamrep}
  \AAA = \Lmult{\FF_{e}}\CCC \,\Lmult{\FF_{e}}^{T} + \Rmult{\SSSSS_{e}}  .
\end{align}
It is worth emphasising that the tensor \( \CCC \) has the full set of classical elastic symmetries,
\begin{align}
  \mathsf{C}_{ijkl} = \mathsf{C}_{jikl} = \mathsf{C}_{ijlk} = \mathsf{C}_{klij}, 
\end{align}
due to the symmetry of \( \CC_{e} \), and so possesses at most 21 independent components.

\subsection{Transformation of the reference configuration}
\label{change_ref_config}
The motion of an elastic body has been described relative to a fixed reference configuration involving material parameters \( \rho \) and \( W \). The same body can, of course, be described using a different choice of reference configuration, and it is natural to ask how the two points of view are related. This question was discussed by \citet{Al_Attar_2016} and \cite{Al_Attar_2018}; here we simply recall  the results relevant to this work.

\subsubsection{Particle-relabelling}
\label{particle_relabelling}
Let \( \phib:\MM \times \mathbb{R} \rightarrow \mathbb{R} \) denote the motion of an
elastic body relative to a given reference configuration.  The same motion
relative to  a different reference configuration will be written
\( \tilde{\phib}:\tilde{\MM} \times \mathbb{R} \rightarrow \mathbb{R} \),
where $\tilde{\MM}$ is the associated reference body that will not, in general, 
be equal to $\MM$. At a time \( t \), the particle labelled by \( \mathbf{x} \in \MM \)
lies at the point \( \phib(\mathbf{x},t) \) in physical space. Relative to the
second description of the motion, there must be a unique point \( \tilde{\mathbf{x}} \in \tilde{\MM} \)
such that
\begin{align}
  \phib(\mathbf{x},t) = \tilde{\phib}(\tilde{\mathbf{x}},t).
\end{align}
This correspondence between \( \xx \) and \(\xxt\) holds for all times,
defining a mapping \( \xib:\tilde{\MM} \rightarrow \MM \)
that relates the two motions through
\begin{align}
  \phib(\mathbf{x},t) = \tilde{\phib}[\xib(\tilde{\mathbf{x}}),t].
\end{align}
It is assumed for simplicity that \( \xib \) is  smooth and has a smooth inverse.
Under such a \textit{particle relabelling transformation} the form of the equations
of motion is clearly left unchanged, while it was shown by \citet{Al_Attar_2016} that
the material parameters \( \tilde{\rho} \) and \( \tilde{W} \) relative
to the second reference configuration can be obtained from those in the first by
\begin{align}
  \tilde{\rho}(\tilde{\mathbf{x}}) &=  J_{\xib}(\tilde{\mathbf{x}}) \, \rho[\xib(\tilde{\mathbf{x}})], \label{eq:ptcl_relabelling_strain_energy_rho} \\
  \tilde{W}(\xxt,\tilde{\FF}) &=
  J_{\xib}(\tilde{\mathbf{x}}) \, W[\xib\brak{\xxt},\tilde{\FF}\,\FF_{\xib}(\xxt) ^{-1}] , \label{eq:ptcl_relabelling_strain_energy_W}
\end{align}
where \( \mathbf{F}_{\xib} = D \xib \) and \( J_{\xib} = \det \mathbf{F}_{\xib} \).

\subsubsection{Natural reference configurations}
\label{natural_reference_configuration}

When considering linearised motions of an elastic body it is conventional to select the reference configuration so that the equilibrium configuration takes the simple form
\begin{align}
  \phib_{e}(\xx) = \xx  .
\end{align}
In this manner, the label for each particle is simply its position in physical
space at equilibrium.
In the terminology of \citet{Al_Attar_2016}, such a  reference configuration
is said to be \textit{natural}.
The equilibrium deformation gradient  then satisfies
\begin{align}
  \FF_{e}(\mathbf{x}) = \id ,
\end{align}
while its Jacobian is everywhere equal to one. Given this choice, the
equilibrium first Piola-Kirchhoff stress is obtained by evaluating the first derivative of the strain-energy at the identity:
\begin{align}
  \PP_{e} = D_{\FF}W\brak{\cdot,\id} .
\end{align}
An attractive feature of natural reference configurations is that the distinction between the different equilibrium stress-tensors vanishes. It is trivial to verify that equation
(\ref{eq:various_eqm_stresses_general}) now reads
\begin{align}
\label{eq:nat_ref_config_stress_tensors}
  \PP_{e} = \SSSSS_{e} = \sig_{e}  .
\end{align}
In particular, it follows that
\begin{align}
  \label{eq:signat}
  \sig_{e} = D_{\FF}W\brak{\cdot,\id} = 2 D_{\CC} V\brak{\cdot,\id} ,
\end{align}
an expression we will dissect in Section \ref{sec:lambda_of_sigma}. 

In the same manner, the elastic tensor takes the simpler form
\begin{align}
\label{eq:Lam_defn_nat_ref}
  \aaaa = D^{2}_{\FF}W\brak{\cdot,\id} ,
\end{align}
which, from eq.(\ref{eq:lamrep}), can be written equivalently as
\begin{align}
  \label{eq:lamrep2}
  \aaaa = \cccc + \Rmult{\sig_{e}}  .
\end{align}
Note that we have denoted the first and second elastic tensors by lower-case \( \aaaa \) and \( \cccc \). This is a notational convention used by \citeauthor{marsden1994mathematical}, the aim of which is to emphasise that these elastic tensors are defined with respect to \wording{(what is here termed)} a natural reference configuration. As noted above, the tensor 
\begin{align}
  \label{eq:adef}
  \cccc = 4 D^{2}_{\CC}V\brak{\cdot,\id} 
\end{align}
possesses all the classical elastic symmetries. In contrast, the second term in eq.(\ref{eq:lamrep2})
has components
\begin{align}
  [\Rmult{\sig_{e}}]_{ijkl} = \delta_{ik}[\sig_{e}]_{jl} 
\end{align}
which, for general \( \sig_{e} \ne \mathbf{0} \), are not invariant under
 the interchange of either \( i \leftrightarrow j\) or \( k \leftrightarrow l\) (although the symmetry of \( \sig_{e} \) ensures that \( \Rmult{\sig_{e}} \) possesses the hyperelastic symmetry). We therefore see that \( \aaaa \) inherits the full complement of classical symmetries \textit{only with respect to a stress-free natural reference configuration}. Moreover, it is only with respect to a natural reference configuration that the propagation directions \( \hat{\mathbf{p}} \) within eq.(\ref{eq:chris}) can be equated with directions in physical space, allowing the slowness surface to be interpreted in a straightforward manner. Eq.\eqref{eq:lamrep2} is precisely equivalent to eq.\eqref{eq:lambda_basic_decomposition}, though it is important to note that this \wording{equivalence} only holds with respect to a natural reference configuration.

\subsection{Material symmetries}
\label{material_symmetries}

We end our review of finite elasticity theory by discussing material symmetries. These results can be found in MH (Chapter 3, Section 3.5) and \citet[Chapter 50]{gurtin2010mechanics}\wording{, though we place additional emphasis on certain points}.

Relative to a fixed reference configuration, the \textit{material symmetry group} of a strain-energy function \( W \) at a point \( \xx\in\MM \) is 
\begin{align}
\label{eq:mat_sym_def_initial}
  \sym(W,\xx) = \left\{ \QQ\in\sl(3)\, |\, W(\xx,\FF\QQ) = W(\xx,\FF),\,  \forall \FF\in\gl(3) \right\} ,
\end{align}
where \( \mathbf{SL}(3) \) denotes the special linear group on \( \RThree \) whose definition is recalled in Appendix \ref{app_sub:groups}. Physically, the symmetry group reflects the invariance of the strain-energy with respect to orientation of stretching. Following \cite{Noll_1974}, the body is said to be fluid at a point if the symmetry group is equal to \( \mathbf{SL}(3) \), and solid if it is a proper subgroup thereof.

Under a change of reference configuration, the  symmetry group is not generally invariant. To see this, let \( \xib:\tilde{\MM} \rightarrow \MM \) be a particle relabelling transformation, and suppose that \( \xib(\xxt) = \xx \). If \( \tilde{\QQ} \in \sym(\tilde{W},\xxt) \) then from eq.(\ref{eq:ptcl_relabelling_strain_energy_W}) we obtain
\begin{align}
  W[\xx,\tilde{\mathbf{F}} \tilde{\QQ} \mathbf{F}_{\xib}(\xxt)^{-1}] =
    W[\xx,\tilde{\mathbf{F}} \mathbf{F}_{\xib}(\xxt)^{-1}] 
\end{align}
for all \( \tilde{\mathbf{F}} \in \mathbf{GL}(3) \). Hence for a unique \( {\QQ} \in \sym({W},\xx) \) we have
\begin{align}
\tilde{\QQ}  \mathbf{F}_{\xib}(\xxt)^{-1} = \mathbf{F}_{\xib}(\xxt)^{-1} \QQ .
\end{align}
This establishes a group isomorphism between \( \sym({W},\xx) \) and \( \sym(\tilde{W},\xxt) \),
which is given  concretely through matrix conjugation:
\begin{align}
\label{eq:sym_group_relation_under_conjugation}
  \sym({W},\xx) \ni \QQ \mapsto  \mathbf{F}_{\xib}(\xxt)^{-1} \QQ  \mathbf{F}_{\xib}(\xxt) \in
  \sym(\tilde{W},\xxt) .
\end{align}

This mapping leaves \( \mathbf{SL}(3) \) invariant, hence our definitions of the material symmetry group itself and of an elastic fluid are sound. \cite{Noll_1965} showed that, up to this isomorphism,  the largest proper subgroup of \( \mathbf{SL}(3) \) is equal to the special orthogonal group \( \mathbf{SO}(3) \). We can, therefore, define an elastic solid to be \textit{isotropic} at a point if its symmetry group relative to an arbitrary reference configuration is isomorphic under matrix conjugation to \( \mathbf{SO}(3) \). Equivalently, it is isotropic if for \textit{some} reference configuration its symmetry group is equal to \( \mathbf{SO}(3) \). If the symmetry group is isomorphic to a proper subgroup of \( \mathbf{SO}(3) \) we say the solid is \textit{anisotropic}, with the extreme case being when this group consists of the identity matrix alone. In between these two end-members can be found, for example, \textit{transversely-isotropic} materials, whose symmetry group is isomorphic under matrix conjugation to \( \mathbf{SO}(2) \). This corresponds physically to the strain-energy being invariant under rotations about a certain fixed axis. 

Importantly, the preceding discussion is independent of any particular choice of reference configuration. It corrects a mistake of \cite{Al_Attar_2016}, who implied that material symmetries can be lost or gained through particle relabelling transformations. Such transformations simply represent a change in our description of the body's deformation. They cannot entail any physical consequences.

As a final concept that we will need later, consider a natural reference configuration for a \textit{stress-free} elastic body in equilibrium. Such a configuration is defined by \( DV\brak{\CC^{*}}= \mathbf{0} \) with \( \CC^{*}=\id \). From eq.(\ref{eq:mat_sym_def_initial}), the material symmetry group acts on \( \CC \) according to \( \CC \mapsto \QQ^{T}\CC\QQ \), so by definition we have
\begin{align}
  V\brak{\QQ^{T}\CC^{*}\QQ} = V\brak{\CC^{*}} ,
\end{align}
from which
\begin{align}
\label{eq:stress_free_nat_ref_min}
  V\brak{\QQ^{T}\QQ} = V\brak{\id}  .
\end{align}
For the equilibrium to be stable, \( \CC^{*}=\id \) must lie at a strict local minimum of the strain-energy function. This allows us to equate the arguments of the left and right hand sides of eq.(\ref{eq:stress_free_nat_ref_min}). The elements of the symmetry group then satisfy
\begin{align}
  \QQ^{T}\QQ = \id ,
\end{align}
from which it is clear that \( \QQ\in\so(3) \). The symmetry group of a stress-free elastic body, described with respect to a natural reference configuration, is therefore \textit{equal} to a subgroup of \( \so(3) \) rather than just isomorphic thereto. For example, in the stress-free case an isotropic body in a natural reference configuration has a material symmetry group equal to the whole of \( \so(3) \), while that of a transversely-isotropic material is equal to \( \so(2) \), having fixed the orientation of the symmetry-axis.

%%%%%%%%%%%%%%%%%%%%%%%%%%%%%%%%%%%%%%%%%%%%%%%%%%%%%%%%%%%%%%%%%%%%%%%%%%%%%%%%%%%%%%%%
\section{Functional dependence of the elastic tensor on equilibrium stress}
\label{sec:lambda_of_sigma}

Having  recalled the necessary results and notations from the theory of elasticity, we now turn to our main question. Namely, we seek to determine the functional dependence of the elastic tensor on the equilibrium Cauchy stress.

\subsection{Parametrised dependence  on the equilibrium configuration}
\label{subsec:lambda_and_sigma_of_f}

For an equilibrium body \( \MM \), the equilibrium Cauchy stress and elastic tensor take the form
\begin{subequations}
\label{eqs:sig_lam_natural}
\begin{align}
  \sig\brak{\xx} &= DW\brak{\xx,\id}	 \label{eq:sig_natural}  \\
  \aaaa\brak{\xx} &= D^{2}W\brak{\xx,\id} \label{eq:lam_natural} ,
\end{align}
\end{subequations}
where \( W \) is the strain-energy function with respect to a natural reference configuration, and for notational simplicity we have dropped the subscript from \( \sig_{e} \). Our hope is to express the elastic tensor as a function of the equilibrium stress. Variations in \(\sig\) arise, of course, though changes to the equilibrium configuration, but the dependence of eqs.(\ref{eqs:sig_lam_natural}) thereon is masked by the use of a natural reference configuration. As a first step, we must reformulate eqs.(\ref{eqs:sig_lam_natural}) in a way that makes fully explicit the dependence of the two equations on the equilibrium configuration.

We consider an arbitrary fixed reference configuration with the associated reference body denoted by \( \tilde{\MM} \), and with strain-energy function \( \tilde{W} \). The correspondence between this reference configuration and the natural reference configuration \( \MM \) is given by a mapping 
\begin{align}
  \Phib:\tilde{\MM} \rightarrow \MM.
\end{align}
This is just the equilibrium configuration of the body relative to our newly introduced reference configuration (see Fig. \ref{fig:setup}). Regarding the inverse mapping \( \Phib ^{-1} \) as a particle relabelling transformation, we can use eq.(\ref{eq:ptcl_relabelling_strain_energy_W}) to relate \( W \) to \( \tilde{W} \):
\begin{align}
\label{eq:W_W_tilde_relationship}
  W\braksq{\Phib(\xxt),\FF'} = J_{\FF}(\xxt)^{-1}\tilde{W}\braksq{\xxt,\FF'\FF\brak{\xxt}}  .
\end{align}
To avoid cluttered notations here and in what follows, we write \( \mathbf{F} \) for the equilibrium deformation gradient \( D\Phib \) and \( J_{\FF} = \det\FF \), while \( \mathbf{F}'\) represents an arbitrary element of \( \gl(3)\). From eq.\eqref{eq:W_W_tilde_relationship} we see that \( W[\Phib(\xxt),\cdot] \) depends only on \( \tilde{W} \) at the fixed point \( \xxt \in \tilde{\MM} \). Furthermore, the relationship is parametrised by \( \FF \) \textit{evaluated at} \( \xxt \). All in all, the two functions are related in a local manner; no generality is lost by focusing on a single, arbitrary point in \( \MM \) and its pre-image in \( \tilde{\MM} \). We do this from now on, dropping all spatial arguments to arrive at the simpler relations
\begin{subequations}
\label{eqs:sig_lam_natural_no_x}
\begin{align}
  \sig &= DW\brak{\id}	   \label{eq:sig_natural_no_x}  \\
  \aaaa &= D^{2}W\brak{\id} \label{eq:lam_natural_no_x} 
\end{align}
\end{subequations}
and
\begin{align}
\label{eq:wparm}
  W\brak{\FF'} &= J_{\FF}^{-1}\tilde{W}\brak{\FF'\FF}  .
\end{align}
\iffair\begin{figure}
  \begin{center}
  \includegraphics[width=0.75\textwidth]{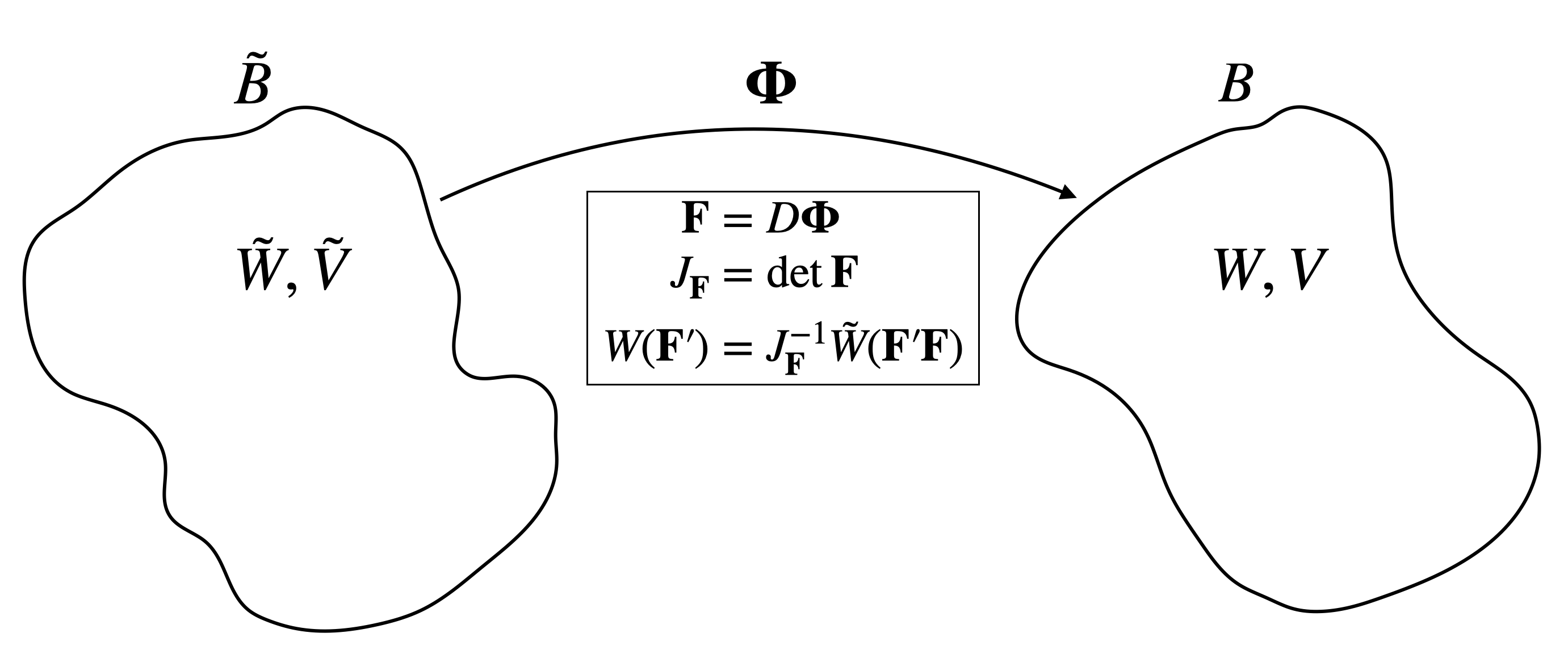}
  \end{center}
  \caption{The setup of our problem. The reference body \( \tilde{\MM} \), with fixed elastic properties, is considered to be a reference configuration for the equilibrium body \( \MM \). The two bodies are related by the equilibrium-mapping \( \Phib \), permitting us to use particle-relabelling transformations to write the strain-energy function \( W \) in terms of \( \tilde{W} \).}
  \label{fig:setup}
\end{figure}\fi

To complete the reformulation of eqs.\eqref{eqs:sig_lam_natural_no_x} we apply the chain rule to eq.(\ref{eq:wparm}) so as to  write \( \sig \) and \( \aaaa \) explicitly in terms of \( \tilde{W} \) and \( \mathbf{F} \). It is in fact preferable to work not with \( W\), but rather with the auxiliary strain-energy function \(V\) that encodes material-frame indifference automatically. Defining \( \CC' = (\mathbf{F}')^{T} \mathbf{F}' \), we recall that \( V \) will satisfy
\begin{align}
\label{eq:W_V_redef}
  W\brak{\FF'} = V\brak{\CC'} ,
\end{align}
and from eqs. \eqref{eq:wparm} and \eqref{eq:W_V_redef} we see that the relationship between \( V \) in \(\MM\) and its counterpart \( \tilde{V}\) in \(\tilde{\MM}\) is given by
\begin{align}
\label{eq:w_trans_v}
V(\CC') = J_{\FF}^{-1} \tilde{V}(\Tmult{\FF}^{T}\cdot\CC'), 
\end{align}
where the operator \( \Tmult{\mathbf{F}} \) is defined in eq.(\ref{eq:tdef}). From Appendix \ref{app:sub:W_V_differentiation} we have
\begin{align}
  DW(\FF') &= 2\Lmult{\FF'}\cdot DV(\CC') \\
  D^{2}W\brak{\FF'} &= 2\Rmult{DV\brak{\CC'}} + 4\Lmult{\FF'} D^{2}V\brak{\CC'} \,\Lmult{\FF'}^{T} ,
\end{align}
and it is readily shown from eq.\eqref{eq:w_trans_v} that
\begin{align}
  D V\brak{\CC'} &= \Tmult{\FF}\cdot D\tilde{V}(\Tmult{\FF}^{T}\cdot\CC') \\
  D^{2}V\brak{\CC'} &= \Tmult{\FF}D^{2}\tilde{V}(\Tmult{\FF}^{T}\cdot\CC')\Tmult{\FF}^{T} .
\end{align}
Evaluating these results at \( \CC' = \mathbf{1}\) -- and noting that \(\Tmult{\FF}^{T}\cdot \mathbf{1} = \CC\) -- we obtain the first of our desired expressions,
\begin{align}
  \label{eq:sigparm}
  \sig = 2 J_{\FF}^{-1} \Tmult{\FF} \cdot D\tilde{V}(\CC) .  
\end{align}
Note that it is equivalent to eq.(\ref{eq:sigid}) but has been restated in a different notation. By considering the second derivative we then obtain the expression
\begin{align}
\label{eq:lamparm}
  \aaaa &=  4 J_{\FF}^{-1} \Tmult{\FF}
    D^{2}\tilde{V}(\CC)\,\Tmult{\FF}^{T} + \Rmult{\sig} ,
\end{align}
which is equivalent to the \wording{expression} of MH (p.217, Box 4.1). We have thus expressed both the equilibrium Cauchy stress and the elastic tensor as explicit functions of \( \FF \), the equilibrium deformation gradient relative to the fixed reference configuration \( \tilde{\MM} \). To emphasise this point we introduce the notation
\begin{subequations}
\label{eqs:sig_Lam_of_F}
\begin{alignat}{3}
  \sig &= \hat{\sig}\brak{\FF} &&\equiv 2 J_{\FF}^{-1}\Tmult{\FF} \cdot D\tilde{V}\brak{\CC} \label{eq:sig_of_F} \\
  \aaaa &= \hat{\aaaa}\brak{\FF} &&\equiv 4 J_{\FF}^{-1}\Tmult{\FF} D^{2}\tilde{V}\brak{\CC} \Tmult{\FF}^{T} + \Rmult{\hat{\sig}\brak{\FF}} \label{eq:Lam_of_F} ,
\end{alignat}	
\end{subequations}
with \(\hat{\sig} \) (resp. \(\hat{\aaaa}\)) the function that takes an equilibrium deformation gradient to the corresponding equilibrium Cauchy stress (resp. elastic tensor). It is worth observing that both of these relations are nonlinear for any non-trivial choice of strain-energy function \(\tilde{V}\). 

Through eqs.\eqref{eqs:sig_Lam_of_F} one can consider a wide variety of problems where a body's elastic properties change as a result of changing its equilibrium configuration. \citet[Section 4]{Tromp_2019} studied just such a problem when they performed their \textit{ab initio} calculations: the (unstrained) sample was subjected to a known strain, and this induced both an incremental Cauchy stress and a change in the elastic tensor. On a more seismological level \citep[e.g.][]{Tromp_2018} one might wish to understand how a given deformation of one of the Earth's regions affects seismic wave propagation and stresses therein. Such a deformation could result from small-scale phenomena such as a cave-in within a gas field, or from large-scale \wording{effects} like tidal loading. The important feature \wording{common to all these problems} is that there is a known initial state \( \tilde{\MM} \) that is deformed in a prescribed way. This produces a new state \( \MM \) whose elastic properties are fully determined in terms of those of \( \tilde{\MM} \) by eqs.\eqref{eqs:sig_Lam_of_F}. As long as \( \tilde{V} \) and its derivatives are \wording{well-behaved} the computations necessitated by these problems can in principle be performed readily. Note that to solve these problems there is in fact no need to find the stress dependence of the elastic tensor.

\subsection{Parametrised dependence on equilibrium stress}
\label{subsec:f_of_sigma}

Seismic inverse theory leads one to a subtly different problem. Say that we wish to study the equilibrium elastic properties of a certain region within the deep Earth \wording{as the equilibrium configuration evolves}, and that we would particularly like to know about the equilibrium stress. We measure that region's properties using seismic data: we make surface observations of \edit{(small-amplitude, high-frequency)} seismic waves that have passed through the region's neighbourhood, construct seismograms, and then use those seismograms to invert for the elastic tensor \( \aaaa \) of the region. Importantly, the equilibrium stress \( \sig \) cannot be measured ``directly" in this way. However, if we knew the stress dependence of the elastic tensor, we would be able to invert \textit{changes} in \( \aaaa \) for \textit{changes} in \( \sig \) as the equilibrium configuration evolves. Our task is therefore to find \( \aaaa \) as an explicit function of \( \sig \). Eqs.\eqref{eqs:sig_Lam_of_F} provide the necessary link between \( \aaaa \) and \( \sig \), but it is not yet in a useful form, parametrised as it is by \( \FF \). Unlike in the previous problem, \( \FF \) cannot be regarded as a known quantity because it could only ever be ``measured" by inverting seismic data for \( \aaaa \) and then using eq.\eqref{eq:Lam_of_F} to invert for \( \FF \). Given this, we will eliminate \( \FF \) from eqs.\eqref{eqs:sig_Lam_of_F} and ask only how \( \aaaa \) varies upon varying \( \sig \). The mathematical problem we are trying to solve can be couched most succinctly (with reference to Figure \ref{fig:setup}) as follows: if we observe a given \( \sig \) in \( \MM \), assume that \( \sig \) was induced by elastically deforming \( \tilde{\MM} \), and parametrise the properties of \( \tilde{\MM} \) by choosing a specific form of \( \tilde{V} \), what elastic tensor \( \aaaa \) will be observed in \( \MM \)?

% TODO maybe give concrete example -- initially isotropic medium; stress produces anisotropy (clear intuitively and from 3.14), will affect seismic observations; can we invert these seismic observations for stress directly w/o inverting for change in equilibrium}

%Seismic inverse theory leads one to a subtly different problem. Let us consider an idealised situation in which seismic data give us access to the elastic tensor \( \aaaa \) at a given point within the Earth through time. As the Earth's equilibrium configuration changes, so will both the elastic tensor and the equilibrium stress. The crucial distinction between this problem and the one discussed above is that now we cannot take \( \FF \) to be \wording{prescribed}; if we are studying the subsurface it is highly unlikely that we will be able to parametrise the change of equilibrium configuration so precisely. Whereas before we could ask how both \( \sig \) and \( \aaaa \) varied upon varying \( \FF \), now we \wording{can only ask} how \( \aaaa \) varies upon varying \( \sig \). Effectively, we are asking if one can parametrise changes in the equilibrium elastic properties in terms of changes in stress. In other words, we wish to find the stress dependence of the elastic tensor. The mathematical problem can be couched most succinctly as follows: if we observe a given \( \sig \) in \( \MM \), and parametrise the properties of \( \tilde{\MM} \) by assuming that \( \tilde{V} \) takes a specific form, what elastic tensor \( \aaaa \) will be observed in \( \MM \)?

The ideal approach to this second problem appears to be to find \( \FF \) as a function of \( \sig \) from eq.\eqref{eq:sig_of_F}, and then substitute the result into eq.\eqref{eq:Lam_of_F} to give \( \aaaa \) as a function of \( \sig \). However, a given equilibrium Cauchy stress can be produced by \wording{many} different deformation gradients, so there is rather a conspicuous mathematical problem. The function \( \hat{\sig} \) maps elements in the nine-dimensional group \( \gl(3) \) into the six-dimensional vector space of symmetric matrices on \( \RThree\), which means that the equation \( \hat{\sig}(\FF) = \sig \) for \(\FF\) is underdetermined. How might we ``invert" \( \hat{\sig} \) for \( \FF \) when \( \sig \) only provides us with six numbers? What does this underdetermination imply for the elastic tensor? A way into the problem is to recall the \textit{polar decomposition theorem} (MH, p.3; see also Fig. \ref{fig:pd}), which shows that any \( \FF\in\gl(3) \) can be written as
\begin{align}
  \FF = \RR\UUU = \VV\RR ,
\end{align}
for unique \( \RR\in\so(3) \) and where \( \UUU \) and \( \VV \) are unique, symmetric, positive-definite matrices.
\iffair\begin{figure}
  \begin{center}
  	\includegraphics[width=0.75\textwidth]{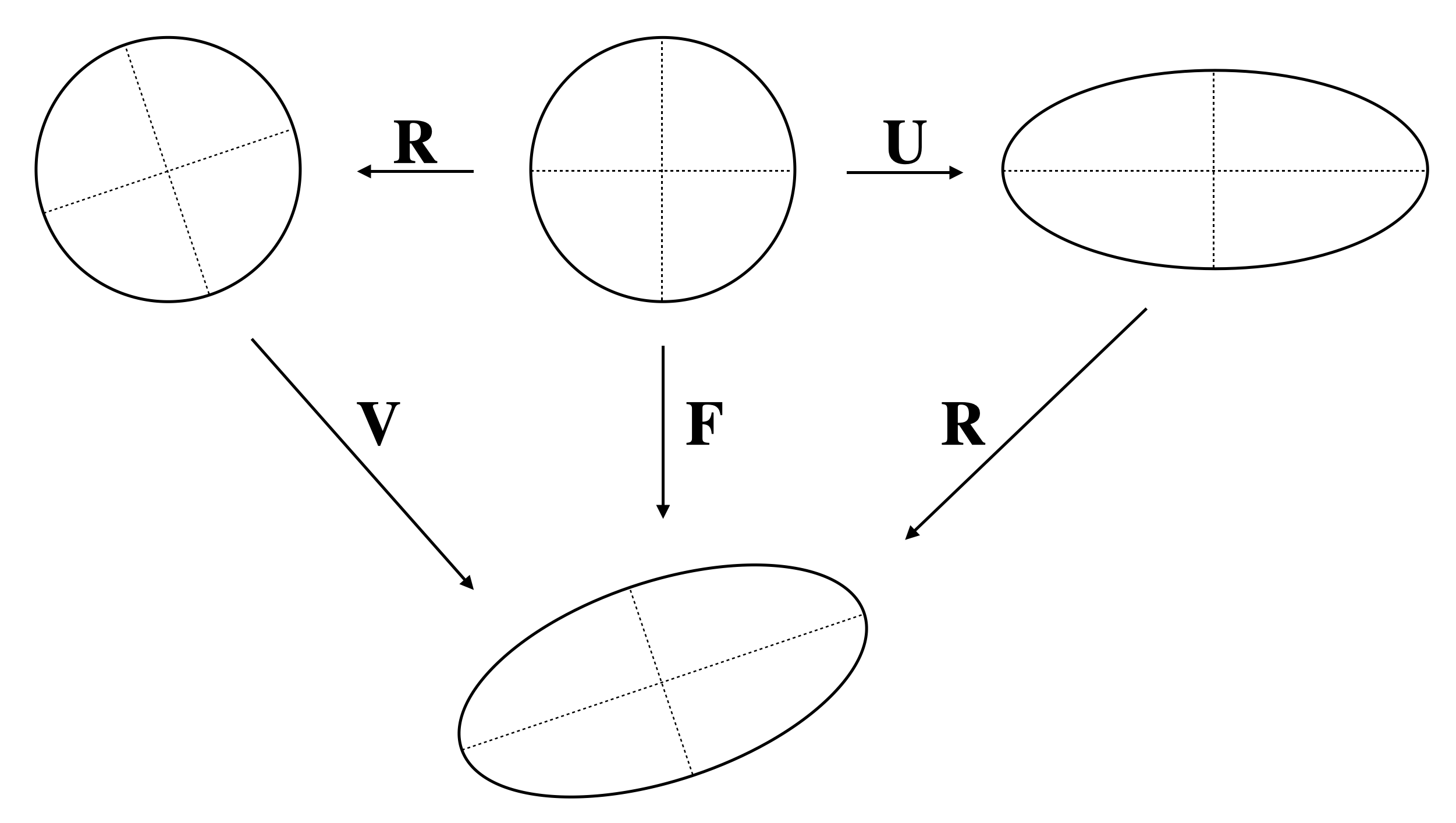}	
  \end{center}
  \caption{The polar decomposition theorem. An arbitrary element \( \FF \) of \( \gl(3) \) can be decomposed as \( \RR\UUU \), corresponding to a stretch followed by a rotation, or vice versa as \( \VV\RR \). Note that \( \UUU \) and \( \VV \) represent stretches along different principal axes. 
}
  \label{fig:pd}
\end{figure}\fi  
The theorem is a rigorous demonstration that any deformation gradient can be considered as ``stretch followed by rotation" (the first equality) or ``rotation followed by stretch" (the second). It thus provides a natural way of factoring the deformation gradient into the product of a rotation matrix and a symmetric matrix. Crucially, \( \UUU \) and \( \sig \) have the same dimensions. Our goal now becomes to investigate how far \( \RR \) and \( \UUU \) ``interact" and whether or not \( \hat{\sig} \) can in some sense be inverted for \( \UUU \). To fully resolve the issue we must examine carefully how the principle of material-frame indifference and material symmetries manifest within eqs.(\ref{eqs:sig_Lam_of_F}).

%\subsection{The role of symmetries}
%\label{subsec:f_of_sigma}

Material-frame indifference concerns the behaviour of the strain-energy function and related quantities under transformations of the deformation gradient of the form \(\FF \mapsto \RR \FF\) with \(\RR \in \so(3)\). Recalling that the value of the right Cauchy--Green deformation tensor is invariant under such a transformation, and using eq.(\ref{eq:tfunc}), we see immediately from eqs.(\ref{eqs:sig_Lam_of_F}) that
\begin{subequations}
\label{eqs:sig_Lam_pmfi}
	\begin{align}
  		\hat{\sig}\brak{\RR\FF} &= \Tmult{\RR}\cdot\hat{\sig}\brak{\FF} \label{eq:sig_pmfi} \\
  		\hat{\aaaa}\brak{\RR\FF} &= \Tmult{\RR}\hat{\aaaa}\brak{\FF}\Tmult{\RR}^{T} \label{eq:lam_pmfi}
	\end{align}	
\end{subequations}
for all \( \RR\in\so(3) \).
% \citep[e.g.][]{marsden1994mathematical}, which shows that  any  \( \FF\in\gl(3) \) can be written as
%\begin{align}
%  \FF = \RR\UUU ,
%\end{align}
%for unique \( \RR\in\so(3) \) and with \( \UUU \) a unique, symmetric, positive-definite matrix. 
Substituting the polar decomposition \( \FF=\RR\UUU \) into eqs.\eqref{eqs:sig_Lam_of_F} and making use of eqs.\eqref{eqs:sig_Lam_pmfi} we then obtain
\begin{subequations}
\label{eqs:sig_Lam_decomposed}
	\begin{align}
	  \hat{\sig}\brak{\FF} &= \Tmult{\RR}\cdot\hat{\Sig}\brak{\UUU} \label{eq:sig_decomposed} \\
	  \hat{\aaaa}\brak{\FF} &= \Tmult{\RR}\hat{\AAA}\brak{\UUU}\Tmult{\RR}^{T} \label{eq:Lam_decomposed} ,
	\end{align}	
\end{subequations}
where we have defined the functions
\begin{subequations}
\label{eqs:sig_hat_lam_hat_restrictions}
\begin{align}
  \hat{\Sig}\brak{\UUU} &= 2 J_{\UUU}^{-1} \Tmult{\UUU}\cdot D\tilde{V}\brak{\UUU^{2}} \label{eq:sig_hat_restriction}\\
  \hat{\AAA}\brak{\UUU}
  &= 4 J_{\UUU}^{-1}\Tmult{\UUU} D^{2}\tilde{V}\brak{\UUU^{2}} \Tmult{\UUU} + \Rmult{\hat{\Sig}\brak{\UUU}} \label{eq:lam_hat_restriction},
\end{align}	
\end{subequations}
whose arguments are \textit{required} to be symmetric, positive-definite matrices. Importantly, the function \( \hat{\Sig} \)  maps symmetric matrices into symmetric matrices. As a result, there is no dimensional obstruction to its being invertible. We will assume that a unique inverse \( \hat{\Sig}^{-1} \) exists wherever required, an assumption that is natural as long as the stress is not too large (see Appendix \ref{Sig_invertibility}). 

We can now examine solutions of the equation \(\hat{\sig}(\FF) = \sig\) for given \(\sig\). We write \( \FF = \RR \UUU\) as above, but now suppose that the value of \(\RR \in \so(3)\) has been fixed arbitrarily. Using eq.(\ref{eq:sig_decomposed}) we then trivially obtain
\begin{align}
\label{eq:U_of_sigma_and_R}
  \UUU &= \hat{\Sig}^{-1}\brak{ \Tmult{\RR}^{T}\cdot \sig } ,
\end{align}
whence it follows that the equilibrium deformation gradient is given by 
\begin{align}
\label{eq:F_hat_initial}
  \FF = \hat{\FF}\brak{\sig,\RR} \equiv \Lmult{\RR} \cdot \hat{\Sig}^{-1}\brak{ \Tmult{\RR}^{T}\cdot \sig }  .
\end{align}
Here we see concretely where the missing three degrees of freedom enter into the inversion of \(\hat{\sig}\). While eq.(\ref{eq:F_hat_initial}) constitutes \textit{one} solution of the given equation, it is not unique. Letting \(\RR\) vary over \( \so(3)\) generates a three-parameter family of solutions, and the uniqueness of the polar decomposition means that every \( \RR \) must correspond to a different \( \FF \). Moreover, \textit{any} \( \FF \) that solves \(\hat{\sig}(\FF) = \sig\) can be polar-decomposed and written in terms of \( \sig \) using eq.(\ref{eq:F_hat_initial}), so we have clearly obtained \textit{all} solutions of the equation. Crucially, the non-uniqueness carries over into  the elastic tensor. Substitution of eq.(\ref{eq:F_hat_initial}) into eq.(\ref{eq:Lam_of_F}) yields
\begin{align}
\label{eq:lambda_of_sigma_final}
  \aaaa = \hat{\aaaa}\!\braksq{\hat{\FF}\brak{\sig,\RR}} \equiv \bar{\aaaa}\brak{\sig,\RR} ,
\end{align}
where the function \( \bar{\aaaa} \) can be written more expansively as
\begin{align}
\label{eq:Lam_bar_full_form}
  \bar{\aaaa}\brak{\sig,\RR} = \Tmult{\RR}\braksq{\brak{\hat{\AAA}\circ\hat{\Sig}^{-1}}\brak{\Tmult{\RR}^{T}\cdot\sig}}\Tmult{\RR}^{T}  .
\end{align}
It follows that we cannot expect to write the elastic tensor as a function of the equilibrium Cauchy stress alone. A definite value for \(\aaaa\) depends on the arbitrary choice of an element of \(\so(3)\).

We can add some nuance to this result by considering material symmetries. Let \( \sym(\tilde{W}) \) denote the material symmetry group (which could be trivial) at the point of interest, whose elements \(\QQ\) act on the deformation gradient on the right through \(\FF \mapsto \FF \QQ\). In terms of the right Cauchy--Green deformation tensor such a transformation takes the form \( \CC \mapsto \QQ^{T} \CC \QQ = \Tmult{\QQ}^{T}\cdot \CC\), and by definition we have
\begin{align}
  \label{eq:vsym}
  \tilde{V}(\Tmult{\QQ}^{T}\cdot \CC ) = \tilde{V}(\CC) 
\end{align}
for all \( \QQ \in \sym(\tilde{W})\). Differentiating this relation in the now standard manner yields
\begin{align}
  D\tilde{V}\brak{\CC} &= \Tmult{\QQ}\cdot D\tilde{V}\brak{\Tmult{\QQ}^{T}\cdot\CC} \\
  D^{2}\tilde{V}\brak{\CC} &= \Tmult{\QQ}D^{2}\tilde{V}\brak{\Tmult{\QQ}^{T}\cdot\CC}\Tmult{\QQ}^{T}  .
\end{align}
Using the properties of \( \Tmult{\FF} \) we then find from eqs.(\ref{eqs:sig_Lam_of_F}) that
\begin{subequations}
\begin{align}
  \hat{\sig}\brak{\FF\QQ} &= 2 J_{\FF}^{-1}\Tmult{\FF} \cdot \braksq{\Tmult{\QQ}\cdot D\tilde{V}\brak{\Tmult{\QQ}^{T}\cdot\CC}} \\
  \hat{\aaaa}\brak{\FF\QQ} &= 4 J_{\FF}^{-1}\Tmult{\FF}\braksq{\Tmult{\QQ}D^{2}\tilde{V}\brak{\Tmult{\QQ}^{T}\cdot\CC}\Tmult{\QQ}^{T}}\Tmult{\FF}^{T} + \Rmult{\hat{\sig}\brak{\FF\QQ}} ,
\end{align}	
\end{subequations}
from which it readily follows that
\begin{subequations}
\label{eqs:sig_lam_mat_sym}
	\begin{align}
  		\hat{\sig}\brak{\FF\QQ} &= \hat{\sig}\brak{\FF} \label{eq:sig_mat_sym} \\
  		\hat{\aaaa}\brak{\FF\QQ} &= \hat{\aaaa}\brak{\FF} \label{eq:lam_mat_sym}
	\end{align}	
\end{subequations}
for any \( \QQ \in \sym(\tilde{W}) \). 

We can now understand \( \bar{\aaaa} \)'s non-unique stress dependence by studying how the structure implied by eqs.(\ref{eqs:sig_lam_mat_sym}) manifests within \( \hat{\Sig}^{-1} \). To that end, it is simplest if we consider \( \tilde{\MM} \) to constitute a natural reference configuration for an equilibrium body that is stress-free. What follows is therefore based on the assumption that there exists \textit{some} stress-free reference-configuration with respect to which we can describe \( \MM \). This way we can take \( \sym(\tilde{W}) \) to be \textit{equal}, rather than just isomorphic, to a subgroup of \( \so(3) \). Because all \( \QQ \in \sym(\tilde{W}) \) are then rotations, the material symmetry group acts through \( (\RR,\UUU) \mapsto (\RR \QQ, \QQ^{T} \UUU \QQ) \) on the level of the polar decomposition. Therefore, using eq.(\ref{eq:sig_decomposed}) we find that eq.(\ref{eq:sig_mat_sym}) requires
\begin{align}
\label{eq:hat_Sig_symmetry}
  \hat{\Sig}\brak{\Tmult{\QQ}\cdot\UUU} = \Tmult{\QQ}\cdot\hat{\Sig}\brak{\UUU}
\end{align}
for arbitrary \( \QQ \in \sym(\tilde{W}) \).
This equation expresses the intuitive notion that a stretch \( \UUU \) and rotation \( \QQ \) will together induce the same stress no matter the order in which they are imposed -- as long as \( \QQ \) is in the material symmetry group. In any case, by acting the inverse function \( \hat{\Sig}^{-1} \) on this expression we obtain the analogous result
\begin{align}
  \hat{\Sig}^{-1}\brak{\Tmult{\QQ}\cdot\Sig} = \Tmult{\QQ}\cdot\hat{\Sig}^{-1}\brak{\Sig}
\end{align}
for arbitrary symmetric \( \Sig \), which we may substitute into eq.\eqref{eq:F_hat_initial} to conclude that
\begin{align}
\label{eq:F_hat_mat_sym}
  \hat{\FF}\brak{\sig,\RR\QQ}
  &= \Lmult{\RR\QQ}\cdot \hat{\Sig}^{-1}(\Tmult{\RR\QQ}^{T}\cdot\sig) \nonumber\\
  &= (\Lmult{\RR}\Lmult{\QQ})\cdot \hat{\Sig}^{-1}[\Tmult{\QQ}^{T}\cdot(\Tmult{\RR}^{T}\cdot\sig)] \nonumber\\
  &= (\Lmult{\RR}\Lmult{\QQ} \Tmult{\QQ}^{T}) \cdot \hat{\Sig}^{-1}(\Tmult{\RR}^{T}\cdot\sig) \nonumber\\
  &= \hat{\FF}\brak{\sig,\RR}\QQ  .
\end{align}
This relationship does not allow us to determine \( \FF \) any more precisely -- it will \textit{always} be known only up to an element of \( \so(3) \) -- but we may trivially write
\begin{align}
  \hat{\aaaa}\braksq{\hat{\FF}\brak{\sig,\RR\QQ}} = \hat{\aaaa}\braksq{\hat{\FF}\brak{\sig,\RR}\QQ}  .
\end{align}
It follows immediately from eq.(\ref{eq:lam_mat_sym}) that
\begin{align}
  \hat{\aaaa}\braksq{\hat{\FF}\brak{\sig,\RR\QQ}} = \hat{\aaaa}\braksq{\hat{\FF}\brak{\sig,\RR}}  .
\end{align}
Hence, using the notation of eq.(\ref{eq:lambda_of_sigma_final}), we have obtained the key identity
\begin{align}
\label{eq:Lam_bar_mat_sym}
  \bar{\aaaa}\brak{\sig,\RR\QQ} = \bar{\aaaa}\brak{\sig,\RR} \quad \forall \QQ\in\sym(\tilde{W})  .
\end{align}

Eq.(\ref{eq:Lam_bar_mat_sym}) implies that two distinct rotation matrices \( \RR,\RR' \in\so(3)\) will lead to the same elastic tensor via eq.\eqref{eq:lambda_of_sigma_final} if \( \RR^{T} \RR' \in \sym(\tilde{W}) \). It is readily verified that this defines an equivalence relation, meaning that \( \so(3) \) can be partitioned into distinct equivalence classes, with the resulting \textit{quotient space} denoted by \( \so(3)/\sym(\tilde{W})\). The function \( \bar{\aaaa}\brak{\sig,\cdot} \) therefore depends not on the rotation matrix \( \RR \) directly, but only on the equivalence class in \( \so(3)/\sym(\tilde{W})\) to which it belongs. In this manner the number of additional parameters required to fix a definite value of the elastic tensor can be reduced.

In summary, we have shown that it is possible to express the elastic tensor as a function of equilibrium stress, but only at the cost of introducing extra arbitrary parameters. Given our initial comments about the form of \( \hat{\sig} \), the presence of these parameters is not surprising. After all, we were essentially trying to fix nine numbers knowing only six. What is pleasing is that we have been able to exploit material-frame indifference to `package' these extra degrees of freedom into a rotation matrix and write down a solution that is still well-defined. In addition, although our argument appeared at first to suggest that the rotation matrix was wholly \wording{arbitrary}, we have shown that the presence of material symmetries can reduce the number of \wording{arbitrary} parameters to be specified. On a physical level, we have shown formally that if one observes a given Cauchy stress in an \wording{arbitrary} (hyperelastic) material, and assumes the material to have reached its present state by some elastic deformation from a prior state with known properties, it is generally impossible to ``disentangle" the effects of the rotation- and stretch-components of the elastic deformation. As a consequence, one cannot in general construct the elastic tensor unambiguously.

\subsection{Linearisation}
\label{subsec:lam_sig_linearisation}

In a geophysical context it will often be convenient to regard the total equilibrium Cauchy stress in the body \( \MM \) as some small perturbation to the equilibrium Cauchy stress of the reference-body \( \tilde{\MM} \). It is therefore useful to linearise expression (\ref{eq:lambda_of_sigma_final}), the calculations for which are laid out in Appendix \ref{app:calculations}. We assume that there exists a zeroth-order equilibrium configuration \( \tilde{\MM} \) possessing Cauchy stress \( \sig^{0} \) and elastic tensor \( \aaaa^{0} = \cccc^{0} +\Rmult{\sig^{0}} \). We can set \( \RR^{0} \) to the identity without loss of generality because this zeroth-order state is taken to be known.

We linearise the system about a small perturbing stress. With \( \epsilon \) a small parameter, the stress is set to
\begin{align}
  \sig = \sig^{0} + \epsilon\,\sig^{1} .
\end{align}
We must also linearise the rotation matrix of eq.(\ref{eq:lambda_of_sigma_final}). It is set to the identity at zeroth-order, so its perturbation satisfies
\begin{align}
  \RR = \id + \epsilon\,\omb^{1}+ \mathcal{O}\brak{\epsilon^{2}},
\end{align}
with \( \omb^{1} \) an antisymmetric matrix. Substituting these into eq.(\ref{eq:lambda_of_sigma_final}) and Taylor expanding, we may write
\begin{align}
  \aaaa = \aaaa^{0} + \epsilon\,\aaaa^{1} + \mathcal{O}\brak{\epsilon^{2}} .
\end{align}
The perturbed elastic tensor \( \aaaa^{1} \) is decomposed as
\begin{align}
  \aaaa^{1} = \cccc^{1} + \Rmult{\sig^{1}},
\end{align}
consistent with eq.(\ref{eq:lamrep2}). Under these definitions, we show in Appendix \ref{app:sub:general_calculation} that \( \cccc^{1} \) is given by
\begin{subequations}
\label{eq:linearisation_general}	
\begin{align}
  &\cccc^{1} = \Chimult{\uu^{1}+\omb^{1}}\cccc^{0} + \cccc^{0}\Chimult{\uu^{1}-\omb^{1}} 
            - \cccc^{0}\tr{\uu^{1}} + 8 D^{3}\tilde{V}\brak{\id}\cdot\uu^{1} , \label{eq:linearisation_general_c} \\[2pt]
  &\text{where \( \uu^{1} \) is a symmetric matrix which satisfies a generalised linear stress-strain relationship} \nonumber\\
  &\sig^{1}-\Chimult{\omb^{1}}\cdot\sig^{0} = \brak{\cccc^{0}+\ChimultHat{\sig^{0}}-\sig^{0}\otimes\id}\cdot\uu^{1}  . \label{eq:linearisation_general_sig}
\end{align}
\end{subequations}
In these expressions we have introduced the tensor product on matrices (eq.\ref{eq:tensor_product_gl3_defn}) and the notations \( \Chimult{} \) (eq.\ref{eq:chimult_defn}) and \( \ChimultHat{} \) (eq.\ref{eq:fourth_rank_sym_defn}). Thus, in order to fully specify the perturbation to the elastic tensor we must provide:
\begin{enumerate}
\renewcommand{\theenumi}{(\arabic{enumi})}
    \item the perturbation to the stress, \( \sig^{1} \);
	\item \( \sig^{0} \) and \( \cccc^{0} \), which encode information about the zeroth-order equilibrium body;
	\item further information about the zeroth-order equilibrium body, via the third derivatives of its strain-energy function at equilibrium;
	\item an arbitrary antisymmetric matrix \( \omb^{1} \).
\end{enumerate}

Eqs.(\ref{eq:linearisation_general}) is a general result whose derivation makes no particular demands on the form of the zeroth-order equilibrium configuration, but we can already make several observations. Firstly, no matter its functional form, \( \cccc^{1} \) can be shown to possess all the classical elastic symmetries, as required by eq.(\ref{eq:lamrep2}). Secondly, given that it is explicitly linear in the stress-perturbation, this theory is superficially rather close to those of \cite{dahlen_1972a}, \cite{dahlen1998theoretical}, \cite{Tromp_2018} and \cite{Tromp_2019}, discussed in Section \ref{sec:intro}. There are some important differences, though. For one, our linearised theory makes explicit reference to third derivatives of the strain-energy function at equilibrium. In fact, those third derivatives can be seen as parametrising the theory. Moreover, eqs.(\ref{eq:linearisation_general}) is parametrised further by the arbitrary degrees of freedom associated with \( \omb^{1} \). One might imagine that terms in \( \omb^{1} \) would drop out when we compute, say, the Christoffel operator, so that it would have no effect on observable phenomena, but we show later, for the particular case of a transversely-isotropic material, that this does not happen. The relationship between the existing theories and this new linearised theory will become clearer as we consider some more concrete examples.

%%%%%%%%%%%%%%%%%%%%%%%%%%%%%%
\section{Examples}
\label{sec:calculations}

We now illustrate how the preceding results apply to a few different physical situations. We will consider both large and small stresses, making use of eqs.(\ref{eq:linearisation_general}) for the latter. The examples discussed here are intended simply to illustrate the general behaviour implied by our theoretical results. For that reason we have used standard, simple strain-energy functions, and all physical quantities are presented suitably nondimensionalised. Henceforward, we will refer to the body \( \MM \) of the previous section simply as `the equilibrium body'. We will describe the fixed reference body \( \tilde{\MM} \) as the \textit{background body}, and similarly for all its associated quantities such as strain-energy function and material symmetry group. All calculations for the following examples were carried out in Mathematica 12 (with the relevant notebook included in the supplementary material). The scenarios involving large stress required numerical inversion of the nonlinear function \( \hat{\Sig} \), for which we used Mathematica's inbuilt `FindRoot' function. 

\subsection{Isotropic materials}

We begin by considering isotropic, stress-free background bodies. In the isotropic special case \( \sig \) alone provides a unique specification of \( \aaaa \). To see this, it is sufficient to return to the identity
\begin{align}
  \bar{\aaaa}\brak{\sig,\RR} = \bar{\aaaa}\brak{\sig,\RR\QQ} ,
\end{align}
for some \( \RR \) and \( \QQ \) both in \( \so(3) \). By the standard group axioms we may write
\begin{align}
  \QQ = \RR^{T}\RR',
\end{align}
where \( \RR' \) is another arbitrary element of \( \so(3) \), whence
\begin{align}
\label{eq:lam_R_R_prime_iso_equal}
  \bar{\aaaa}\brak{\sig,\RR} = \bar{\aaaa}\brak{\sig,\RR'}  .
\end{align}
The matrices \( \RR \) and \( \RR' \) are both arbitrary, so \( \bar{\aaaa}\brak{\sig,\cdot} \) is independent of the choice of rotation matrix. When evaluating \( \bar{\aaaa} \) we may therefore set \( \RR=\id \) without loss of generality. The elastic tensor is then given conveniently by
\begin{align}
\label{eq:Lam_of_sig_iso}
  \aaaa = \bar{\aaaa}\brak{\sig,\id}  .
\end{align}
This argument shows that all rotations are equivalent up to right multiplication by \( \so(3) \); in other words, the quotient space
\begin{align}
  \so(3)/\sym(\tilde{W}) = \so(3)/\so(3)
\end{align}
is a set which contains just one element. With these preliminaries we are in position to investigate the behaviour of isotropic materials under induced stress.

\subsubsection{Exact response to deviatoric stress}

The nonlinearity of expression (\ref{eq:lambda_of_sigma_final}) implies that a general material's response to induced stress is influenced by derivatives of the strain-energy function higher than second-order.
\iffair\begin{figure}
  \includegraphics[width=\textwidth]{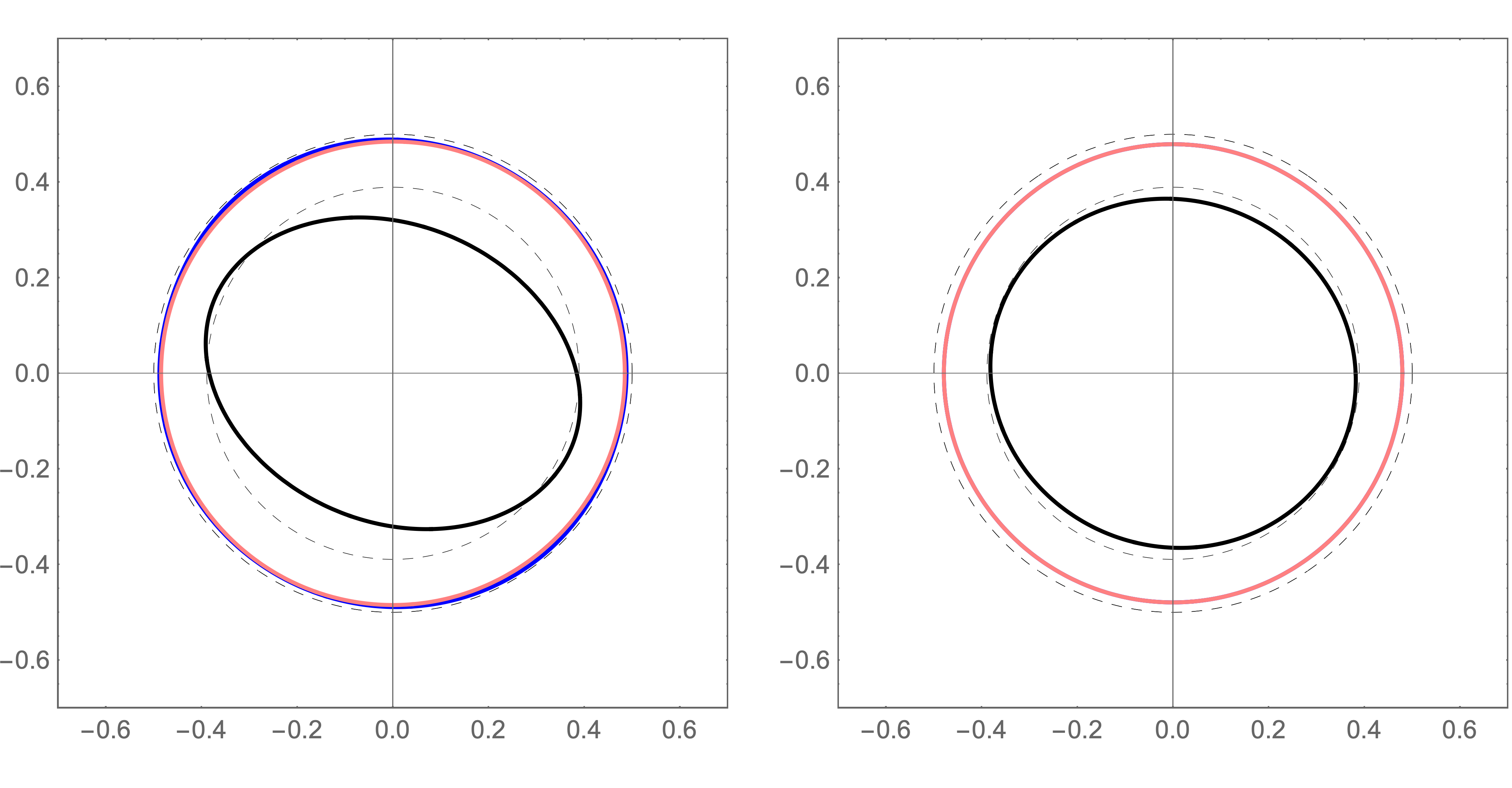}
  \caption{The behaviour of two different isotropic materials subject to the same induced stress. The left panel shows how a material governed by a modified Saint-Venant Kirchhoff strain-energy function reacts when a certain deviatoric stress of magnitude \( \|\sig\| \sim \mu \) is induced. We plot the \( x-y \) slowness surface, with the zero-stress slowness surface shown faintly for reference. P-waves and S-waves are both affected by the stress. We show the same on the right, but for a material governed by a neo-Hookean constitutive relation. Shear-waves are not noticeably split here and the P-wave response is muted. Thus, the two materials behave differently under stress, despite being indistinguishable in its absence.}
  \label{fig:constitutive_relation_comparison_isotropic}
\end{figure}\fi
We demonstrate this in Fig. \ref{fig:constitutive_relation_comparison_isotropic}, contrasting the slowness surfaces of two different isotropic materials under the same induced stress. The strain-energy functions describing the background bodies are \citep[e.g.][]{Holzapfel_2000} \textit{modified Saint-Venant Kirchhoff},
\begin{align}
  \tilde{W}_{\mathrm{MSVK}}\brak{\FF} = \frac{\lambda}{2}\log^{2}{J} + \frac{\mu}{4}\tr{\,\braksq{\CC-\id}^{2}},
\end{align}
and \textit{neo-Hookean},
\begin{align}
  \tilde{W}_{\mathrm{NH}}\brak{\FF} = \frac{\mu}{2}\braksq{\tr{\CC} - 3 + \frac{2\mu}{\lambda}(J^{-\frac{\lambda}{\mu}}-1)} ,
\end{align}
where \( \lambda \) and \( \mu \) are constants. The limit of vanishing induced stress is obtained in both cases by evaluating the background strain-energy functions and their derivatives at \( \FF=\id \). In that case the materials are indistinguishable, each possessing the classical isotropic elastic tensor
\begin{align}
  \mathsf{a}_{ijkl}
  =\mathsf{c}_{ijkl} 
  =\lambda\delta_{ij}\delta_{kl} + \mu\brak{\delta_{ik}\delta_{jl}+\delta_{il}\delta_{jk}}  .
\end{align}
Indeed, any strain-energy function that purports to describe an isotropic solid must give this result in the relevant limit. It is also apparent that \( \lambda \) and \( \mu \) should be interpreted as the standard Lam\'{e} parameters. Under large induced stress, though, the strain-energy functions are evaluated away from the identity, meaning that third- and higher-order derivatives become relevant. As shown in Fig.\ref{fig:constitutive_relation_comparison_isotropic}, where we have induced a deviatoric stress of magnitude \( \|\sig\| \sim \mu \), the materials then display distinct behaviour.

\subsubsection{Exact response to hydrostatic stress}
\label{isotropic_hydrostatic_stress_exact}

When a hydrostatic pressure is induced in a stress-free, isotropic solid, its only effect on the elastic tensor is to change the elastic moduli. Here we derive exact expressions for \( \lambda \) and \( \mu \) as functions of pressure. We return to eqs.(\ref{eqs:sig_hat_lam_hat_restrictions}) and write 
\begin{align}
  \UUU = \phi\id,
\end{align}
for some positive scalar \( \phi \). Given that the system is isotropic and the induced stress hydrostatic, it follows from symmetry considerations and eq.(\ref{eq:hat_Sig_symmetry}) that \( \UUU \) cannot take any other form. The deformation gradient itself can only ever be known up to an arbitrary rotation matrix, so it is given by
\begin{align}
  \FF=\RR\UUU = \phi\RR,
\end{align}
with \( \RR\in\so(3) \), while the right Cauchy--Green deformation tensor is
\begin{align}
  \CC=\phi^{2}\id .
\end{align}
This deformation gradient corresponds physically to a local compression or dilation of the background-body with a rotation superimposed; \( \phi < 1 \) effects a compression and vice versa. If the resulting equilibrium pressure in \( \MM \) is \( p \), the Cauchy stress is \( \sig = -p\id \), so from eqs.(\ref{eqs:sig_Lam_of_F},\ref{eqs:sig_Lam_decomposed},\ref{eqs:sig_hat_lam_hat_restrictions})
\begin{subequations}
\begin{align}
  -p\id &= \frac{2}{\phi}D\tilde{V}\brak{\phi^{2}\id} \\
  \aaaa &= 4\phi D^{2}\tilde{V}\brak{\phi^{2}\id} + \Rmult{\sig}  .
\end{align}	
\end{subequations}
We can set \( \RR \) to the identity in these expressions because the background material is isotropic (see eq.\ref{eq:Lam_of_sig_iso}). 

Now, for the stress-free isotropic medium represented by \( \tilde{\MM} \), the strain-energy function \( \tilde{V} \) is a function of the three \textit{scalar invariants} of \( \CC \) \citep{Holzapfel_2000}. Defining the scalar invariants as
\begin{align}
\label{eq:scalar_invariants_defn_iso}
  I_{i}\brak{\CC} \equiv \tr{\CC^{i}},\quad i=1,2,3 ,
\end{align}
we can write the strain-energy function as
\begin{align}
  \tilde{V}\brak{\CC} \equiv \widehat{V}\braksq{I_{1}\brak{\CC},I_{2}\brak{\CC},I_{3}\brak{\CC}}.
\end{align}
From the chain rule, its first and second derivatives are
\begin{align}
  D\tilde{V} &= \sum_{i}\widehat{V}_{i} \, D I_{i} \\
  D^{2}\tilde{V} &= \sum_{ij}\widehat{V}_{ij} \, D I_{i}\otimes D I_{j} + \sum_{i}\widehat{V}_{i} \, D^{2}I_{i} ,
\end{align}
where we have written the derivatives of \( \widehat{V} \) with respect to its arguments in an obvious way. When the derivatives are evaluated at \( \CC=\phi^{2}\id \), we will write e.g.
\begin{align}
  v_{13}  
  = \widehat{V}_{13}\braksq{I_{1}\brak{\phi^{2}\id}\!,\,
  					    I_{2}\brak{\phi^{2}\id}\!,\,
  					    I_{3}\brak{\phi^{2}\id}} ,
\end{align}
and similarly for the other derivatives. With this, one finds after a little algebra that
\begin{subequations}
\begin{align}
  -p\id &= 
  \frac{2}{\phi}\brak{v_{1} +2v_{2} \phi^{2}+3v_{3} \phi^{4}}\id \\
  \aaaa &= 
  4\phi\braksq{
  \brak{v_{11} +4v_{12} \phi^{2}+\brak{4v_{22} +6v_{13} }\phi^{4} + 12v_{23} \phi^{6}+9v_{33} \phi^{8}}\id\otimes\id
  \right.\nonumber\\
  &\qquad\quad\left.
  +\brak{2v_{2}  + 6v_{3} \phi^{2}}\overline{\idgen}
  } + \Rmult{\sig}  .
\end{align}	
\end{subequations}
where the operator \( \overline{\idgen} \) has been defined in eq.(\ref{eq:id_hat_defn}) and has components
\begin{align}
  (\overline{\idgen})_{ijkl} = \frac{1}{2}\brak{\delta_{ik}\delta_{jl}+\delta_{il}\delta_{jk}} .
\end{align}
We identify the coefficients of the first two operators in the expression for \( \aaaa \) as the Lam\'{e} parameters, given as functions of \( \phi \), so we reach the equations
\begin{subequations}
\begin{align}
  &\lambda = 4\phi\braksq{v_{11} +4v_{12} \phi^{2}+\brak{4v_{22} +6v_{13} }\phi^{4} + 12v_{23} \phi^{6}+9v_{33} \phi^{8}} \label{eq:lam_pressure_iso} \\
  &\mu = 4\phi \brak{v_{2}  + 3v_{3} \phi^{2}} \label{eq:mu_pressure_iso}  .\\
&\text{\( \phi \) is obtained as a function of \( p \) by inverting the relation} \nonumber\\	
&-p = \frac{2}{\phi}\brak{v_{1} +2v_{2} \phi^{2}+3v_{3} \phi^{4}} ,
\end{align}
\end{subequations}
which would be performed numerically for a general strain-energy function. The Lam\'{e} parameters are then given as explicit functions of the equilibrium pressure, parametrised by the derivatives of \( \widehat{V} \).

It is also useful to note that the form of \( \UUU \) considered here, despite producing a stressed configuration from an unstressed one, does not alter the material symmetry group, \( \sym(\tilde{W}) = \so(3) \). Material symmetry groups transform under a particle-relabelling transformation according to eq.(\ref{eq:sym_group_relation_under_conjugation}), and with \( \FF=\phi\RR \)
\begin{align}
  \sym(W) 
  &= \brakbr{\FF^{-1}\QQ\FF,\QQ\in\sym(\tilde{W})} \nonumber\\
  &= \brakbr{\RR^{T}\QQ\RR,\QQ\in\sym(\tilde{W})} \nonumber\\
  &= \sym(\tilde{W})  .
\end{align}
As expected on physical grounds, inducing a pressure in an isotropic body does not break the isotropy. All our conclusions from Section \ref{sec:lambda_of_sigma} therefore apply to an isotropic body even under hydrostatic stress. In particular, we can linearise about a hydrostatically stressed equilibrium given by \( \UUU=\id \) (Appendix \ref{app:sub:general_calculation}) without having to refer the system back to some unstressed state.

\subsubsection{Linearised response to small stress}

Eqs.(\ref{eq:linearisation_general}) simplify dramatically when we consider a small stress induced in a hydrostatically pre-stressed equilibrium. The total stress is written as
\begin{align}
  \sig = -p^{0}\id -p^{1}\id +\DD^{1} ,
\end{align}
with
\begin{align}
  p^{1} &\ll p^{0} \nonumber\\ 
  \|\DD^{1}\| &\ll p^{0} \nonumber\\
  \tr{\DD^{1}} &= 0 ,
\end{align}
and the complete elastic tensor is (Appendix \ref{app:sub:isotropic_calculation})
\begin{align}
\label{eq:linearisation_isotropic}
  \mathsf{a}_{ijkl} 
  &= \brak{\kappa-\frac{2}{3}\mu+p^{1}c}\delta_{ij}\delta_{kl}
   + \brak{\mu+ p^{1}d}\brak{\delta_{ik}\delta_{jl}+\delta_{il}\delta_{jk}}
   -\brak{p^{0}+p^{1}}\delta_{ik}\delta_{jl} \nonumber\\
   &\qquad 
   + a\brak{\delta_{ij}\tau^{1}_{kl}+\delta_{kl}\tau^{1}_{ij}}
   + b\brak{\delta_{ik}\tau^{1}_{jl}+\delta_{jl}\tau^{1}_{ik}+\delta_{il}\tau^{1}_{jk}+\delta_{jk}\tau^{1}_{il}} + \delta_{ik}\tau^{1}_{jl} .
\end{align}
The constants \( a \), \( b \), \( c \) and \( d \) are defined to be
\begin{subequations}
\label{eqs:iso_constants_defns}
\begin{align}
  a &= \frac{\kappa-\frac{2}{3}\mu + \frac{1}{2}\zeta_{2}}{\mu-p^{0}} \label{eq:iso_constants_defns_a} \\
  b &= \frac{\mu + \frac{1}{4}\zeta_{3}}{\mu-p^{0}} \label{eq:iso_constants_defns_b} \\
  c &= -\frac{\kappa-\frac{2}{3}\mu + 3\zeta_{1} + 2\zeta_{2}}{3\kappa+p^{0}} \label{eq:iso_constants_defns_kappa_prime}\\
  d &= -\frac{\mu + \frac{3}{2}\zeta_{2} + \zeta_{3}}{3\kappa+p^{0}} \label{eq:iso_constants_defns_mu_prime},  
\end{align}	
\end{subequations}
with \( \zeta_{1} \), \( \zeta_{2} \) and \( \zeta_{3} \) the \textit{Murnaghan constants} \citep{Murnaghan_1937} which offer a complete characterisation of the third derivatives of an isotropic strain-energy function about equilibrium. Up to third-order accuracy in the deformation gradient, specification of \( p^{0} \), \( \kappa \), \( \mu \), \( \zeta_{1} \), \( \zeta_{2} \) and \( \zeta_{3} \) is sufficient to fix all the elastic properties of the background body. In light of Section \ref{isotropic_hydrostatic_stress_exact}, it should be emphasised that \( \kappa \), \( \mu \), \( \zeta_{1} \), \( \zeta_{2} \) and \( \zeta_{3} \) are constants defined relative to the hydrostatically stressed background. It should be noted that these results first appeared (up to notation) in the nearly forgotten work of \citet[Section 5]{Walton_1974}, which was in part a response to Dahlen's \citeyearpar{dahlen_1972a,dahlen_1972b} papers.

Eq.(\ref{eq:linearisation_isotropic}) is precisely equivalent to eq.\eqref{eq:xi_iso_expression_abcd}. In particular, the four independent components of \( \Pib \) that we identified by symmetry arguments in Section \ref{intro_isotropic} have simply dropped out of the linearisation procedure. Moreover, our consideration of constitutive theory has led to a \wording{more} precise definition of those constants. All four are explicitly determined by the constitutive relation used to describe the background body, emerging from the theory as dimensionless combinations of the equilibrium pressure, shear and bulk moduli, and the three Murnaghan constants. This shows explicitly that they represent just three degrees of freedom. These results reduce to those of \citet{dahlen_1972b} and \citet{Tromp_2018} for certain values of the elastic moduli and Murnaghan constants; we present a detailed comparison between this theory and previous work in Appendix \ref{app:matching_theories}.

\subsection{Transversely-isotropic materials}

Whilst an isotropic material's response to a given induced stress is determined solely by its background strain-energy function, we also need to account for eq.(\ref{eq:lambda_of_sigma_final})'s non-unique stress-dependence when considering materials with smaller symmetry groups. For example, we stated above that the symmetry group of a stress-free, transversely-isotropic material is \( \so(2) \). A definite value of \( \aaaa \) therefore depends upon the choice of an arbitrary element of the quotient space \( \so(3)/\so(2) \). This reflects the fact that \( \bar{\aaaa}\brak{\sig,\cdot} \) cannot distinguish between matrices that only differ in how much rotation they cause about the symmetry axis. Therefore to evaluate \( \bar{\aaaa}\brak{\sig,\RR} \) we should only choose arbitrarily between rotation matrices whose own axes of rotation lie in the plane perpendicular to the symmetry axis. In order to pick such a matrix we must choose a direction for its rotation-axis -- a direction in \( \reals^{2} \) described by an angle \( \phi \) -- and then specify an angle of rotation \( \theta \) about that axis. Careful consideration of the possibility of double-counting shows that
\begin{align}
\begin{aligned}
  \theta &\in [0, \pi) \\
  \phi   &\in [0,2\pi)
\end{aligned}
\end{align}	
(or vice versa). It is clear that we have effectively specified a point on \( \SN{2} \), the unit 2-sphere, or equivalently a direction in \( \RThree \). Indeed, it may be established by rigorous methods that \( \so(3)/\so(2) \cong \SN{2} \).

\subsubsection{Exact response to deviatoric stress}

The effect of \( \aaaa \)'s non-unique stress dependence is illustrated by Fig. \ref{fig:anisotropic_nonuniqueness}, which shows slowness surfaces of a material described by the transversely-isotropic strain-energy function
\begin{align}
  \tilde{W}_{\mathrm{TI}}\brak{\FF} = \tilde{W}_{\mathrm{MSVK}}\brak{\FF} + \braksq{\alpha + 2\beta\log{J} + \gamma\brak{I_{4}-1}}\brak{I_{4}-1} - \frac{\alpha}{2}\brak{I_{5}-1}  .
\end{align}
In this equation \( \alpha \), \( \beta \) and \( \gamma \) are extra material constants required to describe a transversely-isotropic material, while \( I_{4} \) and \( I_{5} \) are the two further scalar invariants in terms of which a transversely-isotropic strain energy function is parametrised \citep{Holzapfel_2000}. They are defined as
\begin{align}
  I_{4} &= \innerproduct{\nub}{\CC\cdot\nub} \\
  I_{5} &= \innerproduct{\nub}{\CC^{2}\cdot\nub} ,
\end{align}
with the unit-vector \( \nub \) pointing along the material's symmetry-axis. This strain-energy function is adapted from \citet{Bonet_1998}, although our definition of \( \beta \) differs from theirs by a factor of two and we have defined the `isotropic part' of the function differently.
\iffair\begin{figure}
\includegraphics[width=\textwidth]{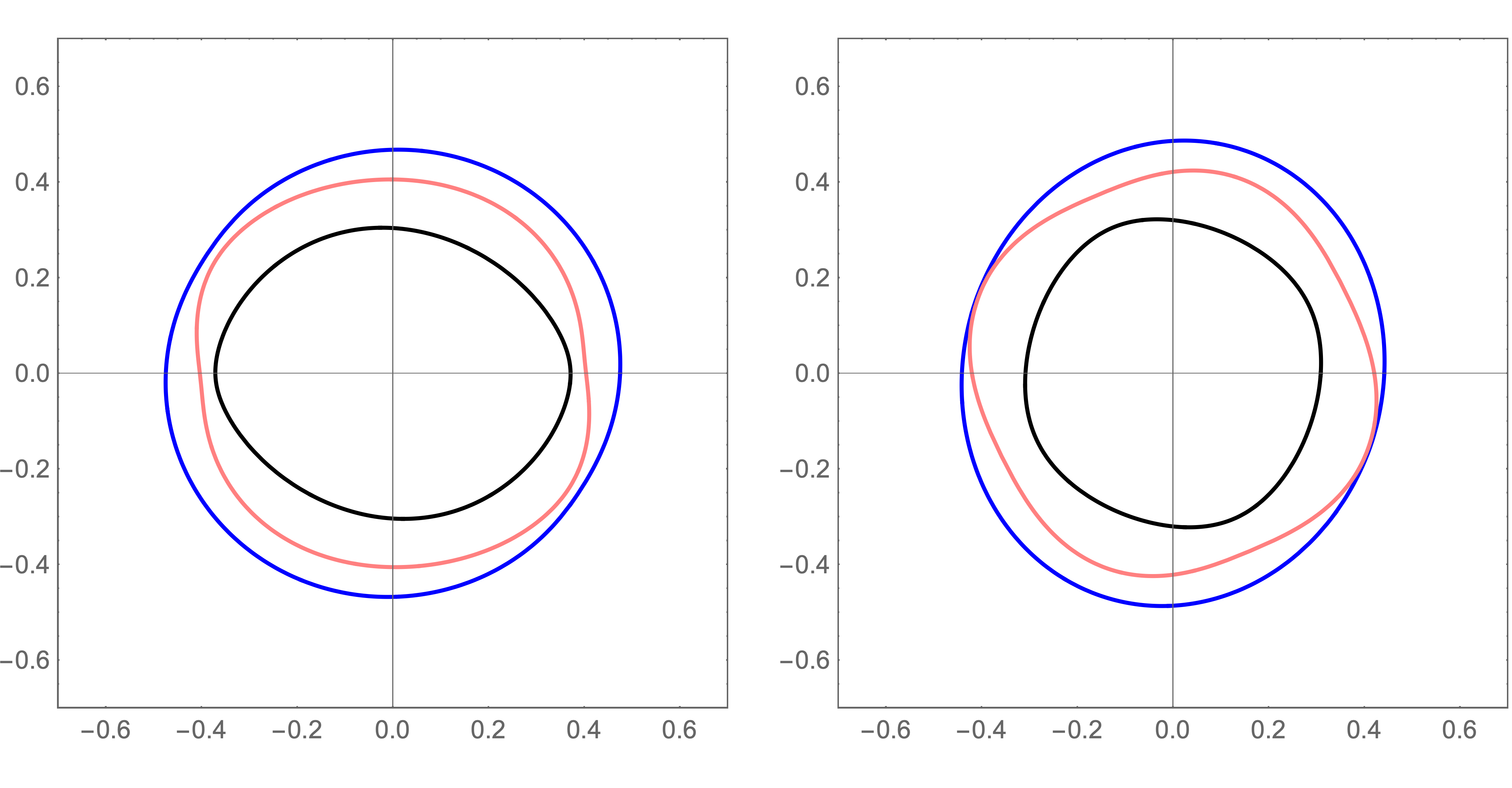}
  \caption{Two \( x \)--\( y \) slowness surfaces of a transversely-isotropic material subjected to \textit{the same} large deviatoric stress, but with different choices of the arbitrary parameters \( \theta \) and \( \phi \). The slow- and fast-directions of P-waves are different, as are the shear-wave splitting patterns. The distinct physical behaviour implied in the two panels is a result of changing the arbitrary parameters in \( \bar{\aaaa}\brak{\sig,\RR} \).}
  \label{fig:anisotropic_nonuniqueness}
\end{figure}\fi
The equilibrium configuration is unstressed, with elastic tensor
\begin{align}
\label{eq:elastic_tensor_ti}
  \repeqLamTI ,
\end{align} 
which can be written alternatively as
\begin{align}
  \aaaa = \cccc
  &= 
    \lambda\id\otimes\id 
  + 2\mu\overline{\idgen} 
  + 8\gamma\NN\otimes\NN 
  + 4\beta\brak{\id\otimes\NN + \NN\otimes\id} 
  - 2\alpha\ChimultHat{\NN} 
\end{align}
if we define \( \NN \) as
\begin{align}
  \NN \equiv \nub\otimes\nub .
\end{align}
We have induced the same stress in the material in both panels of the figure, but selected different \( \brak{\theta,\phi} \) pairs, producing slowness surfaces \textit{of different shapes}. It should be emphasised that the material is described by the same strain-energy function in both panels; that the two slowness surfaces demonstrate distinct physical behaviour is due solely to \( \aaaa \)'s non-unique dependence on \( \sig \) (eq.\ref{eq:lambda_of_sigma_final}).

\subsubsection{Linearised response to deviatoric stress}

Finally, we consider how a transversely-isotropic material responds to a small induced stress. This is the simplest nontrivial example in which one can show analytically how the arbitrary rotation matrix of eq.(\ref{eq:lambda_of_sigma_final}) manifests in the linearised elastic tensor.

In an isotropic material we took the background (zeroth-order) state to be hydrostatic, but in the transversely-isotropic case we should consider a more general zeroth-order stress of the form
\begin{align}
\label{eq:stress_ti}
  \sig^{0} 
  &= -\brak{p^{0}+\frac{q^{0}}{3}}\id + q^{0}\nub\otimes\nub .
\end{align}
The pressure is \( p^{0} \) as before, but the stress now possesses in addition a background deviatoric component consistent with the \( \so(2) \) symmetry. We then induce a small stress \( \sig^{1} \), and a tedious calculation laid out in Appendix \ref{app:sub:transverse_isotropic_calculation} leads to the linearised elastic tensor
\begin{align}
\label{eq:linearisation_ti}
  \cccc^{1} 
  &= 
    \eta_{1}\id\otimes\id 
  + \eta_{2}\overline{\idgen} 
  + \eta_{3}\NN\otimes\NN 
  + \eta_{4}\brak{\id\otimes\NN + \NN\otimes\id} 
  + \eta_{5}\ChimultHat{\NN}
  \nonumber\\&\qquad
  + \eta_{6}\ChimultHat{\sig^{1}} 
  + \eta_{7}\brak{\id\otimes\sig^{1} + \sig^{1}\otimes\id}
  + \eta_{8}\brak{\NN\otimes\sig^{1} + \sig^{1}\otimes\NN}
  + \eta_{9}\braksq{\id\otimes\brak{\Chimult{\NN}\cdot\sig^{1}}+\brak{\Chimult{\NN}\cdot\sig^{1}}\otimes\id}
  \nonumber\\&\qquad
  + \eta_{10}\braksq{\NN\otimes\brak{\Chimult{\NN}\cdot\sig^{1}} + \brak{\Chimult{\NN}\cdot\sig^{1}}\otimes\NN}
  + \eta_{11}\ChimultHat{\brak{\Chimult{\NN}\cdot\sig^{1}}}
  + \eta_{12}\brak{\ChimultHat{\sig^{1}}\ChimultHat{\NN} + \ChimultHat{\NN}\ChimultHat{\sig^{1}}}
  \nonumber\\&\qquad
  + \eta_{13}\ChimultHat{\omb}
  + \eta_{14}\braksq{\id\otimes\brak{\Chimult{\omb}\cdot\NN} + \brak{\Chimult{\omb}\cdot\NN}\otimes\id}
  + \eta_{15}\braksq{\NN\otimes\brak{\Chimult{\omb}\cdot\NN} + \brak{\Chimult{\omb}\cdot\NN}\otimes\NN}
  \nonumber\\&\qquad
  + \eta_{16}\brak{\ChimultHat{\omb}\ChimultHat{\NN} - \ChimultHat{\NN}\ChimultHat{\omb}},
\end{align} 
with the constants \( \brakbr{\eta_{i}} \) defined in  eqs.(\ref{eqs:eta_defn}). The antisymmetric matrix \( \omb \) defines a vector pointing in an arbitrary direction in the plane perpendicular to the unperturbed material symmetry axis. Its components are
\begin{align}
  \omega_{i} = -\epsilon_{ijk}\omega_{jk},
\end{align}
where \( \epsilon_{ijk} \) are the components of the Levi-Civita tensor, and its direction and magnitude are precisely the two arbitrary constants that must be fixed. On the other hand the \( \brakbr{\eta_{i}} \) are unique, scalar-valued functions of:
\begin{enumerate}
\renewcommand{\theenumi}{(\arabic{enumi})}
	\item the constants \( p^{0} \) and \( q^{0} \) which parametrise the zeroth-order equilibrium stress;
	\item the transversely-isotropic constants \( \lambda \), \( \mu \), \( \alpha \), \( \beta \) and \( \gamma \) (which are implicitly functions of \( p^{0} \), \( q^{0} \) and their respective stress-free values);
	\item  \( \tr{\sig^{1}} \) and \( \ip{\nub}{\sig^{1}\cdot\nub} \);
	\item the seven constants \( \brakbr{\zeta_{i}} \) defined in eq.(\ref{eq:strain_energy_third_deriv_param_zeta_defn}) which, analogously to the Murnaghan constants, parametrise the third derivatives of a transversely-isotropic strain-energy function.
\end{enumerate}
The precise functional forms of the \( \brakbr{\eta_{i}} \) are not nearly as important as the fact that they depend on the seven \wording{further} degrees of freedom represented by the \( \brakbr{\zeta_{i}} \), not to mention the two arbitrary parameters \wording{contained within} \( \omb^{1} \). \edit{In contrast, when \citeauthoryearpossessive{Tromp_2019} result is specialised to the transversely-isotropic case it has just five free parameters, namely the pressure derivatives of the elastic moduli (the independent components of their tensor \( \Gam' \)).} We also believe that the term in \( \eta_{11} \) within eq.\eqref{eq:linearisation_ti} is \wording{not present} in \citeauthor{Tromp_2019}'s expression.

As a last point, let us consider the perturbation to the Christoffel operator associated with terms in \( \omb^{1} \). Continuing to ignore spatial arguments, one can show that definition (\ref{eq:christoffel_defn}) is equivalent to
\begin{align}
  \ip{\aaa}{\rho\mathbf{B}\brak{\pp}\cdot\aaa} = \ip{\aaa\otimes\pp}{\aaaa\cdot\brak{\aaa\otimes\pp}}
\end{align}
for arbitrary \( \aaa \). From this it is a matter of algebra to show that the contribution to \( \rho\mathbf{B} \) of the terms in \( \omb^{1} \) is
\begin{align}
  \rho\mathbf{B}
  &=\brak{\eta_{14}+\eta_{16}}\ip{\nub}{\pp}\braksq{\pp\otimes\brak{\omb^{1}\times\nub}+\brak{\omb^{1}\times\nub}\otimes\pp}
  +\brak{\eta_{14}+\frac{1}{2}\eta_{16}}\ip{\omb^{1}\times\nub}{\pp}\brak{\pp\otimes\nub+\nub\otimes\pp}
  \nonumber\\&\qquad
  +\brak{\eta_{15}\ip{\nub}{\pp}^{2}+\eta_{16}\|\pp\|^{2}}\braksq{\nub\otimes\brak{\omb^{1}\times\nub}+\brak{\omb^{1}\times\nub}\otimes\nub}
  +\ip{\nub}{\pp}\ip{\omb^{1}\times\nub}{\pp}\brak{\eta_{15}\nub\otimes\nub+\eta_{16}\id}  .
\end{align}
Each of the coefficients \( \eta_{14} \), \( \eta_{15} \) and \( \eta_{16} \) depends on a different combination of the \( \brakbr{\zeta_{i}} \). But the \( \brakbr{\zeta_{i}} \) parametrise the third-derivatives of the strain energy and can be specified independently both of the other elastic constants and of each other. It is therefore clear that \( \omb^{1} \) will generally contribute nonzero terms to the Christoffel operator.

\section{Discussion}
\label{sec:discussion}

Working under the theory of finite elasticity, we have investigated how 
%large deformations
changes of equilibrium, 
changes in stress and changes in the elastic tensor are interrelated within hyperelastic bodies. Central to the discussion are eqs.\eqref{eqs:sig_Lam_of_F}, which we can restate in the notation of \citet[Sections 2.10 \& 3.6]{dahlen1998theoretical} as 
\begin{subequations}
\label{eqs:sig_Lam_of_F__DT}
\begin{align}
  T^{0}_{ij} &= \frac{\rho^{0}}{\det\FF} F_{ip}F_{jq} \frac{\partial U^{L}}{\partial E^{L}_{pq}} \label{eq:sig_of_F__DT} \\
  \Xi_{ijkl} &= \frac{\rho^{0}}{\det\FF} F_{ip}F_{jq}F_{kr}F_{ls} \frac{\partial^{2} U^{L}}{\partial E^{L}_{pq} \partial E^{L}_{rs}} \label{eq:Lam_of_F__DT} .
\end{align}	
\end{subequations}
These equations tell us how a hyperelastic body's equilibrium Cauchy stress and elastic tensor change when the body is deformed by a motion with deformation gradient \( \FF \), assuming that the body is described by a given constitutive function \( U^{L} \). They can be used to approach both forward- and inverse-problems within geophysics, which we now illustrate by considering briefly the problem of \textit{monsoon loading}, an example that also allows us to contextualise our main results.

In regions that experience heavy monsoons, the rainwater accumulating on the Earth's surface causes the crust to be loaded nontrivially by different amounts at different times of the year \citep[e.g.][]{Fu_2013}. This load causes the crust to deform, and induces associated stresses within it. Given that the deformation occurs over a timescale of months -- a timescale significantly greater than that associated with seismic wave propagation -- we model the deformations to be quasi-static. That is, we take seismic waves propagating through the Earth at a given time to `see' a static equilibrium Earth that does not interact with them \textit{dynamically}. 
Nevertheless, the deformation of the crust does affect the propagation of the seismic waves because 
%the stresses induced by it are a source of anisotropy of the equilibrium configuration. It is expected that this will manifest as seismic anisotropy, and hence as a change in the elastic tensor. 
the equilibrium configuration has changed, and through eq.\eqref{eq:Lam_of_F__DT} there is a consequent change in the elastic tensor. The results of this paper allow us to think about this problem in a number of ways.

A first approach makes direct use of eqs.\eqref{eqs:sig_Lam_of_F__DT} from the perspective of forward-modelling. For concreteness, say that we know the constitutive relation governing the Earth. We then model the deformation induced by the rainfall by solving a quasi-static mapping problem, from which we obtain the change in equilibrium configuration. The \wording{resultant} mapping has a deformation gradient \( \FF \), and we can use eqs.\eqref{eqs:sig_Lam_of_F__DT} to compute the resulting change in both the equilibrium Cauchy stress and the elastic tensor. Hence, we can directly compute the expected seismic anisotropy.

The procedure just outlined is appealing, but it rests on the assumption that
we know the rainfall loading sufficiently well to predict the change in configuration.
%\( \FF \) is known. 
Although this might be realistic in some cases, it is unlikely to be true in general. If anything, it is perhaps more pragmatic to pose the inverse-problem instead: given that an equilibrium deformation leads to seismic anisotropy, and given that we can observe seismic waves readily, what can seismic data tell us about \( \FF \) and, hence, about the change in the equilibrium configuration? For this problem we do not even need eq.\eqref{eq:sig_of_F__DT}; we need only collect seismic data and use eq.\eqref{eq:Lam_of_F__DT} to invert for the nine components of \( \FF \) (taking account of the condition that \( \FF \) be the gradient of a mapping). The equilibrium mapping could then be (partially) reconstructed. 
%\edit{(although 5.1a not needed directly, we can still compute the stress -- basically shift some of the stuff from the next paragraph about computing the stress to this bit)}

%\edit{(possible segue -- might be asked if you can go directly to stress without \( \FF \))}

%\edit{(another thing you could do would just be to invert the seismic observations for the water load...)}

So far in this example, the problem has been couched in terms of the equilibrium mapping: not once have we needed to consider how the elastic tensor depends on equilibrium stress. But in addition to simply \wording{modelling} the seismological \wording{effect of} large deformations, we could also carry out studies that seek to invert seismological observations for changes in the equilibrium stress. 
%Such studies would be relevant if our goal were, for instance, to learn about the constitutive relation obeyed by the material of the given region of the Earth, or, more prosaically, simply to know the stresses in a given region. 
In that case the mapping itself would not (necessarily) be a relevant quantity; we would be more interested in the \wording{direct} stress dependence of the elastic tensor. Now, we could invert seismic data for \( \FF \) (as just described) and then use eq.\eqref{eq:sig_of_F__DT} to compute \( \TT^{0} \), but it would be desirable from the perspective of inversion to find an explicit expression for \( \Xib \) as a function of \( \TT^{0} \), not least because the Cauchy stress has fewer components than \( \FF \). In searching for such an expression we are effectively \wording{asking} if it is possible to parametrise an equilibrium configuration, not in terms of the deformation gradient that gave rise to it, but rather \wording{in terms of} its present Cauchy stress. We are thus led to address the following problem: if we observe a \textit{given} Cauchy stress and assume that it arose due to \textit{some} elastic deformation of a state with \textit{given} constitutive properties, what elastic tensor would we measure?

As discussed at length in Section \ref{sec:lambda_of_sigma}, in order to find \( \Xib \) as a function of \( \TT^{0} \) we must eliminate \( \FF \) from eqs.\eqref{eqs:sig_Lam_of_F__DT}. We solve \eqref{eq:sig_of_F__DT} in order to find \( \FF \) as a function of \( \TT^{0} \), and substitute the result into \eqref{eq:Lam_of_F__DT}. Physically, \eqref{eq:sig_of_F__DT} describes how a deformation alters an elastic body's equilibrium Cauchy stress. But because many different values of \( \FF \) can lead to the same change in \( \TT^{0} \), the mathematical problem of inverting \eqref{eq:sig_of_F__DT} to find \( \FF \) in terms of \( \TT^{0} \) is naturally underdetermined. It turns out that \( \FF \) depends not only on \( \TT^{0} \), but also on an arbitrary rotation matrix. One can then show that the elastic tensor itself is given as a function of both the Cauchy stress and an arbitrary rotation. As emphasised in Section \ref{sec:lambda_of_sigma}, the presence of these arbitrary parameters shows that measurement of the Cauchy stress \textit{alone} is not sufficient to constrain the elastic tensor fully. However, we also found that \wording{the more symmetric} the initial, reference state, the fewer the arbitrary parameters. In fact, for an isotropic reference state one can ignore the rotation matrix entirely; in that case the Cauchy stress does provide enough information to fix the elastic tensor.

Having derived an expression for the nonlinear dependence of \( \Xib \) on \( \TT^{0} \), we proceeded to linearise it. From the point of view of performing inversions this step is not strictly necessary, but it is useful for two reasons. Firstly, the changes in the Earth's equilibrium stress due to quasi-static deformation are likely to be small because the deformations we consider will almost always be small. This will certainly be the case with monsoon loading, and we also refer the reader to a discussion of the effects of equilibrium stress on seismic waveforms in the Groningen gas field by \citet[Section 8.2]{Tromp_2018}. The change in the elastic tensor will therefore be well described by a linearised theory, and such a theory should be marginally quicker to implement computationally. Secondly, the theories discussed in Section \ref{sec:intro} are all linear in the stress, so we must linearise our theory in order to carry out a comparison.

Linearising within a \wording{general} anisotropic material, the elastic tensor depends not only on \( \Delta\TT^{0} \) but also on a small, arbitrary, antisymmetric matrix \( \omb^{1} \) that results from linearising the rotation matrix of eq.\eqref{eq:lambda_of_sigma_final}. Our general linearised expression (\ref{eq:linearisation_general}) can be expanded and rewritten in the notation of Section \ref{sec:intro} to give
\begin{align}
\label{eq:final_xi_eqn_pi_pi_prime}
  \Delta\Xi_{ijkl} = \Pi_{ijklmn}\Delta T^{0}_{mn} + \Pi'_{ijklmn}\omega^{1}_{mn} ,
\end{align}
where \( \Pib \) is the tensor relating changes in \( \Xib \) to \( \Delta \TT^{0} \), just as before, while we have defined another tensor \( \Pib' \) that relates changes in \( \Xib \) to \( \omb^{1} \). The components of \( \Pib \) and \( \Pib' \) are expressed in terms of the third derivative of the strain-energy function evaluated in the background state, and the components of the background elastic tensor \( \Gam \); it is trivial to show that they therefore possess up to 56 independent components besides the elastic moduli.
% Maybe have 56 components further up as well?
These rather complicated expressions are given in Appendix \ref{notation_conventions}, while here we make two general points.
Firstly, \( \Pib' \) will not usually vanish, which indicates that even linearised stress dependence will generally involve \wording{some level of arbitrariness}. Inspection of Section \ref{app:sub:general_calculation} will show that the neglect of \( \Pib' \) is equivalent to assuming that \( \Delta \TT^{0} \) was induced by a \textit{symmetric} deformation gradient \( \FF \), \ie a pure stretch. Secondly, we have shown that \( \Pib \) need not possess the further symmetries,
\begin{align}
  \Pi_{ijklmn} = \Pi_{ijmnkl} = \Pi_{klmnij} = \Pi_{mnijkl} = \Pi_{mnklij} ,
\end{align}
imposed in eq.\eqref{eq:pi_extra_symmetries}. In short, our theory is parametrised differently from that of \citet{Tromp_2019}; we require up to 59 (\(=\) 56 \( + \) 3) components, while \citeauthor{Tromp_2019} need a maximum of 21. The respective theories are therefore unlikely to give the same results when used to perform inversions.

This statement can be substantiated a little further by applying our general linearised expression to a transversely-isotropic material. We found that seven material-dependent parameters and two arbitrary constants (besides the five elastic moduli) were needed to specify the elastic tensor's linearised stress dependence. Our results cannot be equivalent to those of \citeauthor{Tromp_2019} because their expression requires just five further constants, parametrised as it is by pressure-derivatives of the elastic moduli. (We have not presented here the complete expression for transversely-isotropic \( \Pib \); we trust that the reader who has worked through Appendix \ref{app:sub:transverse_isotropic_calculation} will forgive us.)

We have performed the most complete comparison with \citeauthor{Tromp_2019}'s work in the linearised isotropic case -- which is presumably the case of most practical importance. Our expression for the elastic tensor takes precisely the form derived in Section \ref{intro:heuristic} featuring the four constants \( a \), \( b \), \( c \) and \( d \) (eq.\ref{eq:xi_iso_expression_abcd}). Furthermore, we have shown by considering constitutive behaviour that those constants are functions of the \textit{Murnaghan constants} \( \zeta_{1,2,3} \) \citep{Murnaghan_1937}:
\begin{subequations}
\begin{align}
  a &= \frac{\kappa-\frac{2}{3}\mu + \frac{1}{2}\zeta_{2}}{\mu-p^{0}} \\
  b &= \frac{\mu + \frac{1}{4}\zeta_{3}}{\mu-p^{0}} \\
  c &= -\frac{\kappa-\frac{2}{3}\mu + 3\zeta_{1} + 2\zeta_{2}}{3\kappa+p^{0}} \\
  d &= -\frac{\mu + \frac{3}{2}\zeta_{2} + \zeta_{3}}{3\kappa+p^{0}} .
\end{align}	
\end{subequations}
As a result they represent three degrees of freedom. Up to notation, these four expressions first appeared in the work of \citet[eqs.31--34]{Walton_1974}. Taking into account different definitions of the elastic moduli, we have also established in Appendix \ref{app:matching_theories} that the expressions of \citet{dahlen_1972b} and \citet{Tromp_2018} are consistent with this theory for certain parameter choices. Those theories thus apply to a subset of isotropic materials. If one wished to invert seismic data for equilibrium stress using \citeauthor{Tromp_2018}'s theory, one would need to specify (or invert for) two parameters, the pressure-derivatives of \( \kappa \) and \( \mu \). Our theory would require three such parameters. Finally, it is interesting that we have found expressions for the pressure-derivatives of the elastic moduli (\( c \) and \( d \); see Appendix \ref{app:matching_theories}) in terms of the Murnaghan constants. To our knowledge, this result has not appeared in the literature since Walton's work.

Our linearised results would probably be more taxing to apply to observational seismology than those of \citeauthor{Tromp_2019} because we require the measurement and fitting of more parameters. Moreover, the strain-energy function's third-derivatives are at present measured less \wording{readily} than the moduli's pressure-derivatives. It is also evident from \citeauthor{Tromp_2019}'s \textit{ab initio} calculations (carried out for a material with cubic symmetry) that the extra effects we have derived are not particularly large. We would be keen to see if further such calculations can clarify this apparent \wording{``unreasonable effectiveness"} of pressure-derivatives. Nevertheless, our theory should be practical for isotropic reference states. In that case it just requires one more parameter than \citeauthor{Tromp_2019}'s, and the Murnaghan constants of various materials have been measured in the past \citep[e.g.][]{Hughes_1953,Egle_1976,Payan_2009}. 

There are a number of potential geophysical applications of this work, but we must first mention two caveats. Firstly, the Earth is not elastic over geological time-scales, hence it is not reasonable to regard its equilibrium as having arisen through a finite deformation of an elastic material away from some hypothetical stress-free state. Within future work it would therefore be interesting to extend our methods to account for viscoelastic effects. Nevertheless, our framework should give valid descriptions of phenomena that occur over time-scales sufficiently short for the Earth to respond in an elastic -- or only slightly anelastic -- manner. For example, one might consider the effect on elastic wave speeds of processes that are fast relative to viscoelastic relaxation times but slow compared to those of seismic wave propagation, such as body tides, seasonal loading in the hydrosphere, or anthropogenic activity. A second \wording{caveat} is that inverting seismic data for equilibrium stress is already rendered challenging by the fact that the equilibrium stress is not an entirely free parameter, but is constrained to satisfy the equilibrium equations \citep{Backus_1967a,Al_Attar_2010}. Our results make clear that it is also necessary to either provide or simultaneously invert for additional parameters related to third derivatives of the strain-energy function and, in some cases, infinitesimal rotations. Finally, given the importance of symmetry groups to this work, we would also be \wording{interested} to see how our results mesh with the theory of \textit{homogenisation} \citep[e.g.][]{Cupillard_2018,Capdeville_2018} which describes how small-scale structures are `smeared out' to produce effective media with \textit{different symmetry properties} (recall for instance \citeauthor{Backus_1962}'s \citeyearpar{Backus_1962} point that long-wavelength seismic waves passing through a layered isotropic medium `see' a transversely-isotropic \textit{effective} medium).

\section{Conclusions}
\label{sec:conclusions}

We have derived an expression for the elastic tensor as an explicit function of equilibrium Cauchy stress. Our results differ from previous treatments in two main ways: they show that the elastic tensor's dependence on equilibrium stress is generally both nonlinear and non-unique. On account of the nonlinearity alone, knowledge of a material's background elastic-tensor is not sufficient to determine the material's response to an induced stress; we require the information contained within higher-order derivatives of the background strain-energy function. Furthermore, the elastic tensor is a function not only of the equilibrium stress, but also of an arbitrary rotation matrix. As such, even with a definite strain-energy function in hand, the change in the elastic tensor due to an induced stress depends on the non-unique choice of this matrix. The non-uniqueness arises from the fact that the stress is considered to have been induced by an unspecified elastic deformation. However, we have also shown that the degree of non-uniqueness is reduced if the material under study has a nontrivial background material symmetry group.

In the linearised case, our approach shows that the characterisation of a material's response to a small induced stress depends on more parameters than have been made explicit in previous studies \citep{Tromp_2018,Tromp_2019}. We have shown moreover that the previous expressions for the elastic tensor's stress dependence can be obtained as special cases of our linearised results. Our expressions should be seen to extend the work of the previous authors by including some extra terms that were not \wording{captured} by their derivation. The approach of first deriving a nonlinear expression, and only then linearising, has also allowed us to suggest a different interpretation of the parametrisation of the elastic tensor's linearised stress dependence. Whilst \citeauthor{Tromp_2019} suggest that the relevant parameters are the pressure-derivatives of the elastic moduli, we propose the use of a larger set of parameters: the third-derivatives of the strain-energy function, and, for anisotropic materials, infinitesimal rotations. That said, we would like to emphasise once again that the expressions of \citeauthor{Tromp_2019} are able to fit experimental data rather well \edit{for the case of a cubic material}. So whilst it seems theoretically necessary to account for more parameters, on a practical level it might not be required. This is a curious result that is not obvious to the present authors, and we would be keen to see it investigated further.

\begin{acknowledgments}
MM is supported by an EPSRC studentship and a CASE award from BP. We are grateful to Jeroen Tromp, Brent Delbridge and editor Juan Carlos Afonso for their helpful comments and suggestions. We would particularly like to thank Prof. Tromp for taking the time to discuss our initial results with us at length. Those conversations contributed greatly to the shaping of Section \ref{sec:intro}, Section \ref{subsec:lambda_and_sigma_of_f} and Appendix \ref{app:matching_theories}, and they significantly improved our own understanding of the work.
\end{acknowledgments}

%%%%%%%%%%%%%%%%%%%%%%%%%%%%%%%%%%%%%%%%%%%%%%%%%%%%%%%%%%%%%%%%%%%%%%%%%
\bibliographystyle{gji}
\bibliography{my_paper_library}

%\end{document}
%%%%%%%%%% APPENDICES %%%%%%%%%%%%%%
\appendix

\section{Notations and definitions}
\label{app:notations}

\subsection{Groups}
\label{app_sub:groups}

We define \( \gl(n) \), \textit{the general linear group of dimension n}, to be the set of invertible \( n \times n \) matrices under the operation of matrix multiplication. For a general group \( \mathbf{G} \), a \textit{subgroup} \( \mathbf{H} \) of \( \mathbf{G} \) is a subset of the elements of \( \mathbf{G} \) that is itself a group; \( \mathbf{H} \) is described as a \textit{proper subgroup} of \( \mathbf{G} \) if \( \mathbf{H} \neq \mathbf{G} \). With this, we can define \( \sl(n) \), \textit{the special linear group} of \( n \times n \) matrices with \textit{unit} determinant, which is a proper subgroup of \( \gl(n) \). A particularly important proper subgroup of \( \sl(n) \) is \( \so(n) \), \textit{the n-dimensional special orthogonal group} whose elements are rotation matrices in \( n \) dimensions. For any \( \RR\in\so(n) \),
\begin{align}
  \det{\RR} &= 1 \\
  \RR^{-1}  &= \RR^{T} . 
\end{align}

\subsection{Some nonstandard linear operators}
\label{app_sub:linear_ops}

We have found it useful to introduce the left- and right-multiplication operators \( \Lmult{\mathbf{A}} \) and \( \Rmult{\mathbf{A}} \) which act according to
\begin{align}
\label{eq:RL_basic_defn}
  \Lmult{\mathbf{A}}\cdot\mathbf{B} &= \mathbf{A}\mathbf{B} \\
  \Rmult{\mathbf{A}}\cdot\mathbf{B} &= \mathbf{B}\mathbf{A}
\end{align}
for arbitrary matrices \( \mathbf{A},\mathbf{B} \in \gl(3) \). These operators may be represented by fourth-rank tensors. An operator \( \mathbsf{O} \) acts on a matrix \( \AA \) according to
\begin{align}
  \brak{\mathbsf{O}\cdot\AA}_{ij} = \mathsf{O}_{ijkl}A_{kl} ,
\end{align}
where we use a dot to represent the action of a linear operator on its operand (as is the case throughout this paper). Two such operators are composed by writing
\begin{align}
  \brak{\mathbsf{O}^{1}\mathbsf{O}^{2}}_{ijkl} = \mathsf{O}^{1}_{ijpq}\mathsf{O}^{2}_{pqkl} .
\end{align}
The left- and right-multiplication operators are expressed in index-notation as
\begin{align}
  \brak{\Lmult{\mathbf{A}}}_{ijkl} &= A_{ik}\delta_{lj} \\
  \brak{\Rmult{\mathbf{A}}}_{ijkl} &= \delta_{ik}A_{lj} ,
\end{align}
and, as an example,
\begin{align}
  \brak{\Lmult{\mathbf{A}}\cdot\mathbf{B}}_{ij} 
  &= \brak{\Lmult{\mathbf{A}}}_{ijkl} B_{kl} \nonumber\\
  &= A_{ik}\delta_{lj} B_{kl} \nonumber\\
  &= A_{ik}B_{kj} \nonumber\\
  &= \brak{\mathbf{A}\mathbf{B}}_{ij}  .
\end{align}
It is clear that \( \Lmult{\mathbf{A}} \) and \( \Rmult{\mathbf{B}} \) commute for any choice of \( \mathbf{A} \) and \( \mathbf{B} \), and that they satisfy
\begin{align}
  \Rmult{\mathbf{A}\mathbf{B}} &= \Rmult{\mathbf{B}}\Rmult{\mathbf{A}} \\
  \Lmult{\mathbf{A}\mathbf{B}} &= \Lmult{\mathbf{A}}\Lmult{\mathbf{B}} ,
\end{align}
while their inverses have the property that
\begin{align}
  \Lmult{\mathbf{A}}^{-1} &= \Lmult{\mathbf{A}^{-1}} \\
  \Rmult{\mathbf{A}}^{-1} &= \Rmult{\mathbf{A}^{-1}}  .
\end{align}
We also define the operators
\begin{align}
  \Lmult{\mathbf{A}}^{T} &= \Lmult{\mathbf{A}^{T}} \label{eq:lmult_tran}\\
  \Rmult{\mathbf{A}}^{T} &= \Rmult{\mathbf{A}^{T}} \label{eq:rmult_tran}  .
\end{align}
Then, defining the inner-product on matrices by
\begin{align}
\label{eq:inner_product_gl3}
  \innerproduct{\mathbf{A}}{\mathbf{B}} &= \tr{\mathbf{A}\mathbf{B}^{T}}
\end{align}
and introducing \( \CC\in\gl(3) \), it follows quickly that
\begin{align}
\label{eq:gl3_transpose_inner_product}
  \innerproduct{\mathbf{A}}{\Lmult{\mathbf{C}}\cdot\mathbf{B}}
  	= \innerproduct{\Lmult{\mathbf{C}}^{T}\cdot\mathbf{A}}{\mathbf{B}} ,
\end{align}
and similarly for \( \Rmult{\mathbf{C}} \). This is the origin of our suggestive notation for \( \Lmult{\mathbf{A}}^{T} \) and \( \Rmult{\mathbf{A}}^{T} \): with the inner product as defined, they behave superficially as though they were the respective transpose operators of \( \Lmult{\mathbf{A}} \) and \( \Rmult{\mathbf{A}} \). We also define a \textit{norm} on matrices,
\begin{align}
\label{eq:norm_gl3_defn}
    \|\mathbf{A}\| &= \sqrt{\innerproduct{\mathbf{A}}{\mathbf{A}}},
\end{align}
and \textit{tensor-product},
\begin{align}
\label{eq:tensor_product_gl3_defn}
  \brak{\mathbf{A}\otimes\mathbf{B}}\cdot\CC = \innerproduct{\mathbf{B}}{\CC}\mathbf{A}  .
\end{align}
In order to avoid clutter we have -- just as for the inner product -- simply used the standard notation for a tensor-product and norm on \( \RThree \), trusting that context will make our meaning unambiguous. 

A particularly useful operator is
\begin{align}
  \label{eq:tdef}
  \Tmult{\mathbf{A}} &= \Lmult{\mathbf{A}}\Rmult{\mathbf{A}}^{T}  .
\end{align}
From the properties of  \( \Lmult{\mathbf{A}} \) and \( \Rmult{\mathbf{A}} \) discussed above, we clearly have
\begin{align}
  \Tmult{\mathbf{A}}^{-1} &= \Tmult{\mathbf{A}^{-1}}  \label{eq:tprop_inv}   \\
  \Tmult{\mathbf{A}}^{T}  &= \Tmult{\mathbf{A}^{T}}   \label{eq:tprop_trans} ,
\end{align}
as well as the useful \textit{functoral} property
\begin{align}
\label{eq:tfunc}
  \Tmult{\mathbf{A}\mathbf{B}} &= \Tmult{\mathbf{A}}\Tmult{\mathbf{B}} .
\end{align}
For \( \RR\in\so(3) \), 
\begin{align}
  \Tmult{\RR}\cdot\mathbf{A} = \RR\mathbf{A}\RR^{T}
\end{align}
evidently represents a rotation of \( \mathbf{A} \) by \( \RR \). The related operator
\begin{align}
\label{eq:chimult_defn}
  \Chimult{\mathbf{A}} &= \Lmult{\mathbf{A}} + \Rmult{\mathbf{A}}^{T}
\end{align}
is the term in \( \Tmult{\id+\mathbf{A}} \) linear in \( \mathbf{A} \), and is particularly useful when we consider linearisation. Furthermore, if \( \mathbf{A} \) and \( \mathbf{B} \) are both symmetric the operator satisfies
\begin{align}
\label{eq:chi_mult_sym_identity}
  \Chimult{\mathbf{A}}\cdot\mathbf{B} = \Chimult{\mathbf{B}}\cdot\mathbf{A}  .
\end{align}

Finally, we must sometimes be careful when writing these operators in component form. Consider the expression for the quadratic part of an isotropic strain-energy function \citep[e.g.][]{Holzapfel_2000}:
\begin{align}
  V_{2}\brak{\CC} = \frac{\lambda}{8}\tr{\CC}^{2}+\frac{\mu}{4}\tr{\CC^{2}} .
\end{align}
It is easy to show that the stress vanishes at \( \CC=\id \). In such a stress-free equilibrium we have (eq.\ref{eq:lamrep2})
\begin{align}
  \aaaa = 4D^{2}V_{2}\brak{\id} .
\end{align}
We can evaluate the derivative by using the definitions above to rewrite \( V_{2} \) as
\begin{align}
\label{eq:ashjasjghaskjhasaskhj}
  V_{2}\brak{\CC} &= \frac{\lambda}{8}\ip{\id}{\CC}^{2}+\frac{\mu}{4}\ip{\CC}{\idgen\cdot\CC} \nonumber\\
  &= \ip{\CC}{\brak{\frac{\lambda}{8}\id\otimes\id + \frac{\mu}{4}\idgen}\cdot\CC}	, 
\end{align}
where we further define the \textit{identity operator} \( \idgen \), which has components
\begin{align}
\label{eq:id_op_defn}
  \brak{\idgen}_{ijkl} = \delta_{ik}\delta_{jl} .
\end{align}
However, in the present context this is not the correct identity operator to use. When differentiating strain-energy functions that depend on the symmetric matrix \( \CC \) we find ourselves dealing with operators that naturally act on symmetric matrices. Therefore we must remind ourselves that the component-form expressions for these operators will be symmetric on each pair of indices. We denote this by an overline. In general, the process of symmetrisation is defined as follows. The tensor \( \mathbf{M} \) is symmetrised on its \( p \)'th pair of indices by
\begin{align}
\label{eq:symmetrisation_defn}
  M_{i_{1}j_{1}i_{2}j_{2}\dots i_{p}j_{p} \dots i_{n}j_{n}}
  \rightarrow
  M_{i_{1}j_{1}i_{2}j_{2}\dots \brak{i_{p}j_{p}} \dots i_{n}j_{n}}
  \equiv
  \frac{1}{2}\brak{
  	M_{i_{1}j_{1}i_{2}j_{2}\dots i_{p}j_{p} \dots i_{n}j_{n}}
   +M_{i_{1}j_{1}i_{2}j_{2}\dots j_{p}i_{p} \dots i_{n}j_{n}}} .
\end{align}
A fourth-rank tensor \( \mathbsf{M} \), for example, is symmetrised on both pairs of indices by writing
\begin{align}
\label{eq:fourth_rank_sym_defn}
  (\overline{\mathbsf{M}})_{ijkl}
   =\mathsf{M}_{(ij)(kl)}
   =\frac{1}{4}\brak{\mathsf{M}_{ijkl}+\mathsf{M}_{ijlk}+\mathsf{M}_{jikl}+\mathsf{M}_{jikl}}  .
\end{align}
The symmetrised identity operator is thus written as
\begin{align}
\label{eq:id_hat_defn}
  (\overline{\idgen})_{ijkl} = \frac{1}{2}\brak{\delta_{ik}\delta_{jl}+\delta_{il}\delta_{jk}}  .
\end{align}
Substituting this into eq.(\ref{eq:ashjasjghaskjhasaskhj}) will yield the correct component-form expression for the elastic tensor. Note that this process confers the \textit{minor} elastic symmetries on a fourth-rank tensor, but not necessarily the hyperelastic symmetry.

\subsection{Example: Differentiation of the strain-energy functions \( W \) and \( V \)}
\label{app:sub:W_V_differentiation}

All the calculations discussed in this paper could in principle be performed using index notation. However, we found it to be too cumbersome to carry out the calculations of Appendix \ref{app:calculations}. The purpose of this appendix is to explain how the operator-based notation detailed above can be used to carry out the differentiations of strain-energy functions in Sections \ref{theory} and \ref{sec:lambda_of_sigma}. It avoids clutter in the main text and moreover illustrates the manipulation of these operators.

For an arbitrary \( \FF \in \gl(3) \) and \( \CC = \FF^{T}\FF \) we have
\begin{align}
  W\brak{\FF} = V\brak{\CC} ,
\end{align}
neglecting the spatial argument to avoid clutter. We define small perturbations to \( \FF \) and \( \CC \), respectively \( \delta\FF \) and \( \delta\CC \), such that
\begin{align}
  W(\mathbf{F} + \delta \mathbf{F}) - W(\mathbf{F}) 
  &= \innerproduct{DW(\mathbf{F})}{\delta \mathbf{F}}
  + \mathcal{O}(\|\delta \mathbf{F}\|^{2}) \\
  V(\CC + \delta \CC) - V(\CC) 
  &= \innerproduct{DV(\CC)}{\delta \CC}
  +  \mathcal{O}(\|\delta \CC\|^{2})
\end{align}
where \( \innerproduct{\cdot}{\cdot}\) is the inner product for matrices (eq.\ref{eq:inner_product_gl3}) and \( \|\cdot\| \) is the corresponding norm (eq.\ref{eq:norm_gl3_defn}). It follows from the definition of \( \CC \) that
\begin{equation}
  \label{eq:delc}
  \delta \CC = \mathbf{F}^{T}\delta \mathbf{F} + \delta \mathbf{F}^{T}\mathbf{F}
  + \mathcal{O}(\|\delta \mathbf{F}\|^{2}), 
\end{equation}
which allows us to write
\begin{align}
  \innerproduct{DW(\mathbf{F})}{\delta \mathbf{F}} =
  \innerproduct{DV(\CC)}{\mathbf{F}^{T}\delta \mathbf{F} + \delta \mathbf{F}^{T}\mathbf{F}} .
\end{align}
Now, for any matrices \(\AA \) and \(\mathbf{B}\)\ our inner product satisfies \( \innerproduct{\AA}{\mathbf{B}} = \innerproduct{\AA^{T}}{\mathbf{B}^{T}}\). This identity, along with the fact that \( DV(\CC) \) is necessarily symmetric, implies that
\begin{align}
  \innerproduct{DV(\CC)}{ \delta \mathbf{F}^{T}\mathbf{F}}
  = \innerproduct{DV(\CC)}{\mathbf{F}^{T}\delta \mathbf{F}},
\end{align}
and therefore
\begin{align}
  \innerproduct{DW(\mathbf{F})}{\delta \mathbf{F}} =
  2 \innerproduct{DV(\CC)}{\FF^{T}\delta\mathbf{F}}. 
\end{align}
Looking at the right hand side, the left multiplication operator defined in eq.(\ref{eq:RL_basic_defn}) lets us write
\begin{align}
  \mathbf{F}^{T}\delta \mathbf{F} = \Lmult{\mathbf{F}}^{T}\cdot \delta\mathbf{F}, 
\end{align}
where we have made use of eq.(\ref{eq:lmult_tran}). Using this notation we find that
\begin{align}
  \innerproduct{DV(\CC)}{\mathbf{F}^{T}\delta \mathbf{F}}
  = \innerproduct{DV(\CC)}{\Lmult{\mathbf{F}}^{T}\cdot \delta\mathbf{F}}
  = \innerproduct{\Lmult{\mathbf{F}}\cdot DV(\CC)}{ \delta\mathbf{F}} .
\end{align}
Therefore we conclude that
\begin{align}
  \innerproduct{DW(\mathbf{F})}{\delta \mathbf{F}} =
  \innerproduct{2\Lmult{\mathbf{F}}\cdot DV(\CC)}{ \delta\mathbf{F}}. 
\end{align}
Because the perturbation \(\delta \mathbf{F} \) was arbitrary, we have established the identity
\begin{align}
  \label{eq:fderivid}
  DW(\mathbf{F}) = 2\Lmult{\mathbf{F}}\cdot DV(\CC).
\end{align}

The derivation of the second-derivative follows a similar pattern. We take as a starting point the definitions
\begin{align}
  D W(\mathbf{F} + \delta \mathbf{F}) - D W(\mathbf{F}) 
  &= D^{2}W\brak{\FF} \cdot \delta \mathbf{F} + \mathcal{O}(\|\delta \mathbf{F}\|^{2}) \label{eq:some_eq} \\
  D V(\mathbf{C} + \delta \mathbf{C}) - D V(\mathbf{C}) 
  &= D^{2}V\brak{\CC} \cdot \delta \mathbf{C} + \mathcal{O}(\|\delta \mathbf{C}\|^{2}),  \label{eq:some_other_eq}
\end{align}
for the previously defined \( \delta \mathbf{F} \) and \( \delta\CC \). Using eq.\eqref{eq:fderivid} the left hand side of eq.\eqref{eq:some_eq} can be written
\begin{align}
  2 \Lmult{\mathbf{F} + \delta \mathbf{F}}\cdot DV(\CC+\delta \CC)
  - 2 \Lmult{\mathbf{F}}\cdot DV(\CC). 
\end{align}
Expanding this out to first-order it follows that
\begin{align}
  D^{2}W\brak{\FF}\cdot \delta \mathbf{F} = 2\Lmult{\delta \mathbf{F}}\cdot DV\brak{\CC}
   + 2\Lmult{\mathbf{F}} \cdot \braksq{D^{2}V\brak{\CC}\cdot \brak{\mathbf{F}^{T}\delta \mathbf{F}
     + \delta \mathbf{F}^{T}\mathbf{F}}}, 
\end{align}
where we have used eqs. (\ref{eq:delc}) and (\ref{eq:some_other_eq}). To proceed, we first note that \(\Lmult{\delta \mathbf{F}}\cdot DV\brak{\CC} =
\Rmult{DV\brak{\CC}} \cdot \delta \mathbf{F}\) by definition of the left and right
multiplication operators. Secondly,
\begin{align}
  D^{2}V\brak{\CC} \cdot (\mathbf{F}^{T}\delta \mathbf{F}
     + \delta \mathbf{F}^{T}\mathbf{F})
     =
     2 D^{2}V\brak{\CC} \cdot (\mathbf{F}^{T}\delta \mathbf{F})
\end{align}
because \( V \) is a function on symmetric matrices. Given that \(\delta \mathbf{F}\) is arbitrary, this establishes our desired expression for the second-derivative of \( W \):
\begin{align}
  \label{eq:lamrep}
  D^{2}W\brak{\FF} = 2\Rmult{DV\brak{\CC}} + 4\Lmult{\FF} D^{2}V\brak{\CC} \,\Lmult{\FF}^{T}  .
\end{align}

\subsection{Mapping between different notation conventions}
\label{notation_conventions}

As stated above, Sections \ref{sec:intro} and \ref{sec:discussion} are written in a notation close to that of \citet{dahlen1998theoretical} whilst for the middle sections we use a notation heavily inspired by \citet{marsden1994mathematical}. Table \ref{table:notations} summarises the relationship between the two systems. Besides \wording{differing} choices of letter for various quantities, the principal difference between the two systems concerns index placement. All second-rank tensors must have their indices switched to go from one convention to the other, and for tensors of higher (even) rank the same applies to each pair of indices individually; this is illustrated schematically in Table \ref{table:notations}'s last two lines. Note also that we do not follow MH's convention of using upper- and lower-case indices to distinguish between referential and spatial coordinate-systems because we consider both systems to be implicitly Cartesian. Finally, where a quantity of interest does not appear in MH we have notated it with the same letter as DT, although in Sections \ref{theory}--\ref{sec:calculations} we consistently place our indices according to MH's conventions; such quantities do not appear in the table. 

To conclude this section we ``translate" eqs.\eqref{eq:linearisation_general} (reproduced in Table \ref{table:notations}) into the notation of Section \ref{sec:intro}, giving expressions for the components of eq.\eqref{eq:final_xi_eqn_pi_pi_prime}'s tensors \( \Pib \) and \( \Pib' \). Let us first recall the elastic tensor \( \UPS \) that is defined by \citet[eq.3.123]{dahlen1998theoretical} to have components
\begin{align}
  \Upsilon_{ijkl} = \Xi_{ijkl} + T^{0}_{ik}\delta_{jl} + T^{0}_{jk}\delta_{il} - T^{0}_{ij}\delta_{kl} .
\end{align}
We now define the very closely related tensor \( \YYY \), whose components are
\begin{align}
  Y_{ijkl} = \Gamma_{ijkl} + T^{B}_{ik}\delta_{jl} + T^{B}_{jk}\delta_{il} - T^{B}_{ij}\delta_{kl} ,
\end{align}
and we then symmetrise it to form a new tensor \( \overline{\YYY} \), which has components
\begin{align}
  \overline{Y}_{ijkl} = Y_{ij(kl)}
\end{align}
and satisfies the symmetries
\begin{align}
   \overline{Y}_{jikl} = \overline{Y}_{ijkl} = \overline{Y}_{ijlk} .
\end{align}
The inverse tensor \( \overline{\YYY}^{-1} \) is then defined by the relation
\begin{align}
  \overline{Y}_{ijab}\overline{Y}^{-1}_{abkl} = \frac{1}{2}\brak{\delta_{ik}\delta_{jl}+\delta_{jk}\delta_{il}} .
\end{align}
We can now rewrite the generalised stress-strain relationship (eq.\ref{eq:linearisation_general_sig}) as
\begin{align}
  \uu^{1} = \overline{\YYY}^{-1} \!: \brak{\Delta\TT^{0} - \omb^{1}\cdot\TT^{B} - \TT^{B}\cdot\omb^{1}} .
\end{align}
Eq.\eqref{eq:linearisation_general_c} can be dealt with by defining two final tensors:
\begin{align}
  \Theta_{ijklmn} &= \delta_{im}\Gamma_{njkl}
				  	+\delta_{jm}\Gamma_{inkl}
				    +\delta_{km}\Gamma_{ijnl}
					+\delta_{lm}\Gamma_{ijkn}
				  	-\Gamma_{ijkl}\delta_{mn}
				  	+\rho^{0}\frac{\partial^{3} U^{L}}{\partial E^{L}_{ij} \partial E^{L}_{kl} \partial E^{L}_{mn}} , \label{eq:Theta_def} \\
  \Psi_{ijklmn} &= \delta_{im}\Gamma_{njkl}
			  	  +\delta_{jm}\Gamma_{inkl}
			  	  +\delta_{km}\Gamma_{ijnl}
			  	  +\delta_{lm}\Gamma_{ijkn} . 
\end{align}
The components of \( \Pib \) and \( \Pib' \) are then given by
\begin{align}
  \Pi_{ijklmn} &= \Theta_{ijklpq}\overline{Y}^{-1}_{pqmn} , \\
  \Pi'_{ijklmn} &= \Theta_{ijklpq}\brak{\overline{Y}^{-1}_{pqrn}T^{B}_{rm}-\overline{Y}^{-1}_{pqrm}T^{B}_{rn}} + \Psi_{ijklmn} .
\end{align}

\begin{table}
	\caption{The mapping between the two major notation conventions used in this paper. See the main text for a discussion of the components of \( \Pib \) and \( \Pib' \). Note also that the first and second elastic tensors are sometimes represented by \( \AAA \) and \( \CCC \); we have \wording{listed} \( \aaaa \) and \( \cccc \) in the table because, like the elastic tensors of \citeauthor{dahlen1998theoretical}, they are expressed relative to a natural reference configuration.}
	\label{table:notations}
	\begin{tabular}{@{}lll}
	Quantity &  Dahlen \& Tromp  &  Marsden \& Hughes \\ 
	%%%%%%%%%%%%%%%%%%%%%%%%%%%%%%%%%%%%%%%%%%%%%%%%%%%%
	\hline
	%%%%%%%%%%%%%%%%%%%%%%%%%%%%%%%%%%%%%%%%%%%%%%%%%%%%
	First elastic tensor  &  \( \Lam \) & \( \aaaa \) \\
	%%%%%%%%%%%%%%%%%%%%%%%%%%%%%%%%%%%%%%%%%%%%%%%%%%%%
	Second elastic tensor & \( \Xib \) & \( \cccc \) \\
	%%%%%%%%%%%%%%%%%%%%%%%%%%%%%%%%%%%%%%%%%%%%%%%%%%%%
	Background (second) elastic tensor & \( \Gam \) & \( \cccc^{0} \) \\ 
	%%%%%%%%%%%%%%%%%%%%%%%%%%%%%%%%%%%%%%%%%%%%%%%%%%%%
	Incremental second elastic tensor & \( \Delta\Xib \) & \( \cccc^{1} \) \\
	%%%%%%%%%%%%%%%%%%%%%%%%%%%%%%%%%%%%%%%%%%%%%%%%%%%%
	First Piola-Kirchhoff stress & \( \TT^{PK} \) & \( \PP \) \\
	%%%%%%%%%%%%%%%%%%%%%%%%%%%%%%%%%%%%%%%%%%%%%%%%%%%%
	Second Piola-Kirchhoff stress & \( \TT^{SK} \) & \( \SSSSS \) \\
	%%%%%%%%%%%%%%%%%%%%%%%%%%%%%%%%%%%%%%%%%%%%%%%%%%%%
	Eulerian Cauchy stress & \( \TT^{E} \) & \( \sig \) \\
	%%%%%%%%%%%%%%%%%%%%%%%%%%%%%%%%%%%%%%%%%%%%%%%%%%%%
	Equilibrium Cauchy stress & \( \TT^{0} \) & \( \sig_{e} \) (shortened to \( \sig \) in Sections \ref{sec:lambda_of_sigma} and \ref{sec:calculations}) \\
	%%%%%%%%%%%%%%%%%%%%%%%%%%%%%%%%%%%%%%%%%%%%%%%%%%%%
	Background Cauchy stress & \( \TT^{B} \) & \( \sig^{0} \) \\ 
	%%%%%%%%%%%%%%%%%%%%%%%%%%%%%%%%%%%%%%%%%%%%%%%%%%%%
	Incremental Cauchy stress & \( \Delta\TT^{0} \) & \( \sig^{1} \) \\
	%%%%%%%%%%%%%%%%%%%%%%%%%%%%%%%%%%%%%%%%%%%%%%%%%%%%
	Incremental pressure & \( \Delta p^{0} \) & \( p^{1} \) \\
	%%%%%%%%%%%%%%%%%%%%%%%%%%%%%%%%%%%%%%%%%%%%%%%%%%%%
	Incremental deviatoric stress & \( \Delta\DD^{0} \) & \( \DD^{1} \) \\
	%%%%%%%%%%%%%%%%%%%%%%%%%%%%%%%%%%%%%%%%%%%%%%%%%%%%
%	\edit{Referential density} & \( \rho^{0} \) & \( \rho \) \\
	%%%%%%%%%%%%%%%%%%%%%%%%%%%%%%%%%%%%%%%%%%%%%%%%%%%%
	Polar decomposition theorem & \( \FF = \QQ\RR = \mathbf{L}\QQ \) & \( \FF=\RR\UUU=\VV\RR \) \\ 
	%%%%%%%%%%%%%%%%%%%%%%%%%%%%%%%%%%%%%%%%%%%%%%%%%%%%
	  &   &   \\
	%%%%%%%%%%%%%%%%%%%%%%%%%%%%%%%%%%%%%%%%%%%%%%%%%%%%
	Central equations \eqref{eqs:sig_Lam_of_F} & 
	{$\!
	\begin{aligned}
	T^{0}_{ij} &= \frac{\rho^{0}}{\det\FF} F_{ip}F_{jq} \frac{\partial U^{L}}{\partial E^{L}_{pq}} \\
    \Xi_{ijkl} &= \frac{\rho^{0}}{\det\FF} F_{ip}F_{jq}F_{kr}F_{ls} \frac{\partial^{2} U^{L}}{\partial E^{L}_{pq} \partial E^{L}_{rs}}
	\end{aligned}$} & 
	{$\!
	\begin{aligned}
		 \sig &= 2 J_{\FF}^{-1}\Tmult{\FF} \cdot D\tilde{V}\brak{\CC} \\
  		\aaaa &= 4 J_{\FF}^{-1}\Tmult{\FF} D^{2}\tilde{V}\brak{\CC} \Tmult{\FF}^{T} + \Rmult{\sig} 
    \end{aligned}$} \\
	%%%%%%%%%%%%%%%%%%%%%%%%%%%%%%%%%%%%%%%%%%%%%%%%%%%%
	  &   &   \\
	%%%%%%%%%%%%%%%%%%%%%%%%%%%%%%%%%%%%%%%%%%%%%%%%%%%%
	Linearised expressions \eqref{eq:linearisation_general} &
	{$\!
	\begin{aligned}
		\Delta\Xi_{ijkl} &= \Pi_{ijklmn}\Delta T^{0}_{mn} + \Pi'_{ijklmn}\omega^{1}_{mn}
	\end{aligned}$} & 
	{$\!
	\begin{aligned}
		 \cccc^{1} &= \Chimult{\uu^{1}+\omb^{1}}\cccc^{0} + \cccc^{0}\Chimult{\uu^{1}-\omb^{1}} 
            - \cccc^{0}\tr{\uu^{1}} + 8 D^{3}\tilde{V}\brak{\id}\cdot\uu^{1} \\
  		\uu^{1} &= \brak{\cccc^{0}+\ChimultHat{\sig^{0}}-\sig^{0}\otimes\id}^{-1}\cdot\brak{\sig^{1}-\Chimult{\omb^{1}}\cdot\sig^{0}}
    \end{aligned}$} \\
    %%%%%%%%%%%%%%%%%%%%%%%%%%%%%%%%%%%%%%%%%%%%%%%%%%%%
	  &   &   \\
    %%%%%%%%%%%%%%%%%%%%%%%%%%%%%%%%%%%%%%%%%%%%%%%%%%%%
    Any second-rank tensor & \( \braksq{\mathbf{D \& T}}_{ij} \) & \( \braksq{\mathbf{M \& H}}_{ji} \) \\
	%%%%%%%%%%%%%%%%%%%%%%%%%%%%%%%%%%%%%%%%%%%%%%%%%%%%
	Any fourth-rank tensor & \( \braksq{\mathbf{D \& T}}_{ijkl} \) & \( \braksq{\mathbf{M \& H}}_{jilk} \) \\
	%%%%%%%%%%%%%%%%%%%%%%%%%%%%%%%%%%%%%%%%%%%%%%%%%%%%
	\hline
	%%%%%%%%%%%%%%%%%%%%%%%%%%%%%%%%%%%%%%%%%%%%%%%%%%%%
	\end{tabular}
\end{table}

\section{The invertibility of \( \hat{\Sig} \)}
\label{Sig_invertibility}

The existence of the inverse function \( \hat{\Sig}^{-1} \)  depends on the specific choice of constitutive relation; without fixing \(\tilde{V}\) concretely it is difficult to obtain definite results. However, we can show that the inverse mapping exists at least locally in cases of some physical importance. To proceed, we appeal to the \textit{inverse function theorem} \citep[e.g.][]{marsden1994mathematical}. The theorem tells us that the nonlinear mapping \(\hat{\Sig}\) has a well-defined inverse function in some open neighbourhood of \( \hat{\Sig}\brak{\UUU} \) if \(D\hat{\Sig}(\UUU)\), a linear mapping from the vector-space of symmetric matrices into itself, is invertible. 

Let us work in the vicinity of \( \UUU=\id \). We will set \( \RR=\id \) for clarity, but that does not affect the validity of our argument. In that case we have
\begin{align}
  \hat{\Sig}\brak{\id} &= 2 DV\brak{\id}  .
\end{align}
Making use of eq.\eqref{eq:chi_mult_sym_identity} of Appendix \ref{app:notations}, we can expand \( \hat{\Sig} \) about the identity as
\begin{align}
  \hat{\Sig}\brak{\id+\uu}
  &= 2 \braksq{1-\tr\uu}\braksq{\idgen+\Chimult{\uu}}\cdot\braksq{DV\brak{\id}+2D^{2}V\brak{\id}\cdot\uu} \nonumber\\
  &= \hat{\Sig}\brak{\id} - \tr\uu \hat{\Sig}\brak{\id} + \Chimult{\uu}\cdot\hat{\Sig}\brak{\id} + 4 D^{2}V\brak{\id}\cdot\uu \nonumber\\
  &= \hat{\Sig}\brak{\id} + \brak{4 D^{2}V\brak{\id}+\Chimult{\hat{\Sig}\brak{\id}}-\hat{\Sig}\brak{\id}\otimes\id}\cdot\uu ,
\end{align}
with \( \uu \) a small symmetric matrix. We have shown that
\begin{align}
\label{eq:Sig_hat_first_deriv_app}
  D\hat{\Sig}\brak{\id} = 4 D^{2}V\brak{\id}+\ChimultHat{\hat{\Sig}\brak{\id}}-\hat{\Sig}\brak{\id}\otimes\id ,
\end{align} 
where we have written \( \Chimult{\hat{\Sig}\brak{\id}} \) in its explicitly symmetrised form \( \ChimultHat{\hat{\Sig}\brak{\id}} \) (eq.\ref{eq:fourth_rank_sym_defn}). Inspection of expression \eqref{eq:Sig_hat_first_deriv_app} indicates that the invertibility of \( D^{2}V\brak{\id} \) is a sufficient condition for \( D\hat{\Sig}\brak{\id} \) to possess a unique inverse locally, in the absence of equilibrium stress. As long as \( \hat{\Sig}\brak{\id} \) is not too large, invertibility of \( D^{2}V\brak{\id} \) should remain sufficient for the local existence of a unique \( \hat{\Sig}^{-1} \). One can show that \( D^{2}V\brak{\id} \) being invertible is equivalent to the condition of \textit{linearised stability} of the equilibrium \citep[e.g.][]{marsden1994mathematical}. It is essential to enforce linearised stability, for otherwise unphysical motions are permitted in which the strain-energy decreases upon deformation.

In summary, we have argued that linearised stability of the equilibrium is a sufficient condition for \( \hat{\Sig} \) to possess a unique inverse in the neighbourhood of \( \hat{\Sig}\brak{\id} \), as long as the equilibrium stress is not too large. By no means is this a proof of the global invertibility of \( \hat{\Sig} \), but we believe that it makes it plausible that we should be able to write down a well-defined inverse function \( \hat{\Sig}^{-1} \) when required.

\section{Calculation of the linearised elastic tensor}
\label{app:calculations}

\subsection{General expressions}
\label{app:sub:general_calculation}

We assume that the body is initially in a state described by
\begin{align}
  \sig &= \sig^{0} \\
  \aaaa &= \aaaa^{0} .
\end{align}
Both quantities are considered to be the derivatives of some background strain-energy evaluated at the identity, so that \( \aaaa^{0} \) is given uniquely by
\begin{align}
  \aaaa^{0} = \bar{\aaaa}\brak{\sig^{0},\id}  .
\end{align}
Once again, the function \( \bar{\aaaa} \) has the form
\begin{align}
\label{eq:app_Lam_bar}
  \bar{\aaaa}\brak{\sig,\RR} = \Tmult{\RR}\braksq{\brak{\hat{\AAA}\circ\hat{\Sig}^{-1}}\brak{\Tmult{\RR}^{T}\cdot\sig}}\Tmult{\RR}^{T}  .
\end{align}
We then introduce a small change in the stress:
\begin{align}
  \sig = \sig^{0} + \sig^{1}
\end{align}
for small \( \sig^{1} \). The system is assumed to be perturbative, so that we may write
\begin{align}
  \aaaa = \aaaa^{0} + \aaaa^{1}
\end{align}
for small \( \aaaa^{1} \). For the rest of this section we will neglect all terms higher than first order. 

Under the framework of Section \ref{sec:lambda_of_sigma}, the change in stress is considered to be induced by some deformation gradient \( \FF \), so it follows that \( \FF \) should take the form
\begin{align}
  \FF = \id + \mathbf{f}  .
\end{align}
with \( \mathbf{f} \) small. Now, the arbitrary rotation matrix \( \RR \) in eq.(\ref{eq:app_Lam_bar}) is a relic of \( \FF \), having been introduced by the identity
\begin{align}
  \FF = \RR \UUU  .
\end{align}
Therefore it too must be considered to be a small perturbation to its background value:
\begin{align}
  \RR = \id + \omb ,
\end{align}
with \( \omb \) some small antisymmetric matrix. The same is true for \( \UUU \), that is
\begin{align}
  \UUU = \id + \uu ,
\end{align}
but with the small matrix \( \uu \) of course symmetric. It is clear that the elastic tensor satisfies
\begin{align}
  \aaaa^{0} + \aaaa^{1} = \bar{\aaaa}\brak{\sig^{0}+\sig^{1},\id+\omb} ,
\end{align}
which we may expand in a Taylor series to find
\begin{align}
\label{eq:app_Lam_pert}
  \aaaa^{1} = \brak{D_{\sig}\bar{\aaaa}}\brak{\sig^{0},\id}\cdot\sig^{1} + \brak{D_{\RR}\bar{\aaaa}}\brak{\sig^{0},\id}\cdot\omb  .
\end{align}
The \( \RR \)-derivative can conveniently be written in terms of the \( \sig \)-derivative. Using the full form of \( \bar{\aaaa} \) from eq.(\ref{eq:app_Lam_bar}), as well as the operator \( \idgen \) defined in eq.(\ref{eq:id_op_defn}),
\begin{align}
  \bar{\aaaa}\brak{\sig,\id+\omb} 
  &= \brak{\idgen + \Chimult{\omb}}\braksq{\brak{\hat{\AAA}\circ\hat{\Sig}^{-1}}\brak{\sig - \Chimult{\omb}\cdot\sig}}\brak{\idgen - \Chimult{\omb}} \nonumber\\
  &= \brak{\hat{\AAA}\circ\hat{\Sig}^{-1}}\brak{\sig - \Chimult{\omb}\cdot\sig}
    + \Chimult{\omb}\braksq{\brak{\hat{\AAA}\circ\hat{\Sig}^{-1}}\brak{\sig}} - \braksq{\brak{\hat{\AAA}\circ\hat{\Sig}^{-1}}\brak{\sig}}\Chimult{\omb} \nonumber\\
  &= \bar{\aaaa}\brak{\sig - \Chimult{\omb}\cdot\sig,\id} 
    + \Chimult{\omb}\bar{\aaaa}\brak{\sig,\id} - \bar{\aaaa}\brak{\sig,\id}\Chimult{\omb}\nonumber\\
  &= \aaaa^{0}
    - \brak{D_{\sig}\bar{\aaaa}}\brak{\sig,\id}\cdot\brak{\Chimult{\omb}\cdot\sig}
    + \Chimult{\omb}\aaaa^{0} - \aaaa^{0}\Chimult{\omb} ,
\end{align}
from which it is apparent that
\begin{align}
  \brak{D_{\RR}\bar{\aaaa}}\brak{\sig^{0},\id}\cdot\omb
  = \Chimult{\omb}\aaaa^{0} - \aaaa^{0}\Chimult{\omb}
    - \brak{D_{\sig}\bar{\aaaa}}\brak{\sig^{0},\id}\cdot\brak{\Chimult{\omb}\cdot\sig^{0}}  .
\end{align}
Substituting this into eq.(\ref{eq:app_Lam_pert}), the perturbation to the elastic tensor is
\begin{align}
\label{eq:app_Lam_pert_2}
  \aaaa^{1} = \Chimult{\omb}\aaaa^{0} - \aaaa^{0}\Chimult{\omb}
            + \brak{D_{\sig}\bar{\aaaa}}\brak{\sig^{0},\id}\cdot\brak{\sig^{1} - \Chimult{\omb}\cdot\sig^{0}}  .
\end{align}

Calculation of \( D_{\sig}\bar{\aaaa}\brak{\sig^{0},\id} \) requires us to find \( D\hat{\AAA}\brak{\id} \) and \( D\hat{\Sig}\brak{\id} \). They are evaluated most conveniently by forming explicit binomial expansions of \( \hat{\Sig} \) and \( \hat{\AAA} \) about \( \UUU = \id + \uu \). At zeroth order
\begin{alignat}{3}
  &\hat{\Sig}\brak{\id} &&= \sig^{0} &&= 2 DV\brak{\id} \\
  &\hat{\AAA}\brak{\id} &&= \aaaa^{0} &&= 4 D^{2}V\brak{\id} + \Rmult{\hat{\Sig}\brak{\id}} ,
\end{alignat}
and it is useful to define
\begin{align}
  \cccc^{0} = 4 D^{2}V\brak{\id} = \hat{\AAA}\brak{\id} - \Rmult{\hat{\Sig}\brak{\id}}  .
\end{align}
We showed in Appendix \ref{Sig_invertibility} that
\begin{align}
\label{eq:Sig_hat_first_deriv_app_2}
  D\hat{\Sig}\brak{\id} = \cccc^{0}+\ChimultHat{\sig^{0}}-\sig^{0}\otimes\id  .
\end{align}
In the same manner, \( \hat{\AAA} \) expands as
\begin{align}
  \hat{\AAA}\brak{\id+\uu}
  &= 4 \brak{1-\tr\uu}\brak{\idgen+\Chimult{\uu}}\braksq{D^{2}V\brak{\id}+2D^{3}V\brak{\id}\cdot\uu}\brak{\idgen+\Chimult{\uu}} + \Rmult{\hat{\Sig}\brak{\id+\uu}} \nonumber\\
  &= \cccc^{0} 
        + 8 D^{3}V\brak{\id}\cdot\uu
  		+ \Chimult{\uu}\cccc^{0}
  		+ \cccc^{0}\Chimult{\uu}
  		- \cccc^{0}\tr\uu
  		+ \Rmult{\sig^{0}} + \Rmult{D\hat{\Sig}\brak{\id}\cdot\uu} ,
\end{align}
so its derivative at the identity is given by
\begin{align}
\label{eq:lam_hat_first_deriv_app}
  D\hat{\AAA}\brak{\id}\cdot\uu = 
  \Rmult{D\hat{\Sig}\brak{\id}\cdot\uu} + 8 D^{3}V\brak{\id}\cdot\uu
  + \Chimult{\uu}\cccc^{0}
  		+ \cccc^{0}\Chimult{\uu}
  		- \cccc^{0}\tr\uu
\end{align}
for any \( \uu \). 

Now we can compute \( D_{\sig}\bar{\aaaa}\brak{\sig^{0},\id} \) by writing
\begin{align}
  \bar{\aaaa}\brak{\sig^{0}+\Sig,\id}
  &= \hat{\AAA}\braksq{\hat{\Sig}^{-1}\brak{\sig^{0}+\Sig}}\nonumber\\
  &= \hat{\AAA}\braksq{\hat{\Sig}^{-1}\brak{\sig^{0}} + \braksq{D\brak{\hat{\Sig}^{-1}}}\brak{\sig^{0}}\cdot\Sig} \nonumber\\
  &= \hat{\AAA}\brakbr{\id + \braksq{D\hat{\Sig}\brak{\id}}^{-1}\cdot\Sig} \nonumber\\
  &= \hat{\AAA}\brak{\id} + D\hat{\AAA}\brak{\id}\cdot\brakbr{\braksq{D\hat{\Sig}\brak{\id}}^{-1}\cdot\Sig} ,
\end{align}
valid for small symmetric \( \Sig \), where we have used the identity
\begin{align}
  D\brak{\hat{\Sig}^{-1}}\brak{\sig^{0}}
  = \brakbr{D\hat{\Sig}\braksq{\hat{\Sig}^{-1}\brak{\sig^{0}}}}^{-1}
  = \braksq{D\hat{\Sig}\brak{\id}}^{-1} . 
\end{align}
Thus,
\begin{align}
  \brak{D_{\sig}\bar{\aaaa}}\brak{\sig^{0},\id}\cdot\brak{\sig^{1} - \Chimult{\omb}\cdot\sig^{0}}
  = D\hat{\AAA}\brak{\id}\cdot\brakbr{\braksq{D\hat{\Sig}\brak{\id}}^{-1}\cdot\brak{\sig^{1} - \Chimult{\omb}\cdot\sig^{0}}},
\end{align}
and we can combine this with eqs. (\ref{eq:lam_hat_first_deriv_app}), (\ref{eq:Sig_hat_first_deriv_app_2}) and (\ref{eq:app_Lam_pert_2}) to write the perturbation to the elastic tensor as
\begin{align}
\label{eq:perturbed_elastic_tensor_final_preliminary}
  \aaaa^{1} = \Chimult{\omb}\aaaa^{0} - \aaaa^{0}\Chimult{\omb}
            + \Chimult{\uu}\cccc^{0} + \cccc^{0}\Chimult{\uu} 
            - \cccc^{0}\tr\uu + 8 D^{3}\tilde{V}\brak{\id}\cdot\uu
            + \Rmult{\brak{\sig^{1}-\Chimult{\omb}\cdot\sig^{0}}} ,
\end{align}
with
\begin{align}
  \uu = \brak{\cccc^{0}+\ChimultHat{\sig^{0}}-\sig^{0}\otimes\id}^{-1}\cdot\brak{\sig^{1}-\Chimult{\omb}\cdot\sig^{0}}  .
\end{align}
A final simplification is obtained by noting that
\begin{align}
  \Chimult{\omb}\aaaa^{0} - \aaaa^{0}\Chimult{\omb}
  =  \Chimult{\omb} \cccc^{0} - \cccc^{0}\Chimult{\omb}
   + \Chimult{\omb} \Rmult{\sig^{0}} - \Rmult{\sig^{0}}\Chimult{\omb}  .
\end{align}
It is then readily shown that
\begin{align}
  \Chimult{\omb} \Rmult{\sig^{0}} - \Rmult{\sig^{0}}\Chimult{\omb}
  =\Rmult{\Chimult{\omb}\cdot\sig^{0}} ,
\end{align}
which partially cancels with the final term in eq.(\ref{eq:perturbed_elastic_tensor_final_preliminary}). Thus, our final expression for the perturbation to the elastic tensor is
\begin{align}
\begin{aligned}
  \aaaa^{1} &= \cccc^{1} + \Rmult{\sig^{1}} \\
  \cccc^{1} &= \Chimult{\uu+\omb}\cccc^{0} + \cccc^{0}\Chimult{\uu-\omb} 
            - \cccc^{0}\tr\uu + 8 D^{3}\tilde{V}\brak{\id}\cdot\uu \\
       \uu &= \brak{\cccc^{0}+\ChimultHat{\sig^{0}}-\sig^{0}\otimes\id}^{-1}\cdot\brak{\sig^{1}-\Chimult{\omb}\cdot\sig^{0}} .
\end{aligned}
\end{align}
Note that the arbitrary antisymmetric matrix \( \omb \) insinuates itself implicitly via the definition of \( \uu \), as well as appearing explicitly in the expression for \( \cccc^{1} \). When the background material is isotropic \( \omb \) can be set to zero without loss of generality.

\subsection{Isotropic, hydrostatically pre-stressed background}
\label{app:sub:isotropic_calculation}

For a zeroth-order hydrostatic stress and elastic tensor given by
\begin{align}
  \sig^{0} &= -p^{0}\id \nonumber\\
  \cccc^{0}  &= \lambda\id\otimes\id + 2\mu\overline{\idgen} ,
\end{align}
the perturbation to the stress satisfies
\begin{align}
  \sig^{1} 
  &= \brak{\cccc^{0}+\ChimultHat{\sig^{0}}-\sig^{0}\otimes\id}\cdot\uu \nonumber\\
  &= \brak{\lambda\id\otimes\id + 2\mu\overline{\idgen}-2p^{0}\overline{\idgen}+p^{0}\id\otimes\id}\cdot\uu \nonumber\\
  &= \tilde{\lambda}\tr{\uu}\id + 2\tilde{\mu}\uu ,
\end{align}
with
\begin{subequations}
\begin{align}
  \tilde{\lambda} &= \lambda + p^{0} \\
  \tilde{\mu} &= \mu - p^{0}  .
\end{align}	
\end{subequations}
It follows easily that
\begin{align}
  \tr{\sig^{1}} &= \brak{3\tilde{\lambda}+2\tilde{\mu}}\tr{\uu},
\end{align}
from which
\begin{align}
\uu &= \frac{1}{2\tilde{\mu}}\braksq{\sig^{1}-\frac{\tilde{\lambda}}{3\tilde{\lambda}+2\tilde{\mu}}\tr{\sig^{1}}\id}  .
\end{align}
Next, we split the stress into its hydrostatic and deviatoric components. Writing
\begin{align}
  \sig^{1} = -p^{1}\id + \DD^{1},
\end{align}
with
\begin{align}
  \tr{\DD^{1}} = 0,
\end{align}
we find that
\begin{align}
  \uu = -x\id + \frac{1}{2\tilde{\mu}}\DD^{1},
\end{align}
where we define the shorthand
\begin{align}
  x \equiv \frac{p^{1}}{3\tilde{\lambda}+2\tilde{\mu}}  .
\end{align}

Let us now turn to the expression for \( \cccc^{1} \), which is given by
\begin{align}
  \cccc^{1} = \Chimult{\uu}\cccc^{0} + \cccc^{0}\Chimult{\uu} - \tr{\uu}\cccc^{0} + 8 D^{3}\tilde{V}\brak{\id}\cdot\uu
\end{align}
in the isotropic case. For the third-derivative term, we follow the same philosophy as \citet{Murnaghan_1937}, considering a Taylor-expansion of the strain-energy function \( \tilde{V} \) up to third-order in the \textit{scalar-invariants of} \( \CC \). The most general such expression is, in index-notation,
\begin{align}
  [8 D^{3}\tilde{V}\brak{\id}]_{ijklmn}C_{ij}C_{kl}C_{mn}
  &=  \zeta_{1} \tr{\CC}^{3} 
    + 3\zeta_{2}\tr{\CC}\tr{\CC^{2}} 
    + 2\zeta_{3} \tr{\CC^{3}} ,
\end{align}
where we have defined the constants \( \zeta_{1} \), \( \zeta_{2} \) and \( \zeta_{3} \). This expression should be compared with \citet[p.250, 2nd equation from bottom]{Murnaghan_1937}, wherein the Murnaghan constants \( l \), \( m \) and \( n \) are introduced, as well as the bottom equation of p.240 of that paper, where the scalar-invariants of a tensor are defined. Our constants \( \zeta_{1} \), \( \zeta_{2} \) and \( \zeta_{3} \), which we have found convenient to use when performing calculations, are equivalent to Murnaghan's, and we shall simply refer to them as `the Murnaghan constants'. Using our operator notation, the expression for the third-derivative term in \( \cccc^{1} \) is
\begin{align}
  8 D^{3}\tilde{V}\brak{\id}\cdot\uu 
  = \zeta_{1}\tr{\uu}\id\otimes\id + \zeta_{2}\tr{\uu}\overline{\idgen}
  + \zeta_{2}\brak{\id\otimes\uu + \uu\otimes\id} + \zeta_{3}\ChimultHat{\uu},
\end{align}
where we have used hats to mark symmetrisation of certain operators (see eq.\ref{eq:fourth_rank_sym_defn}). With \( \cccc^{0} \) given above, and bearing in mind that
\begin{align}
  \Chimult{\uu}\brak{\id\otimes\id} &= 2 \uu\otimes\id, \\
  \Chimult{\uu}\overline{\idgen} &= \ChimultHat{\uu},
\end{align}
we obtain
\begin{align}
  \cccc^{1} = \brak{\zeta_{1}-\lambda}\tr{\uu}\id\otimes\id + \brak{\zeta_{2}-2\mu}\tr{\uu}\overline{\idgen} + \brak{2\lambda+\zeta_{2}}\brak{\id\otimes\uu+\uu\otimes\id} + \brak{4\mu+\zeta_{3}}\ChimultHat{\uu}  .
\end{align}
We can now substitute in our expression for \( \uu \) in terms of \( p^{1} \) and \( \DD^{1} \), yielding
\begin{align}
  \cccc^{1} 
  &= -\frac{\lambda+3\zeta_{1}+2\zeta_{2}}{3\lambda+2\mu+p^{0}}p^{1}\id\otimes\id
     -\frac{2\mu+3\zeta_{2}+2\zeta_{3}}{3\lambda+2\mu+p^{0}}p^{1}\overline{\idgen}
     +\frac{2\lambda+\zeta_{2}}{2\brak{\mu-p^{0}}}\brak{\id\otimes\DD^{1}+\DD^{1}\otimes\id} 
     +\frac{4\mu+\zeta_{3}}{2\brak{\mu-p^{0}}}\ChimultHat{\DD^{1}}  .
\end{align}
With this, we may write the perturbation to first elastic tensor as
\begin{align}
  \aaaa^{1} = 
  c p^{1}\id\otimes\id + 2 d p^{1}\overline{\idgen} + a\brak{\id\otimes\DD^{1}+\DD^{1}\otimes\id} + 2b\ChimultHat{\DD^{1}}
  - p^{1}\idgen + \Rmult{\DD^{1}} ,
\end{align}
where \( a \), \( b \), \( c \) and \( d \) are given in terms of the Murnaghan constants as
\begin{align}
  a &\equiv \frac{\lambda + \frac{1}{2}\zeta_{2}}{\mu-p^{0}} \\
  b &\equiv \frac{\mu + \frac{1}{4}\zeta_{3}}{\mu-p^{0}} \\
  c &\equiv -\frac{\lambda+3\zeta_{1}+2\zeta_{2}}{3\kappa+p^{0}} \\
  d &\equiv -\frac{\mu + \frac{3}{2}\zeta_{2} + \zeta_{3}}{3\kappa+p^{0}}  .
\end{align}
The first elastic tensor is written in index notation as
\begin{align}
  \mathsf{a}_{ijkl} 
  &= \brak{\kappa-\frac{2}{3}\mu+p^{1}c}\delta_{ij}\delta_{kl}
   + \brak{\mu+ p^{1}d}\brak{\delta_{ik}\delta_{jl}+\delta_{il}\delta_{jk}}
   -\brak{p^{0}+p^{1}}\delta_{ik}\delta_{jl} \nonumber\\
   &\qquad 
   + a\brak{\delta_{ij}\tau^{1}_{kl}+\delta_{kl}\tau^{1}_{ij}}
   + b\brak{\delta_{ik}\tau^{1}_{jl}+\delta_{jl}\tau^{1}_{ik}+\delta_{il}\tau^{1}_{jk}+\delta_{jk}\tau^{1}_{il}} + \delta_{ik}\tau^{1}_{jl} .
\end{align}

\subsection{Transversely-isotropic, `quasi-hydrostatically' pre-stressed background}
\label{app:sub:transverse_isotropic_calculation}

The transversely-isotropic calculation is more intricate than the isotropic one for a number of reasons. Firstly, and perhaps most simply remedied, we can no longer ignore the arbitrary antisymmetric matrix \( \omb \). Secondly, the stress-strain relationship
\begin{align}
\label{eq:stress_strain_ti_app_initial}
  \uu &= \brak{\cccc^{0}+\ChimultHat{\sig^{0}}-\sig^{0}\otimes\id}^{-1}\cdot\brak{\sig^{1}-\Chimult{\omb}\cdot\sig^{0}}
\end{align}
is not as easily inverted for a transversely-isotropic elastic tensor. Thirdly, there are many more terms to keep track of.

Let us begin by recalling the stress-free transversely-isotropic elastic-tensor given in eq.(\ref{eq:elastic_tensor_ti}):
\begin{align}
  \repeqLamTI  .
\end{align}
For the remainder of the calculation we will switch to our `operator-based' notation. We also introduce a shorthand that is standard in many areas of physics, whereby a symmetric expression is written as
\begin{align}
  \brak{\text{symmetric object}} = \brak{\text{not-necessarily-symmetric object}} + h.c. ,
\end{align}
an example being
\begin{align}
  \brak{\id\otimes\DD^{1}+\DD^{1}\otimes\id} = \id\otimes\DD^{1} + h.c. .
\end{align}
It will allow us to avoid significant clutter later on. (Here, \( h.c. \) technically stands for `hermitian conjugate'.) We can now rewrite the zeroth-order elastic tensor as
\begin{align}
  \cccc^{0} = \lambda \id\otimes\id + 2\mu\overline{\idgen} + 8\gamma \NN\otimes\NN + 4\beta\brak{\id\otimes\NN + h.c.} - 2\alpha \ChimultHat{\NN} ,
\end{align}
and the pre-stress (eq.\ref{eq:stress_ti}) as
\begin{align}
  \sig^{0} = -\brak{p^{0}+\frac{q^{0}}{3}}\id + q^{0}\NN  .
\end{align}
\( \NN \) satisfies the relations
\begin{align}
  \tr{\NN} = 1,
\end{align}
and
\begin{align}
  \NN^{n} = \NN
\end{align}
for positive integer \( n \), from which it follows by induction that
\begin{align}
\label{eq:chimulthat_N^n_defn}
  \Chimult{\NN}^{n} = \Chimult{\NN} + \brak{2^{n}-2}\NN\otimes\NN .
\end{align}
It is useful to note that
\begin{align}
  \ChimultHat{\sig^{0}}-\sig^{0}\otimes\id
  &= -2\brak{p^{0}+\frac{q^{0}}{3}}\overline{\idgen} + q^{0}\ChimultHat{\NN} + \brak{p^{0}+\frac{q^{0}}{3}}\id\otimes\id - q^{0}\NN\otimes\id,
\end{align}
for now we can rewrite the stress-strain relationship (eq.\ref{eq:stress_strain_ti_app_initial}) as
\begin{align}
\label{eq:stress_strain_ti_app_second}
  \sig^{1}-\Chimult{\omb}\cdot\sig^{0}
  &= \braksq{\cccc^{0}-2\brak{p^{0}+\frac{q^{0}}{3}}\overline{\idgen} + q^{0}\ChimultHat{\NN} + \brak{p^{0}+\frac{q^{0}}{3}}\id\otimes\id - q^{0}\NN\otimes\id}\cdot\uu \nonumber\\
  &= \tilde{\lambda}\tr{\uu}\id 
   + 2\tilde{\mu}\uu 
   + 8\tilde{\gamma}\innerproduct{\NN}{\uu}\NN 
   - 2\tilde{\alpha}\ChimultHat{\NN}\cdot\uu 
   + 4\tilde{\beta}\innerproduct{\NN}{\uu}\id
   + 4\tilde{\beta}'\tr{\uu}\NN ,
\end{align}
with
\begin{subequations}
\label{eqs:ti_tilde_constants_defns}
\begin{align}
  \tilde{\lambda} &= \lambda + p^{0} + \frac{q^{0}}{3} \\
  \tilde{\mu}     &=     \mu - p^{0} - \frac{q^{0}}{3} \\
  \tilde{\gamma}  &= \gamma \\
  \tilde{\alpha}  &= \alpha - \frac{q^{0}}{2} \\
  \tilde{\beta}   &= \beta \\
  \tilde{\beta}'  &= \beta - \frac{q^{0}}{4}  .
\end{align}	
\end{subequations}

We are now in position to solve for \( \uu \) in terms of \( \sig^{1} \) and \( \omb \). In the isotropic case we could invert for \( \uu \) simply by writing \( \tr{\uu} \) in terms of \( \tr{\sig^{1}} \). Here, by analogy, we take not only the trace of both sides, but also the inner product with \( \NN \), to find that
\begin{align}
\label{eq:tru_Nu_relation}
  \begin{pmatrix}
  	\tr{\sig^{1}}\\
  	\innerproduct{\NN}{\sig^{1}}
  \end{pmatrix}
  =
  \begin{pmatrix}
  	3\tilde{\lambda}+2\tilde{\mu}+4\tilde{\beta}' & 
  	8\tilde{\gamma}-4\tilde{\alpha}+12\tilde{\beta} \\
  	\tilde{\lambda}+4\tilde{\beta}' &
  	2\tilde{\mu}+8\tilde{\gamma}-4\tilde{\alpha}+4\tilde{\beta} 
  \end{pmatrix}
  \begin{pmatrix}
  	\tr{\uu}\\
  	\innerproduct{\NN}{\uu}
  \end{pmatrix} .
\end{align}
Note that \( \omb \) is not present here because
\begin{align}
  \tr{\Chimult{\omb}\cdot\sig^{0}}
  &= \innerproduct{\id}{\Chimult{\omb}\cdot\sig^{0}} 
  = \innerproduct{\Chimult{\omb}^{T}\cdot\id}{\sig^{0}},
\end{align}
which vanishes due to the antisymmetry of \( \omb \). By the same token,
\begin{align}
  \innerproduct{\NN}{\Chimult{\omb}\cdot\sig^{0}}
  &= -\brak{p^{0}+\frac{q^{0}}{3}}\innerproduct{\NN}{\Chimult{\omb}\cdot\id}
  	 +q^{0}\innerproduct{\NN}{\Chimult{\omb}\cdot\NN} = 0,
\end{align}
with the second term vanishing due to the antisymmetry of the operator \( \Chimult{\omb} \) (which obviously follows from that of \( \omb \)). Thus, both \( \tr{\uu} \) and \( \ip{\NN}{\uu} \) can be written in terms of known quantities wherever they appear.

Now, in a further step reminiscent of the isotropic case, we move most of the terms in eq.\eqref{eq:stress_strain_ti_app_second} over to the left hand side, giving
\begin{align}
  \label{eq:stress_strain_ti_app_third}
  2\tilde{\mu}\uu - 2\tilde{\alpha} \Chimult{\NN}\cdot\uu = 2\tilde{\mu}\Sig',
\end{align}
where \( \Sig' \) is a shorthand defined as
\begin{align}
  \Sig' 
  = \frac{1}{2\tilde{\mu}}
  \braksq{\sig^{1}-\Chimult{\omb}\cdot\sig^{0}-\tilde{\eta}\id-\tilde{\theta}\NN} ,
\end{align}
with
\begin{align}
\label{eq:eta_theta_defn}
  \begin{pmatrix}
  	\tilde{\eta}\\ 
  	\tilde{\theta}
  \end{pmatrix}
  =
  \begin{pmatrix}
  	\tilde{\lambda} & 4\tilde{\beta}\\
  	4\tilde{\beta}' & 8\tilde{\gamma}
  \end{pmatrix}
  \begin{pmatrix}
  	\tr{\uu}\\
  	\ip{\NN}{\uu}
  \end{pmatrix} .
\end{align}
We are left to solve the equation
\begin{align}
  \brak{\idgen-\frac{\tilde{\alpha}}{\tilde{\mu}}\Chimult{\NN}}\cdot\uu = \Sig'  .
\end{align}
It is readily verified that the necessary inverse operator is
\begin{align}
  \brak{\idgen-\frac{\tilde{\alpha}}{\tilde{\mu}}\Chimult{\NN}}^{-1}
  &= \idgen + \tilde{P}\Chimult{\NN} + \tilde{Q}\NN\otimes\NN,
\end{align}
with
\begin{subequations}
\label{eqs:PQ_defns}
\begin{align}
  \tilde{P} &= \frac{\frac{\tilde{\alpha}}{\tilde{\mu}}}{1-\frac{\tilde{\alpha}}{\tilde{\mu}}} \\
  \tilde{Q} &= \frac{2\frac{\tilde{\alpha}}{\tilde{\mu}}}{1-2\frac{\tilde{\alpha}}{\tilde{\mu}}}\tilde{P},
\end{align}	
\end{subequations}
from which it follows that
\begin{align}
  \uu = \Sig' + \tilde{P}\Chimult{\NN}\cdot\Sig' + \tilde{Q}\ip{\NN}{\Sig'}\NN .
\end{align}
This procedure seems to break down when \( \tilde{\mu} \) equals \( \tilde{\alpha} \) or \( 2\tilde{\alpha} \), but the physical significance of these cases is not yet clear to the present authors. From here, with the help of the identity
\begin{align}
  \Chimult{\NN}\Chimult{\omb}\cdot\NN = \Chimult{\omb}\cdot\NN ,
\end{align} 
it is a matter of relatively straightforward algebra to find that
\begin{align}
\label{eq:u_of_sigma_and_omega}
  2\tilde{\mu}\uu = \sig^{1}-\tilde{\eta}\id + \tilde{P}\Chimult{\NN}\cdot\sig^{1}+\tilde{R}\NN - q^{0}(1+\tilde{P})\Chimult{\omb}\cdot\NN,
\end{align}
with
\begin{align}
\label{eq:RTilde_defn}
  \tilde{R} = \tilde{Q}\brak{\ip{\NN}{\sig^{1}}-\tilde{\eta}-\tilde{\theta}}-\tilde{\theta}-2\tilde{P}\brak{\tilde{\eta}+\tilde{\theta}}  .
\end{align}
We now have an expression for \( \uu \) in terms of:
\begin{enumerate}
\renewcommand{\theenumi}{(\arabic{enumi})}
	\item the transversely-isotropic constants \( \lambda \), \( \mu \), \( \alpha \), \( \beta \) and \( \gamma \), as well as \( \nub \);
	\item the constants \( p^{0} \) and \( q^{0} \) which define the equilibrium stress-state;
	\item the perturbation to the stress \( \sig^{1} \);
	\item an antisymmetric matrix \( \omb \) `pointing' in an arbitrary direction in the plane perpendicular to the unperturbed symmetry-axis.
\end{enumerate}
This is to be substituted into the expression
\begin{align}
  \cccc^{1} &= \Chimult{\uu+\omb}\cccc^{0} + \cccc^{0}\Chimult{\uu-\omb} 
            - \cccc^{0}\tr\uu + 8 D^{3}\tilde{V}\brak{\id}\cdot\uu
\end{align}
for the perturbed elastic tensor.

In order to make progress we must parametrise the third derivatives of the transversely-isotropic strain-energy function. As stated in the main text, such a strain-energy function will generally depend on \( \CC \) not only through the invariants \( I_{1}\), \( I_{2} \) and \( I_{3} \) defined in eq.\eqref{eq:scalar_invariants_defn_iso}, but also through the terms \citep[e.g.][]{Holzapfel_2000}
\begin{align}
  \ip{\nub}{\CC\cdot\nub} 
  &= \ip{\NN}{\CC} \\
  \ip{\nub}{\CC^{2}\cdot\nub} 
  &= \ip{\NN}{\CC^{2}} = \frac{1}{2}\ip{\CC}{\Chimult{\NN}\cdot\CC}  .
\end{align}
The third order terms of \( \tilde{V} \)'s Taylor-series about the identity therefore satisfy
\begin{align}
\label{eq:strain_energy_third_deriv_param_zeta_defn}
  [8 D^{3}\tilde{V}\brak{\id}]_{ijklmn}C_{ij}C_{kl}C_{mn} 
  &= \zeta_{1}\tr{\CC}^{3} 
   + 3\zeta_{2}\tr{\CC}\tr{\CC^{2}} 
   + 2\zeta_{3}\tr{\CC^{3}} 
   \nonumber\\&\qquad
   + 3\zeta_{4}\tr{\CC}\ip{\CC}{\NN}^{2} 
   + 3\zeta_{5}\tr{\CC}\ip{\CC}{\Chimult{\NN}\cdot\CC} 
   \nonumber\\&\qquad
   + 3\zeta_{6}\ip{\CC}{\idgen\cdot\CC}\ip{\CC}{\NN} 
   + 3\zeta_{7}\ip{\CC}{\NN}\ip{\CC}{\Chimult{\NN}\cdot\CC},
\end{align}
for some material-dependent constants \( \brakbr{\zeta_{i}} \), so that the corresponding term in the elastic tensor is
\begin{align}
  8 D^{3}\tilde{V}\brak{\id}\cdot\uu
  &=
  \zeta_{1}\tr{\uu}\id\otimes\id 
  \nonumber\\&\qquad
  + \zeta_{2}\braksq{\tr{\uu}\overline{\idgen}+\brak{\id\otimes\uu + h.c.}}
  \nonumber\\&\qquad
  + \zeta_{3}\ChimultHat{\uu}
  \nonumber\\&\qquad
  + \zeta_{4}\braksq{\ip{\NN}{\uu}\brak{\id\otimes\NN+h.c.} + \tr{\uu}\NN\otimes\NN}
  \nonumber\\&\qquad
  + \zeta_{5}\braksq{\brak{\id\otimes\brak{\Chimult{\NN}\cdot\uu} + h.c.} + \tr{\uu}\Chimult{\NN}}
  \nonumber\\&\qquad
  + \zeta_{6}\braksq{\brak{\NN\otimes\uu+h.c.}+\ip{\NN}{\uu}\idgen}
  \nonumber\\&\qquad
  + \zeta_{7}\braksq{\brak{\NN\otimes\brak{\Chimult{\NN}\cdot\uu} + h.c.} + \ip{\NN}{\uu}\Chimult{\NN}} \nonumber\\
  &= \zeta_{1}\tr{\uu}\id\otimes\id 
   + \brak{\zeta_{2}\tr{\uu}+\zeta_{6}\ip{\NN}{\uu}}\overline{\idgen}
   + \zeta_{2}\brak{\id\otimes\uu + h.c.}
   + \zeta_{3} \ChimultHat{\uu}
   \nonumber\\&\qquad
   + \zeta_{4} \braksq{\ip{\NN}{\uu}\brak{\id\otimes\NN + h.c.} + \tr{\uu}\NN\otimes\NN}
   + \zeta_{5} \braksq{\id\otimes\brak{\Chimult{\NN}\cdot\uu}+h.c.}
   \nonumber\\&\qquad
   + \zeta_{6} \brak{\NN\otimes\uu + h.c.}
   + \zeta_{7} \braksq{\NN\otimes\brak{\Chimult{\NN}\cdot\uu}+h.c.}
   + \brak{\zeta_{5}\tr{\uu}+\zeta_{7}\ip{\NN}{\uu}}\ChimultHat{\NN}  .
\end{align}
Whilst for an isotropic solid we needed three extra material-dependent constants to parametrise the third derivatives, here we need seven.

With this, we are ready to write down an expression for the elastic tensor's perturbation. Observe that
\begin{align}
  \Chimult{\uu}\cccc^{0} &= \brak{\cccc^{0}\Chimult{\uu}}^{T} \\
  \Chimult{\omb}\cccc^{0} &= \brak{\cccc^{0}\Chimult{\brak{-\omb}}}^{T}  .
\end{align}
%Given this, we can avoid significant clutter by introducing a shorthand that is standard in many areas of physics, whereby a symmetric expression is written as
%\begin{align}
%  \brak{\text{symmetric object}} = \brak{\text{not-necessarily-symmetric-object}} + h.c. .
%\end{align}
%Here, \( h.c. \) technically stands for `hermitian conjugate'. 
Thus,
\begin{align}
\label{eq:ti_elastic_tensor_calc_initial}
  \cccc^{1} 
  &= \tr{\uu}\braksq{
  		\brak{\zeta_{1}-\lambda}\id\otimes\id
  	   +\brak{\zeta_{2}-2\mu}\overline{\idgen}
  	   +\brak{\zeta_{4}-8\gamma}\NN\otimes\NN
  	   +\brak{\zeta_{5}+2\alpha}\ChimultHat{\NN}
  	   -4\beta\brak{\id\otimes\NN+h.c.}
  		}
  \nonumber\\&\qquad
   + \ip{\NN}{\uu}\braksq{
  	    \zeta_{6}\overline{\idgen}
  	   +\zeta_{7}\ChimultHat{\NN}
  	   +\zeta_{4}\brak{\id\otimes\NN+h.c.}
  		}
  \nonumber\\&\qquad
  + \brak{\zeta_{3}+4\mu}\ChimultHat{\uu}
  + \brak{\zeta_{2}+2\lambda}\brak{\id\otimes\uu+h.c.}
  + \brak{\zeta_{6}+8\beta}\brak{\NN\otimes\uu+h.c.}
  \nonumber\\&\qquad
  + \brak{\zeta_{5}+4\beta}\braksq{\id\otimes\brak{\Chimult{\NN}\cdot\uu}+h.c.}
  + \brak{\zeta_{7}+8\gamma}\braksq{\NN\otimes\brak{\Chimult{\NN}\cdot\uu}+h.c.}
  - 2\alpha\brak{\ChimultHat{\uu}\ChimultHat{\NN} + h.c.}
  \nonumber\\&\qquad
  + 4\mu \ChimultHat{\omb}
  + 8\gamma\braksq{\NN\otimes\brak{\Chimult{\omb}\cdot\NN} + h.c.}
  + 4\beta\braksq{\id\otimes\brak{\Chimult{\omb}\cdot\NN} + h.c.}
  - 2\alpha\brak{\ChimultHat{\omb}\ChimultHat{\NN} + h.c.}  .
\end{align}
It is now a matter of tedious algebra to complete the calculation. The nontrivial identities that we need are:
\begin{align}
  \Chimult{\NN}^{2} &= \Chimult{\NN} + 2\NN\otimes\NN \\
  \Chimult{\NN}\Chimult{\omb}\cdot\NN &= \Chimult{\omb}\cdot\NN \\
  \Chimult{\brak{\Chimult{\omb}\cdot\NN}} &= \Chimult{\omb}\Chimult{\NN} + h.c. \\
  \ChimultHat{\Chimult{\NN}\cdot\sig^{1}}\ChimultHat{\NN} + h.c. &=
   \ChimultHat{\Chimult{\NN}\cdot\sig^{1}}
   +2\braksq{\NN\otimes\brak{\Chimult{\NN}\cdot\sig^{1}}+h.c.}
   +2\ip{\NN}{\sig^{1}}\ChimultHat{\NN}  .
\end{align}
Finally, the perturbation to the elastic tensor is given by
\begin{align}
  \cccc^{1} &= 
    \eta_{1}\id\otimes\id
  + \eta_{2}\overline{\idgen} 
  + \eta_{3}\NN\otimes\NN 
  + \eta_{4}\brak{\id\otimes\NN + h.c.} 
  + \eta_{5}\ChimultHat{\NN}
  \nonumber\\&\qquad
  + \eta_{6}\ChimultHat{\sig^{1}} 
  + \eta_{7}\brak{\id\otimes\sig^{1} + h.c.}
  + \eta_{8}\brak{\NN\otimes\sig^{1} + h.c.}
  + \eta_{9}\braksq{\id\otimes\brak{\Chimult{\NN}\cdot\sig^{1}}+h.c.}
  \nonumber\\&\qquad
  + \eta_{10}\braksq{\NN\otimes\brak{\Chimult{\NN}\cdot\sig^{1}}+h.c.}
  + \eta_{11}\ChimultHat{\Chimult{\NN}\cdot\sig^{1}}
  + \eta_{12}\brak{\ChimultHat{\sig^{1}}\ChimultHat{\NN} + h.c.}
  \nonumber\\&\qquad
  + \eta_{13}\ChimultHat{\omb}
  + \eta_{14}\braksq{\id\otimes\brak{\Chimult{\omb}\cdot\NN}+h.c.}
  + \eta_{15}\braksq{\NN\otimes\brak{\Chimult{\omb}\cdot\NN}+h.c.}
  + \eta_{16}\brak{\ChimultHat{\omb}\ChimultHat{\NN}+h.c.} .
\end{align} 
The \( \brakbr{\eta_{i}} \) are defined as
\begingroup
\allowdisplaybreaks
\begin{subequations}
\label{eqs:eta_defn}
\begin{align}
    %%%%%%%%%%%%%%%%%%%%%%%%%%%%%%%%%%%%%%%%%%%%%%%%%%%%%%%%%
	2\tilde{\mu}\eta_{1}  &= 
	2\tilde{\mu}\tr{\uu}\brak{\zeta_{1}-\lambda} 
	- \braksq{2\tilde{\eta}\brak{\zeta_{2}+2\lambda}} \\
	%%%%%%%%%%%%%%%%%%%%%%%%%%%%%%%%%%%%%%%%%%%%%%%%%%%%%%%%%
	2\tilde{\mu}\eta_{2}  &= 
	2\tilde{\mu}\braksq{\brak{\zeta_{2}-2\mu}\tr{\uu}+\zeta_{6}\ip{\NN}{\uu}} 
	- \braksq{2\tilde{\eta}\brak{\zeta_{3}+4\mu}} \\
	%%%%%%%%%%%%%%%%%%%%%%%%%%%%%%%%%%%%%%%%%%%%%%%%%%%%%%%%%
	2\tilde{\mu}\eta_{3}  &= 
	2\tilde{\mu}\tr{\uu}\brak{\zeta_{4}-8\gamma} 
	+ \braksq{2\tilde{R}\brak{\zeta_{6}+8\beta}-8\alpha\tilde{R}+\brak{4\tilde{R}-4\tilde{\eta}+4\tilde{P}\ip{\NN}{\sig^{1}}}\brak{\zeta_{7}+8\gamma}} \\
	%%%%%%%%%%%%%%%%%%%%%%%%%%%%%%%%%%%%%%%%%%%%%%%%%%%%%%%%%
	2\tilde{\mu}\eta_{4}  &= 
	2\tilde{\mu}\braksq{\zeta_{4}\ip{\NN}{\uu}-4\beta\tr{\uu}} 
	+ \braksq{\tilde{R}\brak{\zeta_{2}+2\lambda}-\tilde{\eta}\brak{\zeta_{6}+8\beta}+\brak{2\tilde{P}\ip{\NN}{\sig^{1}}+2\tilde{R}-2\tilde{\eta}}\brak{\zeta_{5}+4\beta}}  \\
	%%%%%%%%%%%%%%%%%%%%%%%%%%%%%%%%%%%%%%%%%%%%%%%%%%%%%%%%%
	2\tilde{\mu}\eta_{5}  &= 
	2\tilde{\mu}\braksq{\brak{\zeta_{5}+2\alpha}\tr{\uu}+\zeta_{7}\ip{\NN}{\uu}}
	+\braksq{\tilde{R}\brak{\zeta_{3}+4\mu}-2\alpha\brak{2\tilde{R}-4\tilde{\eta}+2\tilde{P}\ip{\NN}{\sig^{1}}}}\\
	%%%%%%%%%%%%%%%%%%%%%%%%%%%%%%%%%%%%%%%%%%%%%%%%%%%%%%%%%
	2\tilde{\mu}\eta_{6}  &= \zeta_{3}+4\mu \\
	2\tilde{\mu}\eta_{7}  &= \zeta_{2}+2\lambda \\
	2\tilde{\mu}\eta_{8}  &= \zeta_{6}+8\beta \\
	%%%%%%%%%%%%%%%%%%%%%%%%%%%%%%%%%%%%%%%%%%%%%%%%%%%%%%%%%
	2\tilde{\mu}\eta_{9}  &= 
    \tilde{P}\brak{\zeta_{2}+2\lambda}+\brak{1+\tilde{P}}\brak{\zeta_{5}+4\beta} \\
	%%%%%%%%%%%%%%%%%%%%%%%%%%%%%%%%%%%%%%%%%%%%%%%%%%%%%%%%%
	2\tilde{\mu}\eta_{10} &= 
	\tilde{P}\brak{\zeta_{6}+8\beta-4\alpha}+\brak{1+\tilde{P}}\brak{\zeta_{7}+8\gamma} \\
	%%%%%%%%%%%%%%%%%%%%%%%%%%%%%%%%%%%%%%%%%%%%%%%%%%%%%%%%%
	2\tilde{\mu}\eta_{11} &= \tilde{P}\brak{\zeta_{3}+4\mu-2\alpha} \\
	2\tilde{\mu}\eta_{12} &= -2\alpha \\
    2\tilde{\mu}\eta_{13} &= 8\mu\tilde{\mu} \\
    %%%%%%%%%%%%%%%%%%%%%%%%%%%%%%%%%%%%%%%%%%%%%%%%%%%%%%%%%
	2\tilde{\mu}\eta_{14} &= 
	8\beta\tilde{\mu} - q^{0}\brak{1+\tilde{P}}\brak{\zeta_{2}+\zeta_{5}+2\lambda+4\beta} \\
	%%%%%%%%%%%%%%%%%%%%%%%%%%%%%%%%%%%%%%%%%%%%%%%%%%%%%%%%%
	2\tilde{\mu}\eta_{15} &= 
	16\gamma\tilde{\mu} - q^{0}\brak{1+\tilde{P}}\brak{\zeta_{6}+\zeta_{7}-4\alpha+8\beta+8\gamma} \\
	%%%%%%%%%%%%%%%%%%%%%%%%%%%%%%%%%%%%%%%%%%%%%%%%%%%%%%%%%
	2\tilde{\mu}\eta_{16} &= -\braksq{
	4\alpha\tilde{\mu}+q^{0}\brak{1+\tilde{P}}\brak{\zeta_{3}+4\mu-2\alpha}} ,
\end{align}
\end{subequations}%
\endgroup
definitions which should in turn be combined with the earlier definitions eqs.(\ref{eqs:ti_tilde_constants_defns},\ref{eq:tru_Nu_relation},\ref{eq:eta_theta_defn},\ref{eqs:PQ_defns},\ref{eq:RTilde_defn},\ref{eq:strain_energy_third_deriv_param_zeta_defn}). The perturbation to the elastic tensor is given in terms of
\begin{enumerate}
\renewcommand{\theenumi}{(\arabic{enumi})}
	\item the transversely-isotropic constants \( \lambda \), \( \mu \), \( \alpha \), \( \beta \) and \( \gamma \), as well as \( \nub \);
	\item the constants \( p^{0} \) and \( q^{0} \) which parametrise the zeroth-order equilibrium stress;
	\item the seven constants \( \brakbr{\zeta_{i}} \) defined in eq.(\ref{eq:strain_energy_third_deriv_param_zeta_defn}), which parametrise the third derivatives of a transversely-isotropic strain-energy function about equilibrium;
	\item the small induced stress \( \sig^{1} \);
	\item the arbitrary 2-parameter antisymmetric matrix \( \omb \).
\end{enumerate}

\section{The isotropic elastic moduli, their pressure-derivatives, and wave speeds}
\label{app:matching_theories}

Working under the assumption of isotropic background states, this section compares our linearised theory with the theories of \citet{dahlen_1972b} and \citet{Tromp_2018}. We demonstrate that the theories are compatible for certain choices of parameter and, moreover, that our parameters \( c \) and \( d \) \wording{are not readily comparable} with \citeauthor{Tromp_2018}'s \( \kappa' \) and \( \mu' \). For this section we will work in the notation of Section \ref{sec:intro}, using the Lam\'{e} parameter \( \lambda \) \wording{interchangeably} with \( \kappa \).

Recall from Section \ref{sec:intro} that the most general isotropic elastic tensor \( \Xib \) is written, upon inducing an incremental stress, as
\begin{align}
\label{eq:hopefully_last_label}
  \Xi_{ijkl} &= \brak{\kappa-\frac{2}{3}\mu+\Delta p^{0}c}\delta_{ij}\delta_{kl}
  			  +\brak{\mu+\Delta p^{0}d}\brak{\delta_{ik}\delta_{jl}+\delta_{il}\delta_{jk}}
  			  \nonumber\\
  			  &\qquad
  			  +a\brak{\delta_{ij}\Delta\tau^{0}_{kl}+\delta_{kl}\Delta\tau^{0}_{ij}}
  			  +b\brak{\delta_{ik}\Delta\tau^{0}_{jl}+\delta_{il}\Delta\tau^{0}_{jk}
  			  		 +\delta_{jk}\Delta\tau^{0}_{il}+\delta_{jl}\Delta\tau^{0}_{ik}}
\end{align}
and that we then obtained \citeauthor{Tromp_2018}'s \( \Xib \) by setting
\begin{align}
  c &= -2a \\
  d &= -2b
\end{align}
and
\begin{align}
  a &= \frac{1}{2}+\frac{1}{3}\mu'-\frac{1}{2}\kappa' \\
  b &= -\frac{1}{2}-\frac{1}{2}\mu'
\end{align}
with the constants \( \kappa' \) and \( \mu' \) interpreted as adiabatic pressure-derivatives of the elastic moduli. These equations can be considered as a four-dimensional system
\begin{subequations}
\label{eqs:t_t_matching_eqns}
	\begin{align}
	  c &= \lambda'-1  \label{eq:t_t_matching_eqns_1} \\
	  d &= \mu'+1         \label{eq:t_t_matching_eqns_2} \\
	  c &= -2a            \label{eq:t_t_matching_eqns_3} \\
	  d &= -2b                \label{eq:t_t_matching_eqns_4}
	\end{align}
\end{subequations}
parametrised by \( \lambda' \) and \( \mu' \), where
\begin{align}
  \lambda' \equiv \kappa'-\frac{2}{3}\mu' .
\end{align}
But in eqs.\eqref{eqs:iso_constants_defns} we showed that \( a \), \( b \), \( c \) and \( d \) are given in terms of the \textit{three} Murnaghan constants. Under what conditions are the two theories compatible?

Eqs.(\ref{eq:t_t_matching_eqns_3},\ref{eq:t_t_matching_eqns_4}) just constrain eqs.(\ref{eqs:iso_constants_defns}), effectively fixing two of the Murnaghan constants. Let us define two shorthands
\begin{align}
  y &= \frac{\lambda+p^{0}}{\mu-p^{0}} \\
  z &= \frac{2\lambda+\mu+p^{0}}{\mu-p^{0}} .
\end{align} 
The explicit forms of eqs.(\ref{eqs:iso_constants_defns}) can then be used to write eqs.(\ref{eq:t_t_matching_eqns_3},\ref{eq:t_t_matching_eqns_4}) as
\begin{align}
  \begin{pmatrix}
  	1 & -y & 0 \\
  	0 &  1 & -y
  \end{pmatrix}
  \begin{pmatrix}
  	\zeta_{1} \\
  	\zeta_{2} \\
  	\zeta_{3}
  \end{pmatrix}
  =
  z
  \begin{pmatrix}
  	\lambda \\
  	2\mu
  \end{pmatrix} ,
\end{align}
after which we may solve for \( \zeta_{2,3} \) in terms of \( \zeta_{1} \), finding that
\begin{align}
  \begin{pmatrix}
  	\zeta_{2} \\
  	\zeta_{3}
  \end{pmatrix}
  =
  \frac{1}{y^{2}}
  \begin{pmatrix}
  	-y & 0 \\
  	 1 & -y
  \end{pmatrix}
  \begin{pmatrix}
  	z & 0  & -1 \\
  	0 & 2z &  0
  \end{pmatrix}
  \begin{pmatrix}
  	\lambda \\
  	\mu \\
  	\zeta_{1}
  \end{pmatrix} .
\end{align}
This expression is written in a rather strange way, but we have found it highly amenable to algebraic manipulation. One free parameter, \( \zeta_{1} \), remains with which to satisfy eqs.(\ref{eq:t_t_matching_eqns_1},\ref{eq:t_t_matching_eqns_2}). Using eqs.(\ref{eqs:iso_constants_defns}) once again, after some algebra we find from eqs.(\ref{eq:t_t_matching_eqns_1},\ref{eq:t_t_matching_eqns_2}) that the constants \( \lambda' \) and \( \mu' \) are given by
\begin{align}
\label{eq:t_t_theory_matching_constraint}
  \begin{pmatrix}
  	\lambda' \\
  	\mu'
  \end{pmatrix}
  =
  \begin{pmatrix}
  	1 \\
  	-1
  \end{pmatrix}
  -\frac{1}{3\kappa+p^{0}}
  \begin{pmatrix}
  	1-2\frac{z}{y}                    & 0 & 3+\frac{2}{y} \\
  	-\brak{\frac{3}{2}y+1}\frac{z}{y^{2}} & 1-2\frac{z}{y} & \brak{\frac{3}{2}y+1}\frac{1}{y^{2}}
  \end{pmatrix}
  \begin{pmatrix}
  	\lambda \\
  	\mu \\
  	\zeta_{1}
  \end{pmatrix} .
\end{align}
As \( \zeta_{1} \) is arbitrary, this shows that there exists a one-parameter family of values of \( \lambda' \) and \( \mu' \) whereby both eqs.(\ref{eqs:t_t_matching_eqns}) and eqs.(\ref{eqs:iso_constants_defns}) are satisfied and \citeauthor{Tromp_2018}'s theory thus agrees with ours. Moreover, this will hold for any \wording{physically allowed} \( \lambda \), \( \mu \) and \( p^{0} \). It formalises our earlier assertion that their theory is valid for a certain class of isotropic materials. The theory of \citet{dahlen_1972b} is \wording{formally} obtained from that of \citet{Tromp_2018} by setting \( \kappa'=\mu'=0 \). However, now that two free parameters have been removed, eq.(\ref{eq:t_t_theory_matching_constraint}) reduces to the expression
\begin{align}
  f\brak{\lambda,\mu,p^{0}} = 0 
\end{align}
for some real-valued function \( f \), and the elastic moduli themselves are constrained.

One small puzzle remains. From looking at eq.(\ref{eq:hopefully_last_label}), our two constants \( c \) and \( d \) might be thought to represent the pressure-derivatives of the elastic moduli, but eqs.(\ref{eq:t_t_matching_eqns_1},\ref{eq:t_t_matching_eqns_2}) make clear that they cannot be equivalent to \citeauthor{Tromp_2018}'s \( \kappa' \) and \( \mu' \). In order to better understand this discrepancy we should focus on the pressure dependence of P- and S-wave speeds. We can do that by neglecting deviatoric stress increments and considering the Christoffel operator (defined in eq.\ref{eq:christoffel_defn} and written here with \( \hat{\mathbf{k}} \) substituted for \( \hat{\mathbf{p}} \))
\begin{align}
  \rho B_{jl} &=
  \braksq{\brak{\kappa+\frac{1}{3}\mu}+\Delta p^{0}\brak{c+d}}\hat{k}_{j}\hat{k}_{l}
  + \braksq{\mu-\Delta p^{0}\brak{1-d}}\delta_{jl}
  + \brak{\text{deviatoric}} .
\end{align}
This gives wave speeds of 
\begin{align}
  	\rho c_{P}^{2} &= \braksq{\brak{\lambda + \Delta p^{0} c} 
  	                       + 2\brak{\mu + \Delta p^{0} d}} 
  	                       - \Delta p^{0} \\
    \rho c_{S}^{2} &= \braksq{\mu + \Delta p^{0} d} - \Delta p^{0} .
\end{align} 
The wave speeds' dependence on \( \Delta p^{0} \) comes not only from terms in \( c \) and \( d \) that belong to \( \Xib \), but also from the term \( \Delta T^{0}_{ik}\delta_{jl} \) that is added to \( \Xib \) in order to form \( \Lam \) (and hence the Christoffel operator). The construction of the theories of Dahlen, Tromp and Trampert is such that the hydrostatic contribution of \( \Delta T^{0}_{ik}\delta_{jl} \) is always cancelled out. For example, the components of \citeauthor{Tromp_2018}'s \( \Lam \) are found to be
\begin{align}
\label{eq:Lambda_moduli_derivs}
  \Lambda_{ijkl} = \brak{\lambda+\Delta p^{0}\lambda'}\delta_{ij}\delta_{kl}
  +\brak{\mu+\Delta p^{0}\mu'}\brak{\delta_{ik}\delta_{jl}+\delta_{il}\delta_{jk}} + \Delta p^{0}\brak{\delta_{il}\delta_{jk}-\delta_{ij}\delta_{kl}}
      +\brak{\text{deviatoric}} .
\end{align}
The third term vanishes upon contraction with \( \hat{k}_{i}\hat{k}_{k} \) and can therefore be neglected when forming the Christoffel operator. This leaves us with the neatly defined elastic moduli and derivatives within the first two terms, and we ultimately arrive at \citeauthor{Tromp_2018}'s \wording{``quasi-classical"} expressions for the wave speeds
\begin{align}
  	\rho c_{P}^{2} &= \brak{\lambda+\lambda'\Delta p^{0}} + 2\brak{\mu+\mu'\Delta p^{0}} \\ 
    \rho c_{S}^{2} &= \brak{\mu+\mu'\Delta p^{0}} .
\end{align}  
In effect, by using this definition of the elastic moduli's pressure derivatives one is ``redefining" the elastic moduli in order to subsume the pressure-dependence introduced by \( \Delta T^{0}_{ik}\delta_{jl} \).

\end{document}